\documentclass[pra,twocolumn,showpacs,amsmath,amssymb,superscriptaddress,nofootinbib]{revtex4-2}
\usepackage{diagbox}
\usepackage{braket}
\usepackage{cases}
\usepackage{color}
\usepackage{graphicx,subfigure}%
\usepackage{float}
\usepackage{bm}
\usepackage{adjustbox}
\usepackage{hyperref}
\hypersetup{
    colorlinks=true, 
    linkcolor=cyan,
    citecolor=magenta, 
    filecolor=magenta, 
    urlcolor=cyan,
    runcolor=cyan
}
\usepackage[capitalise]{cleveref}

\usepackage{commath}
\usepackage{color}
\usepackage{amsmath}

\def\be{\begin{equation}}
\def\ee{\end{equation}}

\newcommand{\bes} {\begin{subequations}}
\newcommand{\ees} {\end{subequations}}
\newcommand{\beq}{\begin{equation}}
\newcommand{\eeq}{\end{equation}}

\def\a{\alpha}

\def\s{\sigma}

\def\o{\omega}

\def\>{\rangle}
\def\<{\langle}

\newcommand{\ketb}[2]{|{#1}\>\!\<#2|}

\begin{document}

\title{Modeling low- and high-frequency noise in transmon qubits with resource-efficient measurement}

\author{Vinay Tripathi}
\affiliation{Department of Physics \& Astronomy, University of Southern California,
Los Angeles, California 90089, USA}
\affiliation{Center for Quantum Information Science \& Technology, University of
Southern California, Los Angeles, California 90089, USA}
\author{Huo Chen}
\affiliation{Center for Quantum Information Science \& Technology, University of
Southern California, Los Angeles, California 90089, USA}
\affiliation{Department of Electrical \& Computer Engineering, University of Southern California,
Los Angeles, CA 90089, USA}
\author{Eli Levenson-Falk}
\affiliation{Department of Physics \& Astronomy, University of Southern California,
Los Angeles, California 90089, USA}
\affiliation{Center for Quantum Information Science \& Technology, University of
Southern California, Los Angeles, California 90089, USA}
\author{Daniel A. Lidar}
\affiliation{Department of Physics \& Astronomy, University of Southern California,
Los Angeles, California 90089, USA}
\affiliation{Center for Quantum Information Science \& Technology, University of
Southern California, Los Angeles, California 90089, USA}
\affiliation{Department of Electrical \& Computer Engineering, University of Southern California,
Los Angeles, CA 90089, USA}
\affiliation{Department of Chemistry, University of Southern California, Los Angeles,
CA 90089, USA}

\date{\today}

\begin{abstract}
Transmon qubits experience open system effects that manifest as noise at a broad range of frequencies. We present a model of these effects using the Redfield master equation with a hybrid bath consisting of low and high-frequency components. We use two-level fluctuators to simulate $1/f$-like noise behavior, which is a dominant source of decoherence for superconducting qubits. By measuring quantum state fidelity under free evolution with and without dynamical decoupling (DD), we can fit the low- and high-frequency noise parameters in our model. We train and test our model using experiments on quantum devices available through IBM quantum experience. Our model accurately predicts the fidelity decay of random initial states, including the effect of DD pulse sequences. We compare our model with two simpler models and confirm the importance of including both high-frequency and $1/f$ noise in order to accurately predict transmon behavior.


\end{abstract}

\maketitle
\section{Introduction}

Quantum computing based on superconducting qubits has made significant progress. Starting with the first implementations of superconducting qubits~\cite{Nakamura:99, Mooij:99}, the field has developed several flavors of qubits, broadly classified as charge, flux, and phase qubits~\cite{Clarke2008}. However, the real workhorse behind many of the recent critical developments~\cite{Hacohen2016, Kandala:2017aa, Minev2019, Arute:2019aa, Havlicek2019, Kandala2019, Campagne2020, Andersen2020, Arute2020, wu2021strong} in gate-based quantum computing is the transmon qubit~\cite{transmon-invention}. Transmons are designed by adding a large shunting capacitor to charge qubits, the result being that they are almost insensitive to charge noise. Transmon-based cloud quantum computers (QCs) are now widely available to the broad research community for proof of principle quantum computing experiments~\cite{Devitt2016, Wootton:2018aa, Vuillot2018, Roffe:2018aa, Willsch2018,  Pokharel2018, Harper:2019aa,Maslov:2021aa,Zhang_2021,pokharel2022demonstration,Pokharel:better-than-classical-Grover}. 

Quantum computers in their current form have high error rates. 
This includes coherent errors (originating from imperfect gates), state preparation and measurement errors (SPAM), and incoherent errors (environment-induced noise). The latter, which results in dephasing and relaxation errors, is a pernicious problem in quantum information processing. Characterizing and modeling these open quantum system effects is crucial for advancing the field and improving the prospects of fault-tolerant quantum computation~\cite{Aliferis:05, Chao:2017ab, Campbell:2017aa}. Various procedures for modeling decoherence and control noise affecting idealized qubits have been discussed~\cite{souza2020process, Harper2020, Georgopoulos2021}. Still, modeling noise effects from first principles, i.e., starting at the circuit level of transmons and including $1/f$ noise, is relatively unexplored~\cite{McCourt2022}.  

In this work, we develop a framework to model environment-induced noise effects on a transmon qubit using the master equation formalism. We use a hybrid quantum bath with an Ohmic-like noise spectrum to model dephasing and relaxation processes in multi-level transmons. We also include classical fluctuators and use a hybrid Redfield model to account for both low- ($1/f$) and high-frequency noise. We develop a simple noise learning procedure relying on dynamical decoupling (DD)~\cite{Viola:98, Duan:98e, Vitali:99, Zanardi:1999fk,Viola:99} to obtain the noise parameters (see Ref.~\cite{Byl11a} for early experimental work in this area). Our procedure relies only on measurements of quantum state fidelity with and without a single type of DD sequence, and so is quite resource-efficient compared to protocols requiring full quantum state tomography or DD-based spectral analysis. We test our noise model via fidelity preservation experiments on IBMQE processors for random initial states and find that the model can correctly capture these experiments. The model is, moreover, capable of reproducing the effects of time-dependent dynamical decoupling pulses on the main qubit. Finally, we compare the predictions based on our model with two simpler models using ideal two-level qubits, excluding the fluctuators and assuming ideal, zero-width DD pulses. In contrast to our complete model, these simpler models fail to capture noise simultaneously in both the low and high-frequency regimes. As a result, whether with or without DD, they underperform in capturing fidelity preservation experiments.


This paper is organized as follows. In \cref{Sec-numerics}, we develop our numerical method focusing on simulating multi-level transmon qubit and single-qubit gates, which form the DD sequences. Next, we discuss our open quantum system in \cref{Sec-OQS} and describe our noise learning method in \cref{sec-methodology}. We then test our learned model on Quito using DD experiments with random initial single-qubit states in \cref{sec:results}. We extend our method to Lima, which relies on a different calibration procedure, in \cref{sec:lima-results}. We conclude in \cref{Sec-conclusion}. The appendix provides additional details and calculations in support of the main text.

\section{Numerical model of transmons}
\label{Sec-numerics}

In this section, we focus on the circuit-level description of the transmon qubit that we use in our model. We start with the transmon Hamiltonian and find an effective Hamiltonian to simulate single-qubit time-dependent microwave gates. We include the Derivative Removal of Adiabatic Gates (DRAG)~\cite{Motzoi2009} technique in our numerical model. DRAG is a state-of-the-art technique used to improve the performance of single-qubit gates by suppressing leakage and phase errors. The former refers to non-zero population of non-computational levels at the end of a pulse, while the latter is a type of coherent error that results from the non-zero population in non-computational levels \emph{during} the pulse: the admixture of such levels leads to a phase shift of the computational levels, resulting in a net phase error at the end of the pulse. By including the DRAG technique (used in the IBMQE devices) and considering the residual errors it is unable to suppress, we more accurately model the transmon behavior.

\subsection{Transmon Hamiltonian}
The Hamiltonian of a fixed-frequency transmon qubit is~\cite{transmon-invention}: 
\begin{equation}
H_{\rm{trans}}=4 E_{C}\left(\hat{n}-n_{g}\right)^{2}-E_{J} \cos \hat{\varphi}
\label{eq:trans-Ham}\ .
\end{equation}
We work in units where $\hbar=1$.
$E_C = e^2/(2C)$ is the charging energy ($C$ is the capacitance, and $e$ is the electron charge), $E_J = I_C/(2e)$ is the potential energy of the Josephson junction ($I_C$ is the critical current of the junction) and $n_g$ represents the charge offset number which can result in charge noise. In the operating regime of a transmon qubit, i.e., $E_J/E_C\gg 1$, the lowest few energy levels of the transmon are almost immune to charge noise, in which case $n_g$ can be safely ignored. The two operators $\hat{n}$ and $\hat{\varphi}$ are a canonically conjugate pair analogous to momentum and position. They satisfy the commutation relation $\left[\hat{n},\hat{\varphi}\right] = i$; $\hat{n}$ is the number operator for the Cooper pairs transferred between the superconducting islands of the Josephson junction, and $\hat{\varphi}$ is the gauge invariant phase difference across the Josephson junction, i.e., between the islands.

\subsection{Time-dependent drives}
\label{sec:time-dependent-drives}

To numerically simulate the time-dependent drive-pulses or gates, we start with \cref{eq:trans-Ham} and write it in the charge basis (the eigenbasis of $\hat{n}$) such that the number of Cooper pairs takes values from $-n_{\max}$ to $n_{\max}$. Eq.~(\ref{eq:trans-Ham}) thus reduces to
\begin{align}
H_{\rm{trans}}=~& 4 E_{C}\sum_{-n_{\max}}^{n_{\max}}{n^2}|n\rangle\langle n| \nonumber\\
& -\frac{E_{J}}{2} \sum_{-n_{\max}}^{n_{\max}}\left(|n\rangle\langle n+1| + |n+1\rangle\langle n|\right)
\label{eq:trans-Ham-chargebasis}\ 
\end{align}
where we have taken $n_g = 0$ since we are in the transmon regime. We truncate to $n_{\max}$ (later we set  $n_{\max}= 50$) and diagonalize the resulting Hamiltonian:

\begin{equation}
H_{\mathrm{trans}}^{\mathrm{eigen}}=S H_{\rm{trans }} S^\dag = \sum_{k\ge 0}\omega_k \ketb{k}{k}\ ,
\label{eq:eig-transmon}
\end{equation}
where $\omega_{k}$ for $k= 0,1,...$ represents the energy of the $k^{\rm{th}}$ 
level in the transmon eigenbasis, and $S$ is the unitary similarity transformation. The eigenfrequencies are $\o_{ij} \equiv \o_i - \o_j$. The \emph{bare qubit frequency} is $\omega_{q} \equiv \omega_{10}$ and the \emph{anharmonicity} is $\eta_{q} \equiv \omega_{10} - \omega_{21}$. Since $\omega_{q}$ and $\eta_{q}$ are the two quantities accessible via experiments, we use these values to obtain the fitting parameters $E_C$ and $E_J$ in \cref{eq:trans-Ham}, which is the starting point of our transmon model.\footnote{In more detail, we try different values of $E_J$ and $E_C$ in \cref{eq:trans-Ham} by diagonalizing the corresponding $H_{\rm{trans}}$ and comparing the result with the experimental values of $\omega_q$ and $\eta_q$. Once we find the values of $E_J$ and $E_C$ yielding the closest match, we proceed to \cref{eq:eig-transmon} to find the full spectrum.}

Next, we add the coupling to the microwave drive, which couples to the transmon charge operator.  
The total system Hamiltonian can then be written as 
\begin{equation}
H_{\mathrm{sys}} = H_{\mathrm{trans}}^{\mathrm{eigen}} + \varepsilon(t) \cos \left(\omega_{\rm d} t+\phi_{\rm d}\right) \hat{n}\ ,
\label{eq:transmon-drive}
\end{equation}
where $\varepsilon(t)$ is the pulse envelope, $\omega_{\rm d}$ is the drive frequency, and $\phi_{\rm d}$ is the phase of the drive. 
We can simplify \cref{eq:transmon-drive} by first writing the charge operator in the transmon eigenbasis of \cref{eq:eig-transmon}, i.e., $\hat{n} = \sum_{k,k'}\bra{k}\hat{n}\ket{k^{\prime}}\ketb{k}{k'}$ and considering the charge coupling matrix elements. Only nearest-level couplings $\bra{k}\hat{n}\ket{k\pm 1}$ are found to be non-negligible, allowing us to ignore all higher order terms:
\beq
\hat{n} \approx \sum_{k\ge 0}\bra{k}\hat{n} \ket{k+1}\ketb{k}{k+1} + {\rm h.c.}
\label{eq:hatp}
\eeq
Transforming into a frame rotating with the drive and employing the rotating wave approximation (RWA), we obtain, for $\phi_{\rm d} = 0$, the effective Hamiltonian
\begin{align}
\label{eq:Xgate}
\tilde{H}_{\mathrm{sys}}&=\sum_{k\ge 0}\left(\omega_{k}-k \omega_{\rm d}\right)\ketb{k}{k} \\
&\ \ +\frac{ \varepsilon(t)}{2} \sum_{k\ge 0} g_{k, k+1}(\ketb{k}{k+1} + \ketb{k+1}{k})\ ,\nonumber
\end{align}
where 
$g_{k, k+1} \equiv \bra{k}\hat{n}\ket{k+1} $. By tuning $\phi_{\rm d}$, we can implement a rotation about any axis in the $(x,y)$ plane of the qubit subspace (after an additional projection). In particular, taking $\phi_{\rm d} = 0$ or $\pi/2$ corresponds to a rotation about the $x$ or $y$ axis, respectively. 
Appendix \ref{app:derivation} provides a derivation of \cref{eq:Xgate} from \cref{eq:transmon-drive}.

The pulse envelope $\varepsilon(t)$ plays a vital role in the final implementation of the gate. Since we are interested mainly in applying $\pi$ pulses, we choose
\begin{equation}
\int_{0}^{t_{g}} \varepsilon(t) d t=\pi \ ,
\label{eq:pi}
\end{equation}
where $t_g$ is the pulse or gate duration. For our numerical simulations, we choose Gaussian-shaped pulses with envelopes given by 
\begin{equation}
\varepsilon(t)=\varepsilon\left[G(t,t_g,\s) - G(0,t_g,\s)\right]\left(\Theta(t)-\Theta\left(t-t_{g}\right)\right)\ ,
\label{eq:pulse}
\end{equation}
where
\begin{equation}
G(t,t_g,\s) = \exp \left(-\frac{\left(t-t_{g} / 2\right)^{2}}{2 \s^{2}}\right) .
\end{equation}
Here $\varepsilon$ is the maximum drive amplitude during the pulse, $\Theta(t)$ is the step function, and $\s$ is the standard deviation of the Gaussian pulse.

\begin{figure}[t]
	\includegraphics[width=0.9\linewidth]{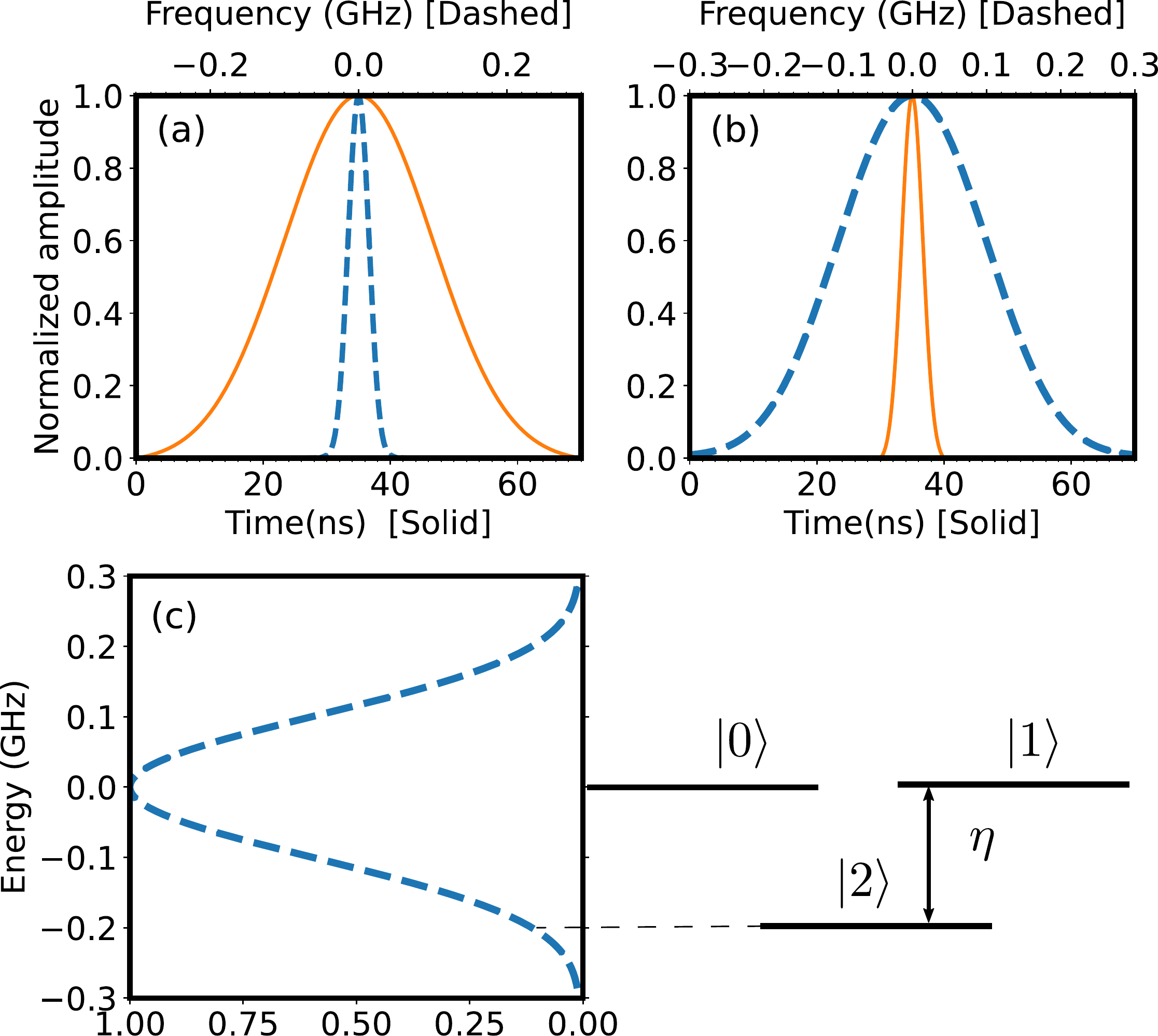}
	\caption{Gaussian pulse envelope (solid, orange) [see \cref{eq:pulse}] and its Fourier transform (dashed, blue) with amplitude $\varepsilon$ chosen to keep both in the range $[0,1]$.  (a) and (b) show the pulse with gate time $t_g=70$ and $10~$ns, respectively, and $\sigma=t_g/6$. The bottom horizontal axis represents time in ns, and the top horizontal axis represents frequency in GHz. Shorter gate times result in a larger frequency spread of the spectrum, with associated larger leakage, as illustrated in (c), which shows the frequency spectrum corresponding to a $t_g=10$~ns gate (left), compared to the energy levels (right) $\ket{0}$, $\ket{1}$ and $\ket{2}$ of the transmon. The energy levels are shown in the rotating frame such that $E_{\ket{0}} = E_{\ket{1}}$ and $E_{\ket{2}} = -\eta_q = -200$ MHz. As indicated by the dashed horizontal line, the spectrum overlaps with level $\ket{2}$, resulting in leakage into this level from the $\{\ket{0},\ket{1}\}$ qubit subspace. The sampling frequency used to compute the Fourier transform is $10$GS/s, which is state-of-the-art in experiments; the pulses that control the IBM processors used in this work have a sampling frequency of $\sim 5$GS/s.}  
\label{fig-pulse}
\end{figure}

An essential aspect of gate design is that population should not leak to higher levels of the transmon, i.e., the drive pulses should be bandwidth-limited (adiabatic).
An accurate measure of these off-resonant excitations is the Fourier transform of the pulse envelope at the detuning frequencies~\cite{freeman1998spin,Motzoi2013-sx}.
For example, consider a Gaussian pulse with standard deviation $\s$. Its Fourier transform has a standard deviation proportional to $1/\s$, which means that the drive pulse applied at the qubit frequency $\o_q$ has a frequency spread close to $1/\s$ about $\o_q$. If $1/\s$ is of the order of the anharmonicity of the transmon, the pulse spectral width will overlap with some of the higher-level transitions. \cref{fig-pulse} shows the Gaussian pulse envelope and its Fourier transform, and illustrates how choosing a shorter gate time results in a larger frequency spread and vice versa. The use of DRAG pulses mitigates this leakage, as discussed further below.

In the two-state (qubit) subspace, \cref{eq:Xgate} reduces to 
\begin{equation}
H_{X}(t) = \frac{\varepsilon(t)}{2}\left(\ketb{0}{1} + \ketb{1}{0} \right) = \frac{\varepsilon(t)}{2} \s^x,
\end{equation}
where $g_{0,1} = g$ has been absorbed into $\varepsilon(t)$ and $\s^x$ is the Pauli $X$ matrix. When $H_{X}(t)$ is evolved for a time $t_g$ such that \cref{eq:pi} is satisfied, the resulting unitary is an ideal $X_{\pi}$ gate. To include the effect of higher levels, we first use the full gate Hamiltonian from \cref{eq:Xgate} and then project the result to the qubit subspace.

The gate fidelity averaged over all input states in the qubit Hilbert space can be written as the average over the six polar states (i.e., the six eigenstates of $\sigma^x,\sigma^y$, and $\sigma^z$)~\cite{Bowdrey:2002aa,Motzoi2009}:
\begin{equation}
F_{g}=\frac{1}{6} \sum_{j=\pm x, \pm y, \pm z} \operatorname{Tr}\left[U_{\rm{ideal }} \rho_{j}^{1 \mathrm{q}} U_{\rm{ideal }}^{\dagger} \Pi[\rho(t_g)]\right]\ ,
\end{equation}
where $\rho_{j}^{1 \mathrm{q}}$ is the single qubit density matrix, and $U_{\rm ideal}$ represents the ideal unitary corresponding to the gate we wish to study.  $\Pi[\rho(t_g)]$ is the projection of the full density matrix onto the single qubit subspace. $F_g$ compares the density matrix $\rho$ after application of the gate (i.e., at $t=t_g$) with the expected density matrix in the qubit subspace. 

To reduce phase errors caused by the presence of additional levels, a commonly used trick to implement single qubit gates such as $X_\pi$ is to break the gate into two halves where each half performs a $\pi/2$ rotation, accompanied by some virtual $Z$ rotations~\cite{Mckay2017, Mckay2018}. Numerically, we observe that with four levels included in the transmon Hamiltonian and a total gate duration $t_g = 70 \;{\rm ns}$, with $\s=t_g/6$, the average single-qubit gate error $1-F_g$ can be suppressed by around $20 \%$ if we use two such pulses instead of a single long pulse. The exact quantitative improvement depended on other model parameters. We observed this error reduction in a closed system setting with no environmental coupling, and so any fidelity improvement may be counteracted by open-system effects. For a detailed numerical study on the error of time-dependent gates with transmon qubits in the open system settings, see Ref.~\cite{Babu2020}.

\subsection{DRAG}
Derivative Reduction by Adiabatic Gate (DRAG)~\cite{Motzoi2009, Gambetta2011} is a useful technique to reduce both the leakage and the phase errors which accumulate during the operation of single-qubit gates. The most standard DRAG technique is to add the derivative of the pulse envelope to the quadrature component such that the final form of the pulse envelope $\tilde{\varepsilon}(t)$ is given by 
\begin{equation}
\tilde{\varepsilon}(t)=\varepsilon(t)+{ i} \alpha \frac{\dot{\varepsilon}(t)}{\eta_{q}}\ ,
\label{eq:drag}
\end{equation}
where $\eta_{q}$ is the transmon anharmonicity and $\alpha$ is a constant that can have different values depending on which errors need to be suppressed, e.g., $\alpha=1$ to suppress leakage and $\alpha=1/2$ to suppress coherent phase errors~\cite{Zijun2016}. \cref{eq:drag} implies that if we need to apply a single $X$ gate whose pulse envelope is given by $\varepsilon(t)$, then $\alpha {\dot{\varepsilon}(t)}/{\eta_{q}}$ needs to be applied along the $y$-axis. 

IBMQE devices use DRAG, but the exact pulse parameters are not available to users. Instead, the value of $\a$ in our simulations can be optimized numerically to match the experimentally reported gate fidelity. This is the approach we take here, with the goal being to model experiments on IBMQE devices with single-qubit gate errors of the order of $10^{-3}$. This is the value reported using randomized benchmarking~\cite{Emerson:2005sf}, 
which equals the average gate infidelity ($1-F_g$) when the gate set has gate-independent errors~\cite{Magesan:2011kx,Proctor:2017uq}, an assumption we make here to justify the use of $10^{-3}$ as our target gate infidelity.

We perform closed-system simulations and find that without DRAG, the gate infidelity (due to leakage and phase errors) is in the range $10^{-2}-10^{-3}$. With DRAG, we find that varying $\alpha$ from $1/2$ to $1$ increases the gate infidelity from $\sim 10^{-6}$ to $10^{-3}$, respectively. We thus choose $\a=1$ to match the reported fidelity.  This suggests, as described in Ref.~\cite{Zijun2016}, that the remaining errors are mostly phase error. Indeed, we find that even without DRAG, single-qubit $X$ and $Y$ gates (both implemented using two $\pi/2$ pulses as explained above) have leakage errors well below $10^{-5}$. This is unsurprising, as these long gates are quite narrowband compared to the transmon anharmonicity. We note that this attributes all errors to coherent closed-system effects rather than decoherence. We expect incoherent errors to be of the order of $t_g / T_2 \approx 5\times 10^{-4}$, suggesting that this choice is defensible (see Table \ref{table1}). Furthermore, given that both gate fidelity and coherence times drift over hour-long timescales, we focus only on matching the correct order of magnitude for fidelity with our coherent error model.

\section{Open quantum system simulation}
\label{Sec-OQS}

This section describes the noise model and discusses the hybrid Redfield equation used for the open quantum system simulations.
For all simulations we truncate to the lowest four levels of the transmon qubit.

\subsection{Interaction Hamiltonian}
\label{sec-int-hamiltonian}

The single-qubit system bath interaction Hamiltonian in the lab frame can be written as 

\begin{equation}
{H}_{\mathrm{SB}} =\sum_{i=x,y,z}g_i{A}_{i} \otimes {B}_{i}\ ,
\end{equation}
where ${A}_i$ and ${B}_i$ represent the dimensionless system and bath coupling operators, respectively, and the coupling strengths $g_i$ have dimensions of energy. There are several contributions to decoherence and noise for a multi-level transmon circuit. With fixed-frequency architectures, charge noise and fluctuations in the critical current contribute most to decoherence. In contrast, in the flux-tunable variants of transmon qubits, the largest contribution comes from flux noise~\cite{transmon-invention}. These considerations determine which coupling operators are needed to describe the noise model for a given architecture. In the IBMQE processors used, the transmons are fixed-frequency. We therefore choose appropriate noise operators below.

We consider the following system-bath interaction Hamiltonian: 
\begin{equation}
{H}_{\mathrm{SB}} = g_x {A}_x \otimes {B}_x +{A}_z \otimes \left( g_z{B}_z + \sum_{k}b_k\chi_k \left(t\right) I_{\mathrm{B}}\right)\ ,
\label{eq:H_int}
\end{equation}
where the coupling operators ${A}_x$ and ${A}_z$ correspond to the charge coupling operator and to the Josephson energy operator and are defined as

\begin{subequations}
\label{eq:A_operators}
\begin{align}
{A}_x &= {\rm c}_1\hat{n}  \\
{A}_{ z} &= {\rm c}_2{\rm cos}\hat{\varphi}\ ,  
\end{align}
\end{subequations}
where $c_1$ and $c_2$ are fixed constants that depend on the charge energy $E_c$ and the Josephson energy $E_J$ of the transmon qubit, respectively. We expect, based on the discussion in \cref{sec:time-dependent-drives} -- and observe in our simulations -- that ${A}_x$ and ${A}_z$ act like $\s^x$ and $\s^z$ when projected into the qubit subspace. We find numerically that \cref{eq:H_int} is an adequate model accounting for the nearly equal decay of the $\ket{+}$ and $\ket{i}$ states, which is why we do not include a separate $\s^y$ coupling term. Note, however, that a (dependent) $\s^y$ component appears when we transform \cref{eq:H_int} from the lab frame into a frame rotating with the drive. 

Previous studies have found that noise in the superconducting circuit can be separated into high and low-frequency components~\cite{Quintana:2017aa}. To account for this observation, we combine two noise models. We choose the bath operators $B_x$ and $B_z$ in \cref{eq:H_int} to be bosonic bath operators, which generally represent the high-frequency component of the noise. However, this is not always the case, as we argue in \cref{sec-methodology}. 

To account for the low-frequency noise component, which is a dominant noise source for superconducting qubits~\cite{RevModPhys.86.361}, we include a sum over classical fluctuators in \cref{eq:H_int}, via the term proportional to the bath identity operator $I_{\rm B}$. This semiclassical contribution, when parameterized properly, can simulate the behavior of $1/f$ noise. 
We model the fluctuators as having equal coupling strengths, i.e., we set $b_k=b$ (with dimensions of energy) for $k=1,\cdots,10$. Each fluctuator can be characterized by a stochastic process $\chi_k(t)$ that switches between $\pm 1$ with a frequency $\gamma_k$, which is log-uniformly distributed between $\gamma_{\rm min}$ and $\gamma_{\max}$~\cite{KaWa-Yip-PhDthesis}. 

\begin{figure*}[t]
	\includegraphics[width=0.7\linewidth]{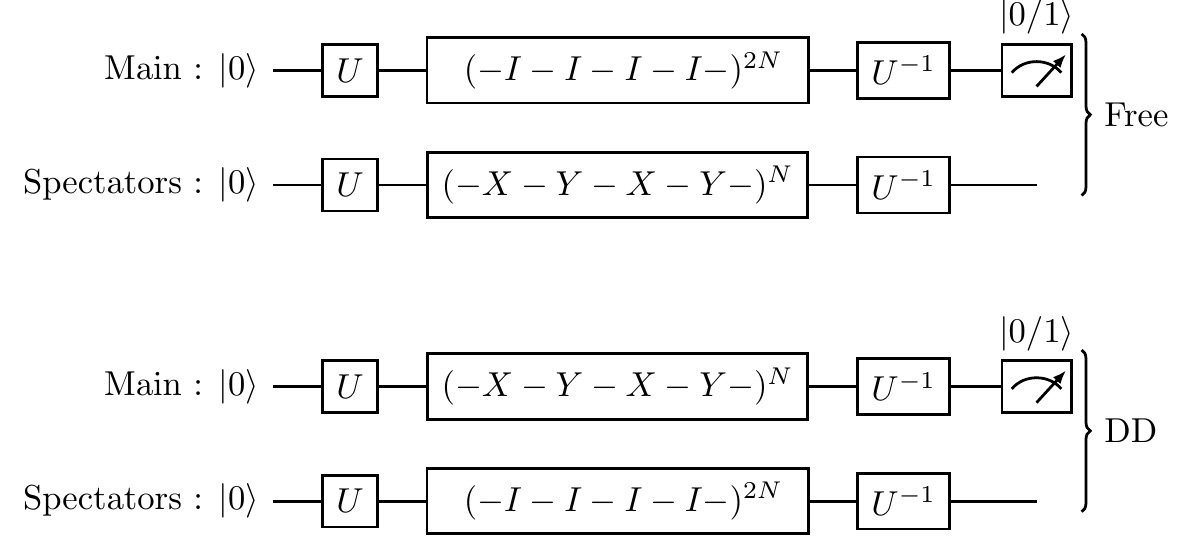}
	\caption{The circuit schematics for the free-evolution and DD-evolution types of experiments.  For the free-evolution case, we apply $N$ cycles of the XY4 dynamical decoupling (DD) sequence on all the spectator qubits and $2N$ cycles of the $I_4$ sequence (here $I_4$ means four identity gates) on the main qubit, which suppresses crosstalk errors~\cite{tripathi2021suppression}. Note that an $X$ or $Y$ gate is twice as long as an identity gate on the IBM cloud quantum devices, hence the extra factor of $2$. For the DD-evolution case, we apply the DD sequence only to the main qubit and apply identity gates to all the spectator qubits. This suppresses both crosstalk and environment-induced noise. We measure only the main qubit.}
	\label{fig-sequences}
\end{figure*}

\subsection{Hybrid Redfield model}\label{sec-redfield}
To simulate the reduced system dynamics of the interaction Hamiltonian in \cref{eq:H_int}, we use a hybrid form of the Redfield (or TCL$2$) master equation~\cite{Redfield:66,chen2020hoqst}. We first define the standard bath correlation function
\begin{equation}
    C_{ij}(t) = \mathrm{Tr}\{U_\mathrm{B}(t)B_iU^\dagger_\mathrm{B}(t) B_j\rho_\mathrm{B}\}\ ,
    \label{eq:C_ij}
\end{equation}
where $U_\mathrm{B}(t)=e^{-iH_{\rm B}t}$ is the unitary evolution operator generated by the bath Hamiltonian $H_\mathrm{B}$, and the reference state $\rho_\mathrm{B}$ is the Gibbs state of $H_\mathrm{B}$:
\begin{equation}
    \rho_\mathrm{B}=e^{-\beta H_\mathrm{B}}/\mathrm{Tr} \big(e^{-\beta H_\mathrm{B}}\big)\ ,
\end{equation}
where $\beta = 1/T$ is the inverse temperature.
Assuming the bath operators $B_x$ and $B_z$ are uncorrelated, i.e., $C_{xz}(t)=C_{zx}(t)=0$, we construct the following hybrid Redfield equation
\begin{equation}
    \label{eq:hybrid_redfield}
    \frac{\partial\rho_\mathrm{S}}{\partial t}= -i[H_{\mathrm{sys}}+A_{z}\sum_{k=1}^{10} b_k \chi_k(t), \rho_\mathrm{S}] + \mathcal{L}_{\mathrm{R}}(\rho_\mathrm{S}) \ ,
\end{equation}
where $\mathcal{L}_\mathrm{R}$ is the Redfield Liouville superoperator
\begin{equation}
    \mathcal{L}_\mathrm{R}(\rho_\mathrm{S}) = -\sum_{i=x,z} [A_i, \Lambda_i(t)\rho_\mathrm{S}(t)] + {\rm h.c.} \ ,
\end{equation}
and
\begin{equation}\label{eq:redfield_lambda}
    \Lambda_i(t) = \int_0^{t} C_{i}(t-\tau)U_\mathrm{sys}(t, \tau)A_iU_\mathrm{sys}^\dagger(t, \tau) \mathrm{d}\tau \ ,
\end{equation}
where $C_j(\tau) \equiv C_{jj}(\tau)$ [from \cref{{eq:C_ij}}] and $U_{\mathrm{sys}}(t)$ is the unitary evolution operator generated by the system Hamiltonian $H_\mathrm{sys}$. The reduced system dynamics are obtained by averaging the solution of \cref{eq:hybrid_redfield} over all the realizations of $\chi_k(t)$ for $k=1,...,10$.

The bath component correlation functions $C_j(\tau)$ are the Fourier transforms of the bath component noise spectra
\begin{equation}
    \gamma_j(\omega) = \int_{-\infty}^{\infty} C_j(\tau) e^{i\omega\tau} \mathrm{d}\tau \ .
\end{equation}
 We choose the bath to be Ohmic, which means that the component noise spectra have the form
\begin{equation}
    \gamma_j(\omega) =2 \pi \eta g_j^{2} \frac{\omega \mathrm{e}^{-|\omega| / \omega^c_j}}{1-\mathrm{e}^{-\beta \omega}}\ ,
    \label{eq:Ohmic}
\end{equation}
where $\omega^c_j$ is the cut-off frequency for bath operator $B_j$, and $\eta$ is a positive
constant with dimensions of time squared arising in the specification of the Ohmic spectral function.

Lastly, the hybrid Redfield equation~\eqref{eq:hybrid_redfield} can be transformed into a  frame rotating with the drive frequency $\omega_{\rm d}$ (see \cref{app:derivation} for details) by replacing every operator with the interaction-picture one [specifically, the $A_i$ operator in \cref{eq:redfield_lambda} needs to be replaced by $A_i(\tau)$]. We simulate the Redfield master equation in this rotating frame in the methodology and results we discuss next.

\section{Methodology and Fitting Results}
\label{sec-methodology}

This section discusses our methodology for modeling a transmon qubit's open quantum system behavior in a multi-qubit processor. We refer to the qubit of interest as the main qubit and all the others as spectator qubits. The goal is to extract the bath parameters in our open quantum system model and then use this model to predict the outcomes of experiments on the main qubit, including dynamical decoupling sequences. We treat qubit 1 (Q1) of the Quito processor as our main qubit. We are interested only in the main qubit's behavior here; hence, we measure only the main qubit. \cref{app:data-analysis} describes the procedure to extract and analyze the experimental data.

\subsection{Free and DD Evolution Experiments}
\label{subsec-free-and-DD}
Our procedure involves two types of experiments, as shown in \cref{fig-sequences}. The first type, which we call a \emph{free-evolution} experiment, consists of initializing all the qubits in a given state by applying a particular unitary operation $U3(\theta,\phi,\lambda)$ \cite{IBMQ-U3} (denoted as $U$ in \cref{fig-sequences}) to each of the qubits. We then apply a sequence of identity gates on the main qubit, which we vary in number. Simultaneously we also apply the XY4 DD sequence to all the other (spectator) qubits (i.e., $X f_\tau Y f_\tau X f_\tau Y f_\tau$, where $f$ denotes free-evolution in the absence of pulses for a duration of $\tau$ \cite{Viola:99}) for the same total duration as that of the identity gates on the main qubit. As shown in~\cite{tripathi2021suppression}, DD sequences applied to spectator qubits suppress unwanted $ZZ$-interactions, i.e., $ZZ$-crosstalk between the main qubit and the spectator qubits. Without crosstalk suppression, we observe oscillations in the probability decay as a function of time~\cite{Pokharel2018, souza2020process}; see \cref{app:B}. With the crosstalk suppression scheme, i.e., DD applied to the spectator qubits, these oscillations disappear, and the main qubit is now primarily affected only by environment-induced noise. Finally, we apply the inverse of $U3(\theta,\phi,\lambda)$ and measure in the $Z$-basis. The result is how we compute the initial states' decay probability. 

Everything remains the same in the second type of experiment, which we call \emph{DD-evolution}, except that we now apply the XY4 sequence to the main qubit and identity gates on the spectator qubits. As discussed in~\cite{tripathi2021suppression}, when we apply the XY4 sequence only to the main qubit, we suppress the $ZZ$-crosstalk between the latter and all the spectator qubits and also decouple unwanted interactions between the main qubit and the environment. We note that, in contrast to experiments using DD to perform noise spectroscopy, here we use only a single type of DD sequence and do not vary any of its parameters.

\begin{figure*}[t]
	\centering
		\begin{tabular}{ccc}
		\includegraphics[width=0.3\linewidth]{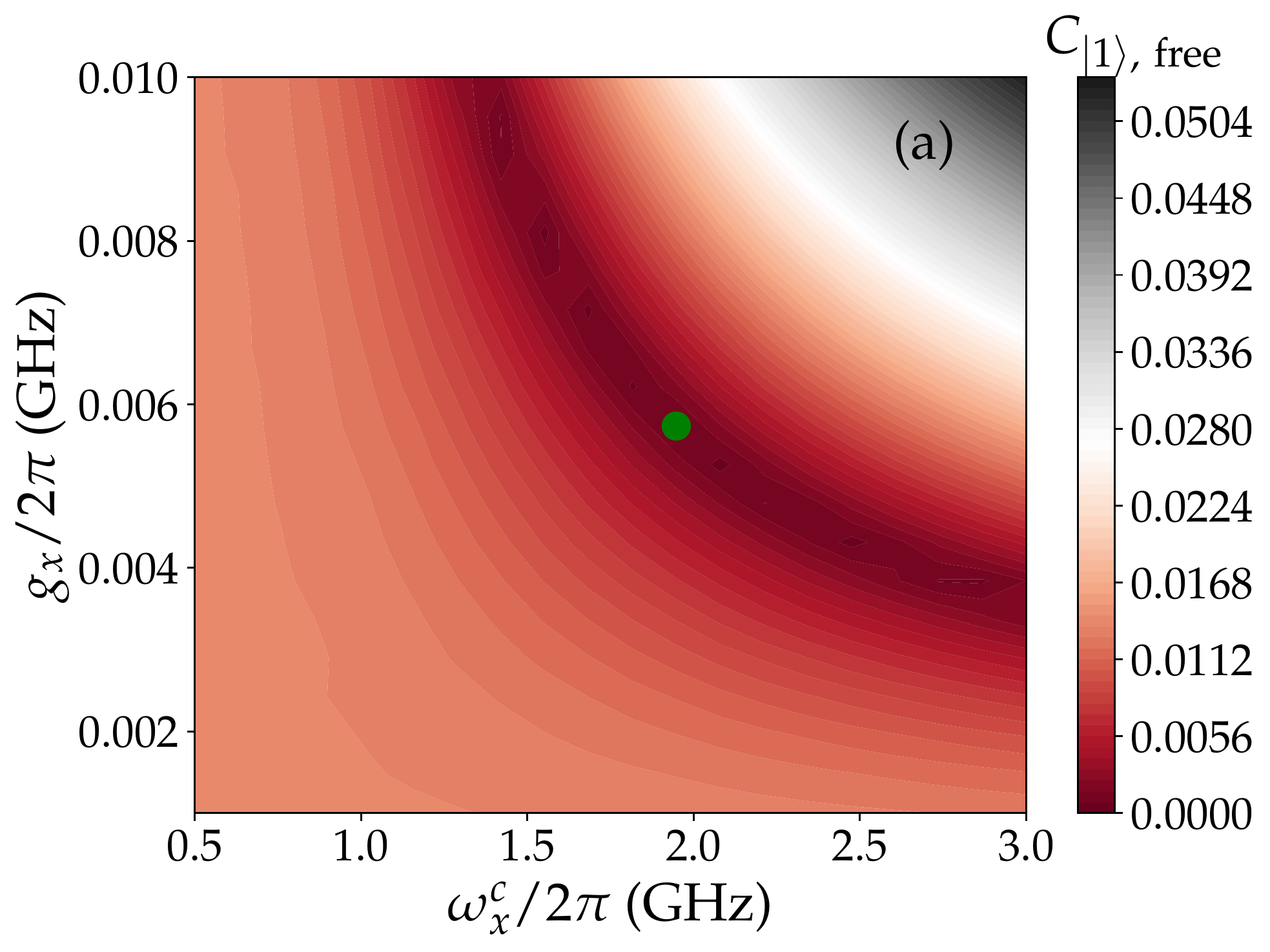}& 
		\includegraphics[width=0.3\linewidth]{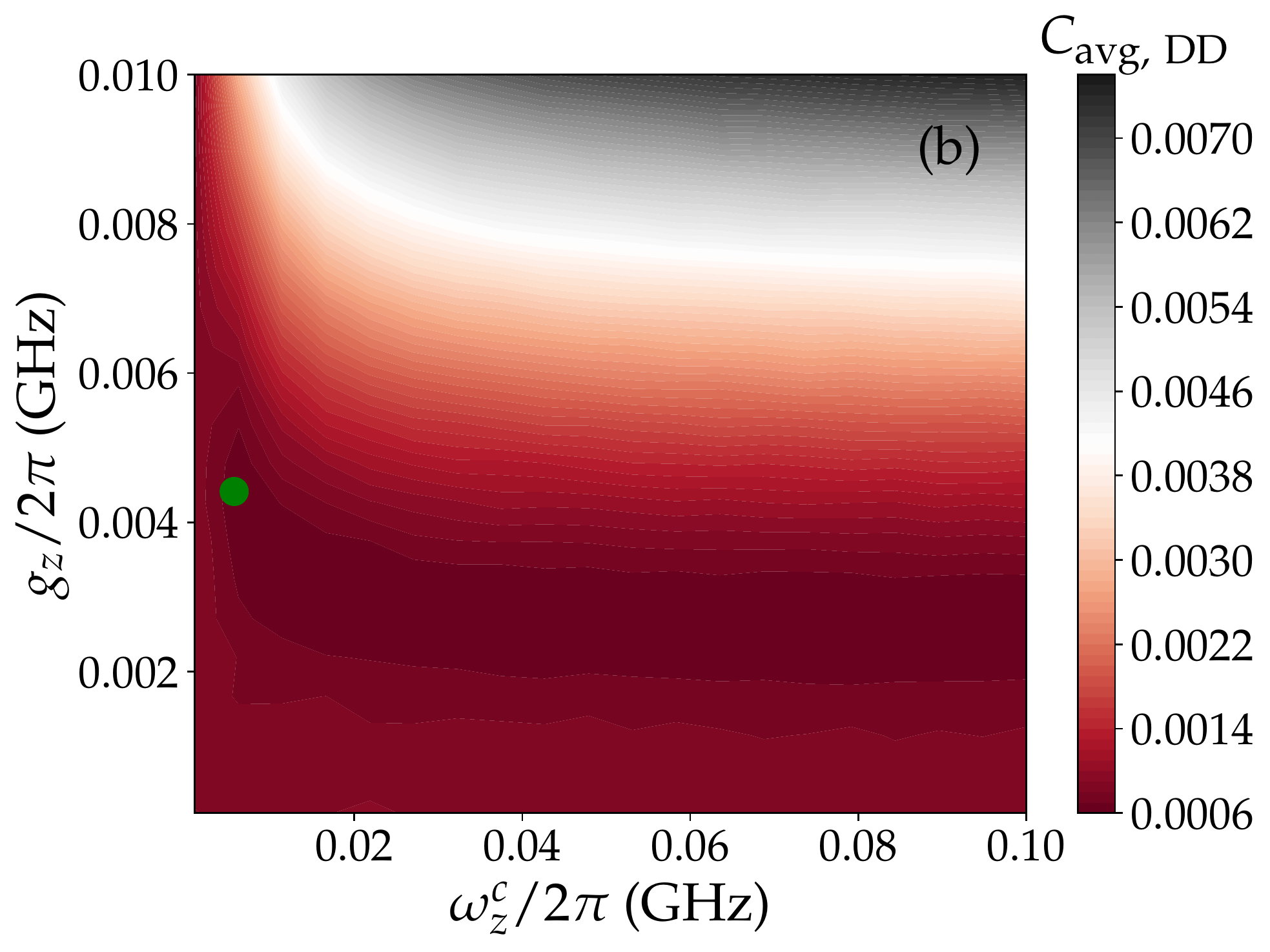}&
		\includegraphics[width=0.3\linewidth]{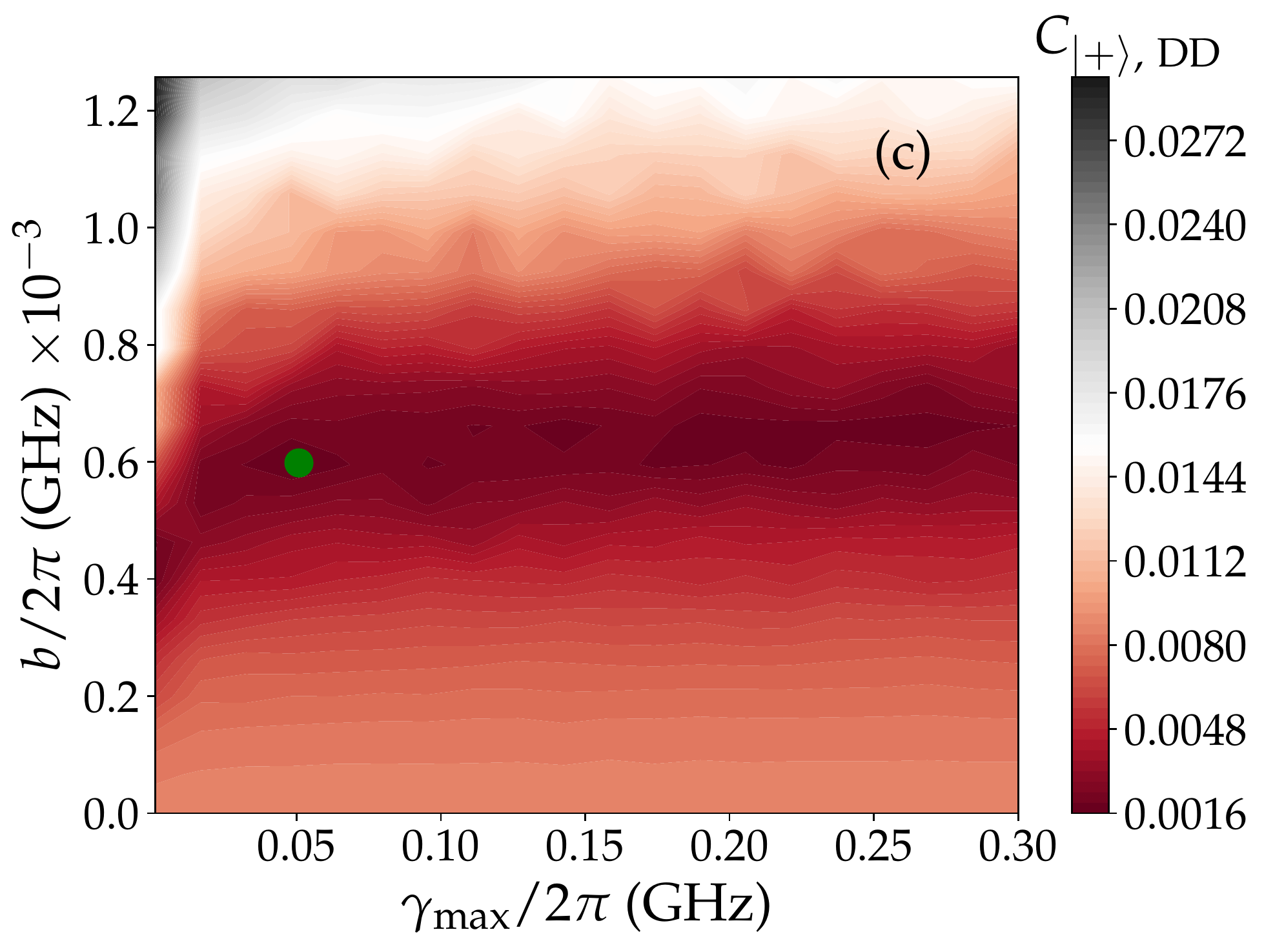}\\
		\includegraphics[width=0.3\linewidth]{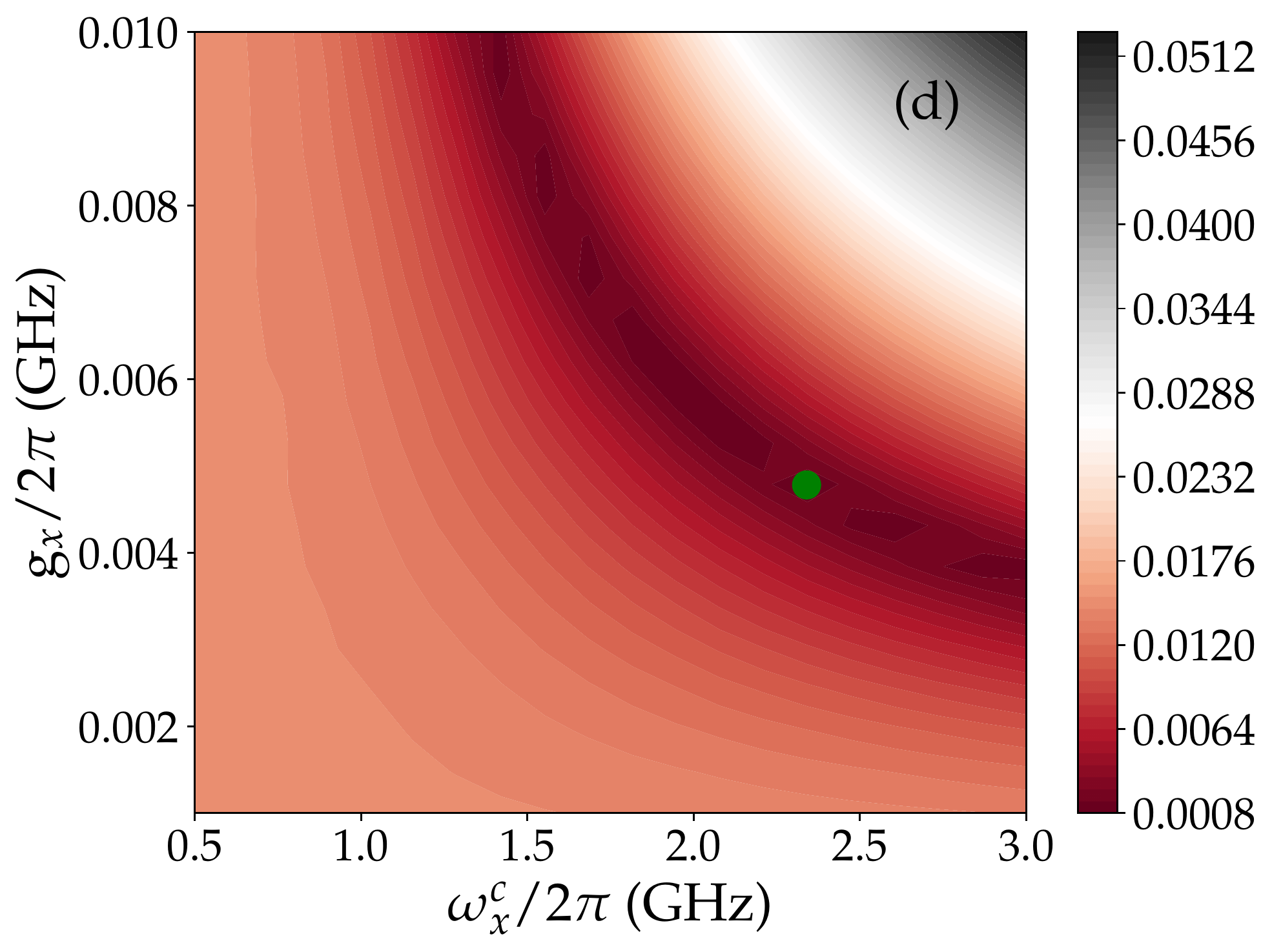}& 
		\includegraphics[width=0.3\linewidth]{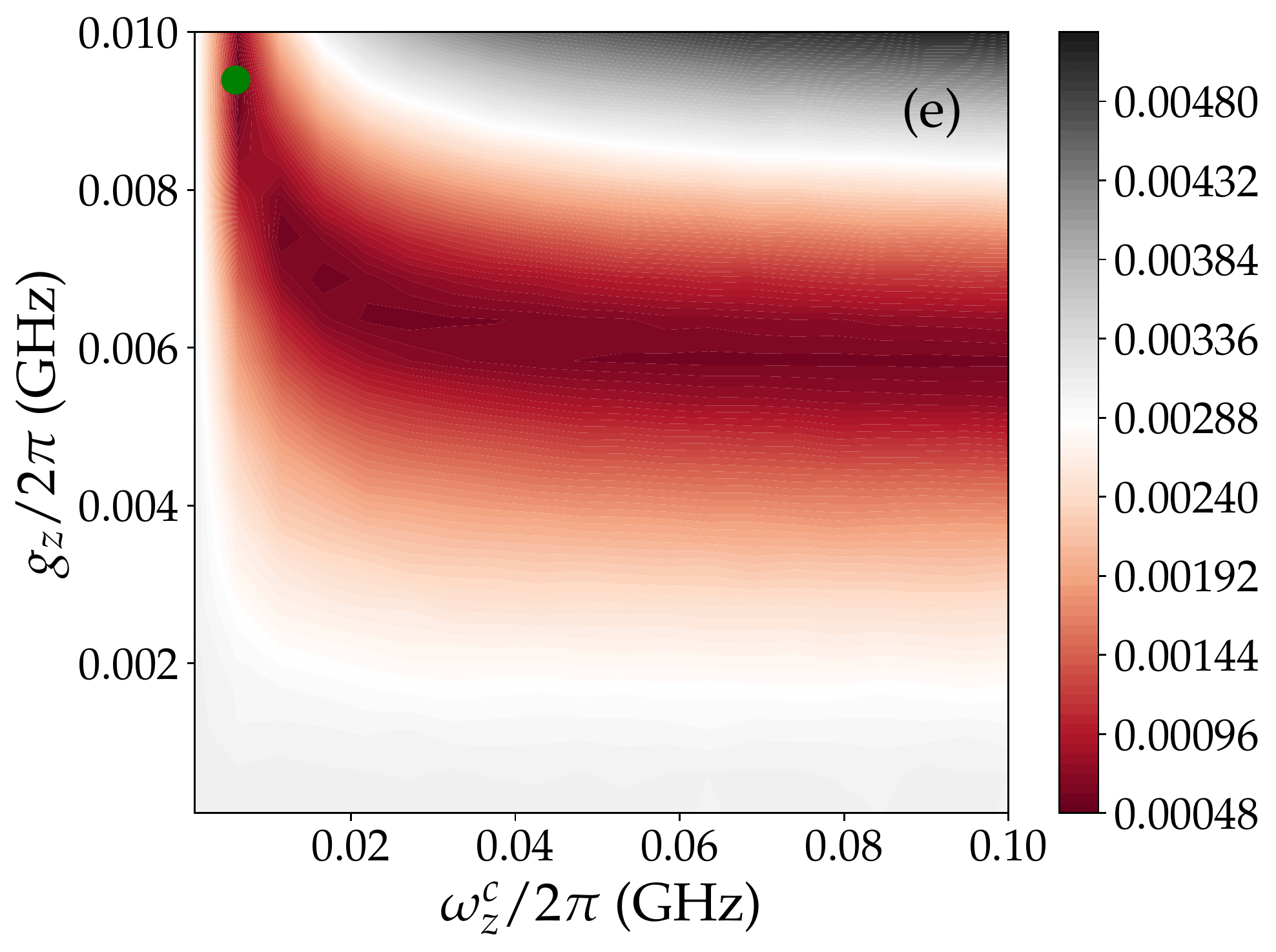}&
		\includegraphics[width=0.3\linewidth]{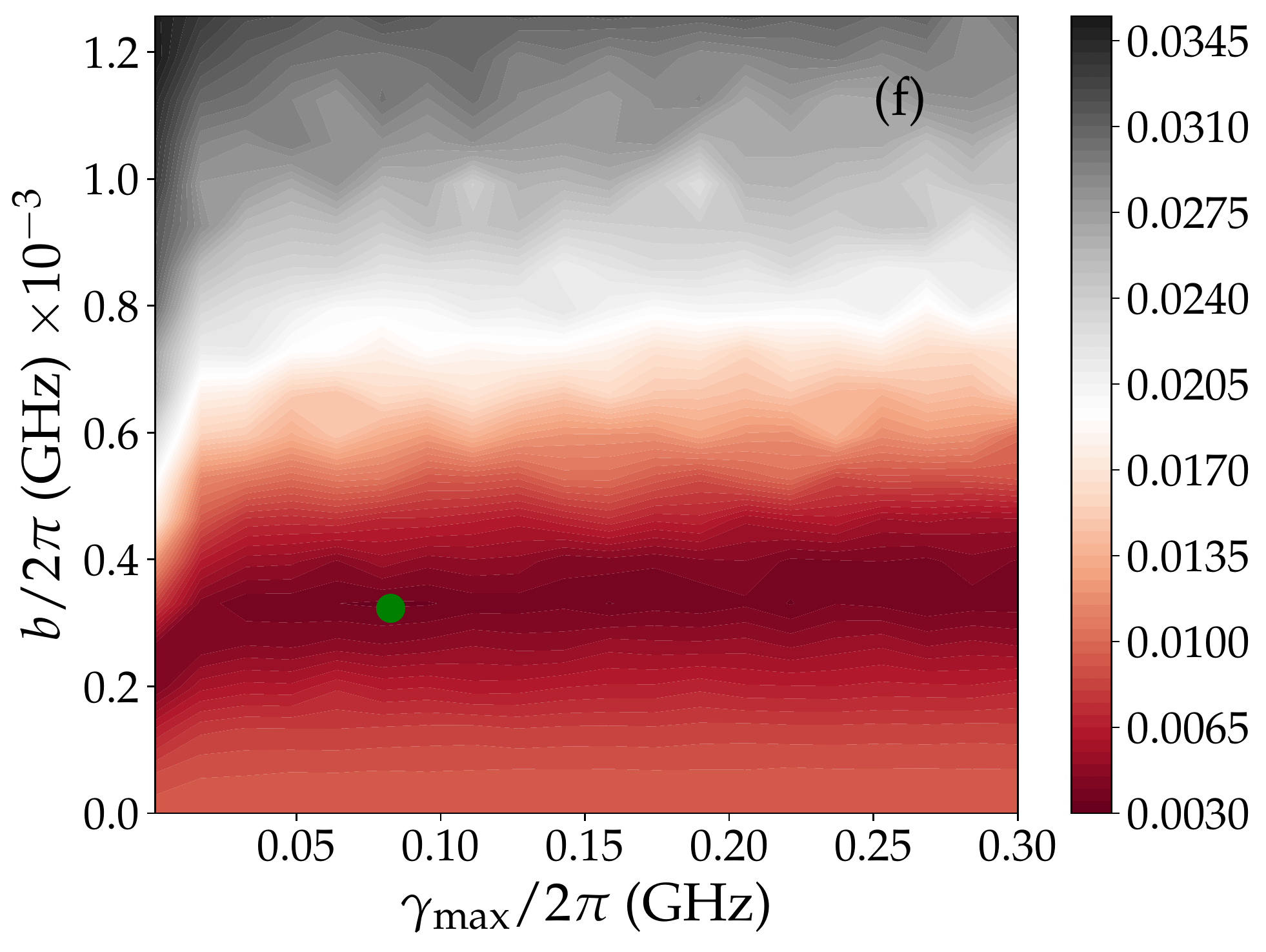}\\
	\end{tabular}
		\caption{Top: Results for the Quito processor. Bottom: Results for the Lima processor. \\
		Left: The cost function defined as the $l_2$ norm distance between the experimental and simulation results [\cref{eq:cost_function}], averaged over $N=70$ time instants, as a function of the bath parameters $\omega^c_x$ and $g_x$ for free-evolution of the $\ket{1}$ initial state.\\
		Middle: The average of the cost function over the six Pauli states for DD-evolution as a function of $\omega^c_z$ and $g_z$.\\
		Right: The cost function for free-evolution of the $\ket{+}$ initial state, as a function of $\gamma_{\max}$ and $b$. \\
		The green circles indicate the positions of the global minima in all the panels.}
		\label{fig-step_1}
\end{figure*}

\subsection{Fitting Procedure}
\label{sec:fit}

We perform the free-evolution and DD-evolution experiments for the six Pauli states as initial states, i.e., we choose $U3(\theta,\phi,\lambda)$ to prepare $\ket{0}$, $\ket{1}$, $\ket{+}$, $\ket{-}$, $\ket{i}$ and $\ket{-i}$, and use the hybrid Redfield equation described in \cref{sec-redfield} to simulate the dynamics of these experiments. We sweep over different values of the bath parameters and obtain the simulated probability decay as a function of time. To identify the simulation parameters that optimally match the experimental results, we define a cost function $C$ for a given initial state $\ket{\psi}$ as the $l_2$ norm distance between the experimental probabilities $P_{\ket{\psi},s}^{\rm Exp}(t_i)$ and the simulation probabilities $P_{\ket{\psi},s}^{\rm Sim}(t_i)$ for every instant $t_i$:
\begin{equation}
        C_{\ket{\psi},s} = \frac{1}{N}\sqrt{\sum_{i=0}^{N-1}\left(P_{\ket{\psi},s}^{\rm Sim}(t_i)-P_{\ket{\psi},s}^{\rm Exp}(t_i)\right)^2}\ ,
        \label{eq:cost_function}
\end{equation}
where $s\in \{\rm{free}, \rm{DD}\}$ and $N$ is the total number of instants.
Note that we compensate for state preparation and measurement (SPAM) errors by shifting the experimental results such that in all cases, $P_{\ket{\psi},s}^{\rm Exp}(0)=1$. 

We limit the number of free parameters requiring fitting to six: the coupling strengths $g_x$, $g_z$, and $b_k\equiv b$ [\cref{eq:H_int}], and the cutoff frequencies $\gamma_{\max}$ (for $1/f$ noise), $\omega^c_x$, and $\omega^c_z$ [\cref{eq:Ohmic}].
We set the bath temperature $T=20\;$mK ($\sim$ the fridge temperature), $\gamma_{\min}=10^{-4}\;$GHz, and $\eta=10^{-4}\;{\rm GHz}^{-2}$; these values are the same as in our previous work~\cite{tripathi2021suppression}, which showed strong agreement between open system simulations and experiments using other IBMQE devices, and remain fixed throughout our fitting procedure. This procedure consists of three steps, which we detail next.

\subsubsection{Step I: free-evolution for $\ket{1}$}
     We first focus on the free-evolution experiment for the initial state $\ket{1}$, the first excited state in the transmon eigenbasis. Since the free-evolution for this state is only affected by charge noise, i.e., noise along the $x$-axis, the only contribution to the decay of $\ket{1}$ should come from the $g_x A_x \otimes B_x$ term in \cref{eq:H_int}. Thus, we consider only this term in our numerical simulations for this initial state. For a given set of values of the coupling strength $g_x$ and bath cutoff frequency $\omega^c_x$, we compute the cost function $C_{\ket{1},\rm{free}}$ using \cref{eq:cost_function}, and obtain the contour plot shown in \cref{fig-step_1}(a). 
     
     In our simulations, we vary $\omega^c_x/(2\pi)$ from $0.5$ to $3$~GHz and $g_x$ from $0$ to $10^{-2}$~GHz, each with $20$ equidistant points so that the contour plot has a total of $400$ data points. We take the position of the global minimum of the cost function on this grid as the optimal set of bath parameter values. To reduce the resulting discretization uncertainty, we interpolate the contour plot and use the Nelder-Mead optimization method to locate the minima. 
     We numerically find the global minimum at $\omega^c_x/(2\pi)= 1.948$\;GHz and $g_x/(2\pi) = 0.573\times 10^{-2}$~GHz, denoted by the green circle in \cref{fig-step_1}(a). With this, we have two out of the six bath parameters, and we use these learned parameters in the subsequent steps.

 \subsubsection{Step II: DD-evolution for all six Pauli states}
    The second step involves the DD-evolution experiment for all six Pauli states. This requires including the term $ g_z A_z \otimes B_z$ in \cref{eq:H_int}, along with the first term whose bath parameters we already obtained. We do not include the semiclassical term in \cref{eq:H_int} consisting of fluctuators since it is expected to be strongly suppressed when DD is applied to the main qubit. We simulate time-dependent gates with DRAG corrections to model the DD pulses, as discussed in \cref{Sec-numerics}. We are again left with just two bath parameters to optimize: $\omega^c_z$ and $g_z$. \cref{fig-step_1}(b) shows the average of the cost function [\cref{eq:cost_function}] over the six Pauli states. The global minimum is found at $\omega^c_z/(2\pi) = 0.569 \times 10^{-2} $~GHz and $g_z/(2\pi) = 0.441\times10^{-2}$~GHz.

\subsubsection{Step III: free-evolution for $\ket{+}$}
    The final step requires optimizing the two remaining free parameters associated with the fluctuators: $\gamma_{\max}$ and $b$. Here we focus on the free-evolution experiment for initial state $\ket{+}$. We now employ the full system-bath Hamiltonian in \cref{eq:H_int} with the optimal parameters found in Steps I and II. \cref{fig-step_1}(c) shows the contour plot for the cost function [\cref{eq:cost_function}], where, as in Step I, we again use $20$ different values of $\gamma_{\max}$ and $b$ each. The global minimum is found at $\gamma_{\max}/(2\pi)=0.051$~GHz and $b/(2\pi) = 0.598\times 10^{-3}$~GHz.

\subsection{Methodology wrap-up}

Let us briefly summarize our methodology and add a few technical details. As explained above, we extract the bath parameters by performing free-evolution experiments for two initial states ($\ket{1}$ and $\ket{+}$) and DD-evolution experiments for up to six initial states (the Pauli states). Our optimization procedure is iterative and is thus not guaranteed to yield the globally optimal values of all the bath parameters, but this is by design: we choose initial states that allow us to isolate the bath parameters one pair at a time, which renders the optimization problem tractable.

This methodology is quite general and can be used to characterize all the transmon qubits on a given transmon processor or, much more broadly, to characterize single qubits on any quantum information processing platform capable of supporting individual qubit gates and measurements, provided a sufficiently accurate and descriptive model of the qubits and the system-bath interaction is available. Our procedure inherently suppresses the effects of crosstalk due to the neighboring qubits via DD applied either to the spectator qubits (free-evolution experiments) or the main qubit (DD-evolution experiment), which reduces the number of free parameters of the noise model by eliminating the need to model crosstalk.

To obtain the contour plots shown in \cref{fig-step_1}, we solve the Redfield master equation (\cref{sec-redfield}) for each point (i.e., each set of model parameters), requiring a total of $400$ simulation runs for each optimization. In the final step, including classical fluctuators to obtain \cref{fig-step_1}(c),  we use the trajectory version of the Redfield model introduced in \cref{sec-redfield} to simulate a total of $600$ trajectories at each point. This is large enough to yield negligible error bars ($< 2\times 10^{-2}$). The experimental results are obtained using the standard bootstrap method (see \cref{app:data-analysis}). In defining the cost function [\cref{eq:cost_function}], we use the mean value of the experimental fidelity obtained after bootstrapping (see \cref{app:B}) and ignore the associated tiny error bars ($\le 6 \times 10^{-3}$). These error bars are much smaller than the error induced by the discrete nature of our $40 \times 40$ parameter grid, and so we can safely ignore them. We confirmed that varying the probabilities to the extremes of the error bars does not affect the values of the bath parameters we have extracted to the least significant digit we report. Table~\ref{tab:params} summarizes the extracted values and the parameters we have fixed.

\begin{figure*}[t!]	
	\centering
		\begin{tabular}{cc}
		\includegraphics[width=0.5\linewidth]{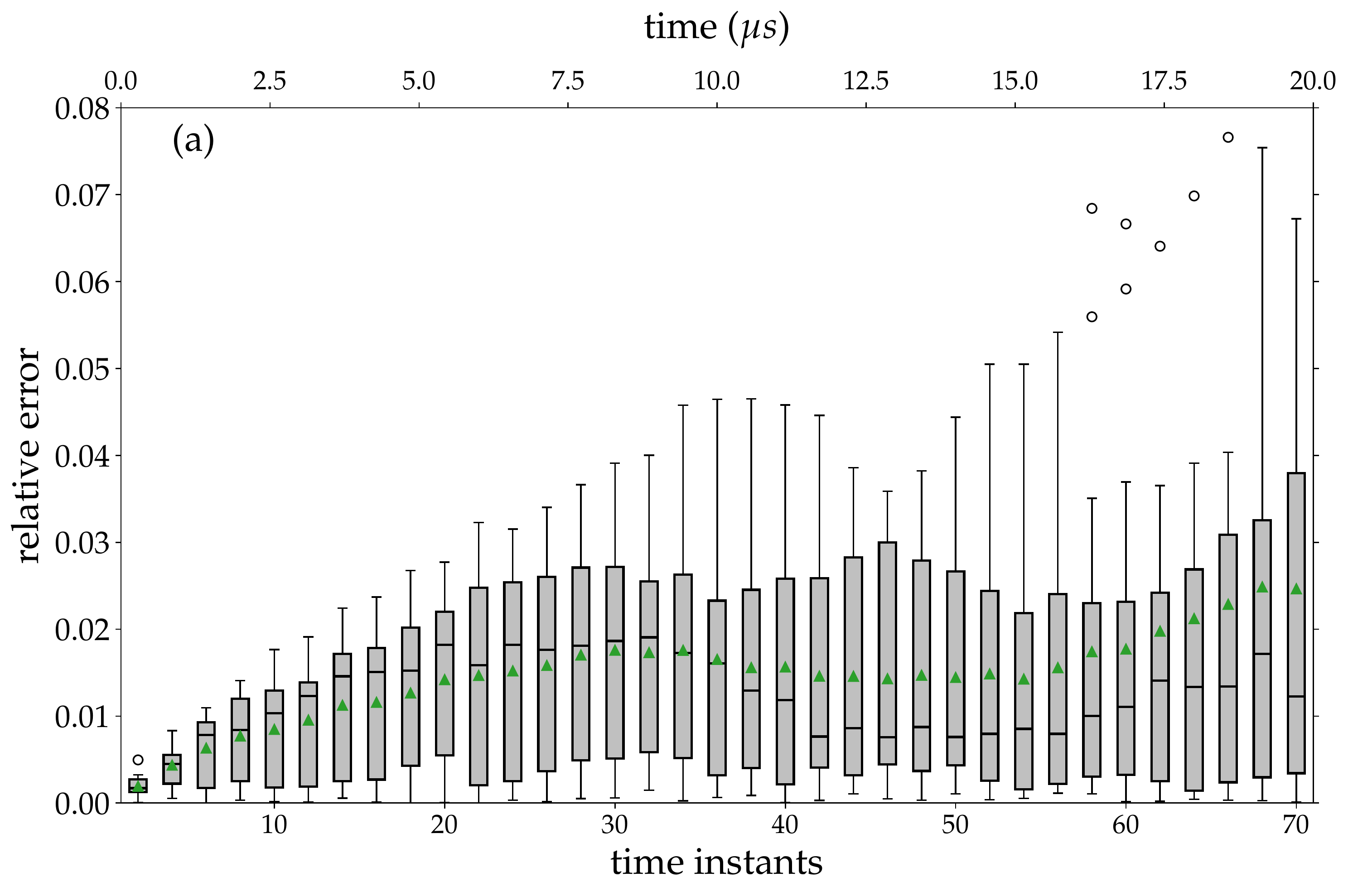}&	
		\includegraphics[width=0.5\linewidth]{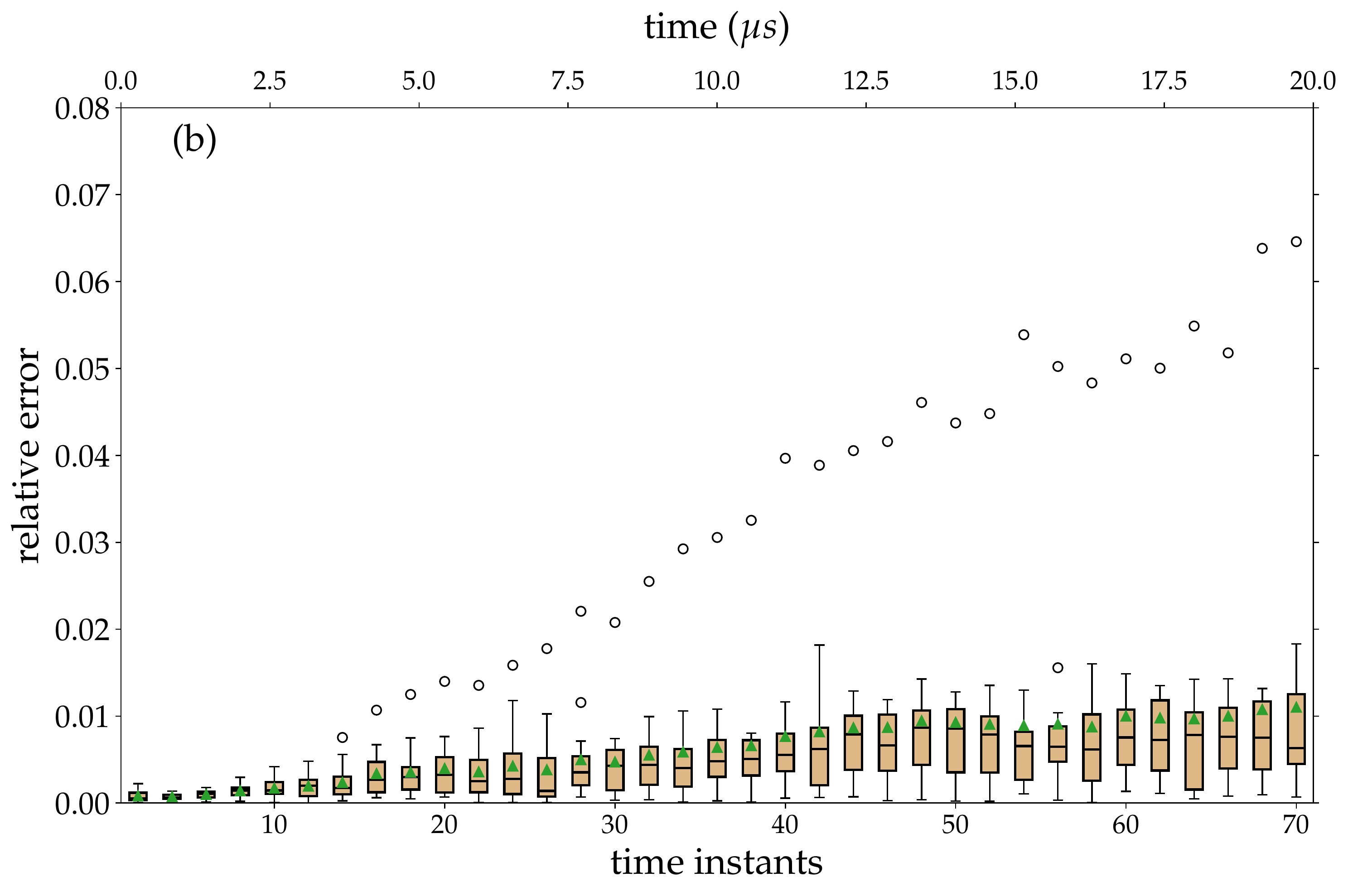}\\
		\includegraphics[width=0.5\linewidth]{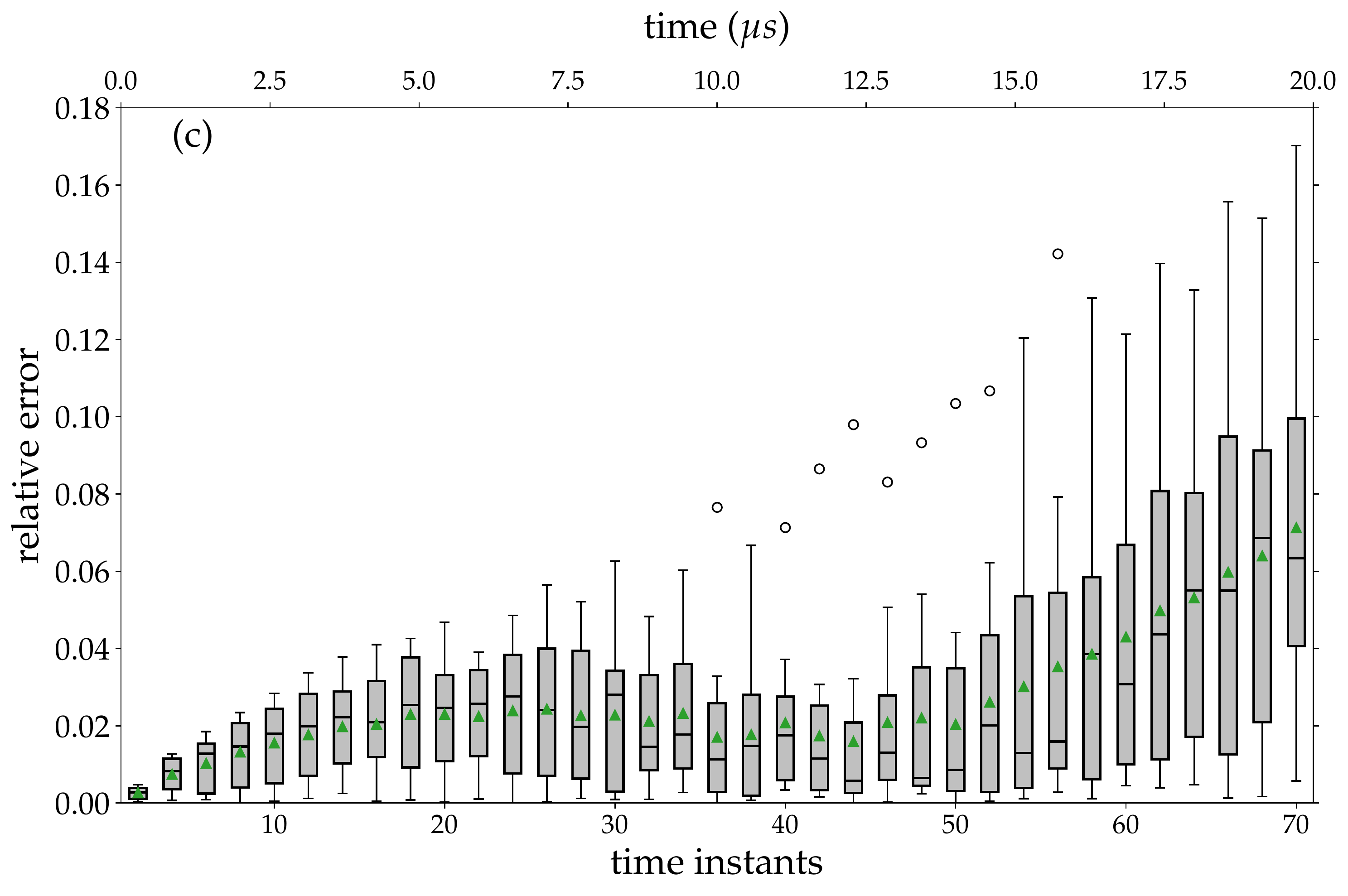}&	
		\includegraphics[width=0.5\linewidth]{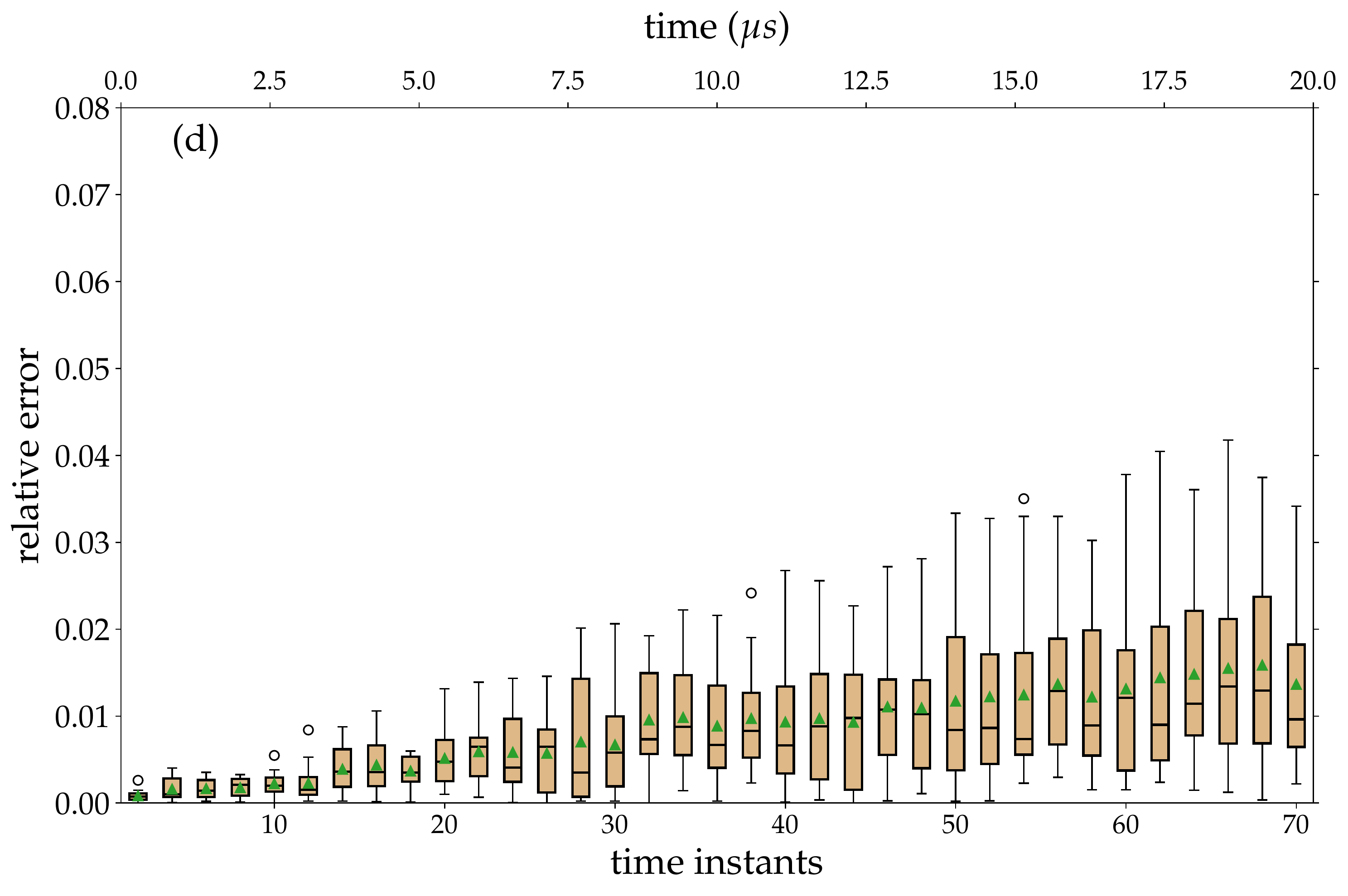}
	\end{tabular}
\caption{Results for the Quito (top row) and Lima (bottom row) processors. Left: Box plots showing the relative error of our model for the free-evolution experiments as a function of time for $16$ different initial states containing six Pauli states and ten Haar-random states. Right: the same as on the left, but for the DD-evolution experiments. We measured a total of $70$ time instants, up to a total evolution time of $19.6~\mu$s, but only display every other instant to avoid overcrowding. Green triangles indicate the mean over the $16$ initial states, black horizontal lines are the median, gray boxes represent the $[25,75]$ percentiles, the whiskers (black lines extending outside the boxes) represent the $[0,25]$ and $[75,100]$ percentiles, and circles are outliers.}
		\label{fig-results}
\end{figure*}

The accuracy of our results depends on the number of points in the contour plots in \cref{fig-step_1}  (we used a $20\times 20$ grid for each panel). Even though we interpolate the otherwise discrete contour plots and find the minima over the resulting smooth surface, the limited number of points affects the precision of the learned bath parameters. Increasing this precision requires more sophisticated optimization techniques to speed up the process of obtaining the bath parameters. This becomes especially acute when extending the model to learning a multi-qubit system-bath Hamiltonian with correlated noise, as in this case, the number of bath parameters increases significantly. Here, we aim to demonstrate the model and methodology and illustrate both via the example of a single transmon qubit, and so we perform a simple brute-force search of the parameter space. Note that our methods for extracting the bath parameters also work with density matrices (from state tomography) instead of just probabilities. In that case, the $l_2$ norm distance in the cost function of \cref{eq:cost_function} can be replaced by the trace-norm distance between the density matrices obtained from the simulation and the experiment. However, quantum state tomography imposes a much higher cost in terms of the number of required experiments and is thus less practical to scale up with a larger number of qubits. Our protocol requires only fidelity measurements and so is more resource-efficient.

\begin{figure}[t]
	\includegraphics[width=1\linewidth]{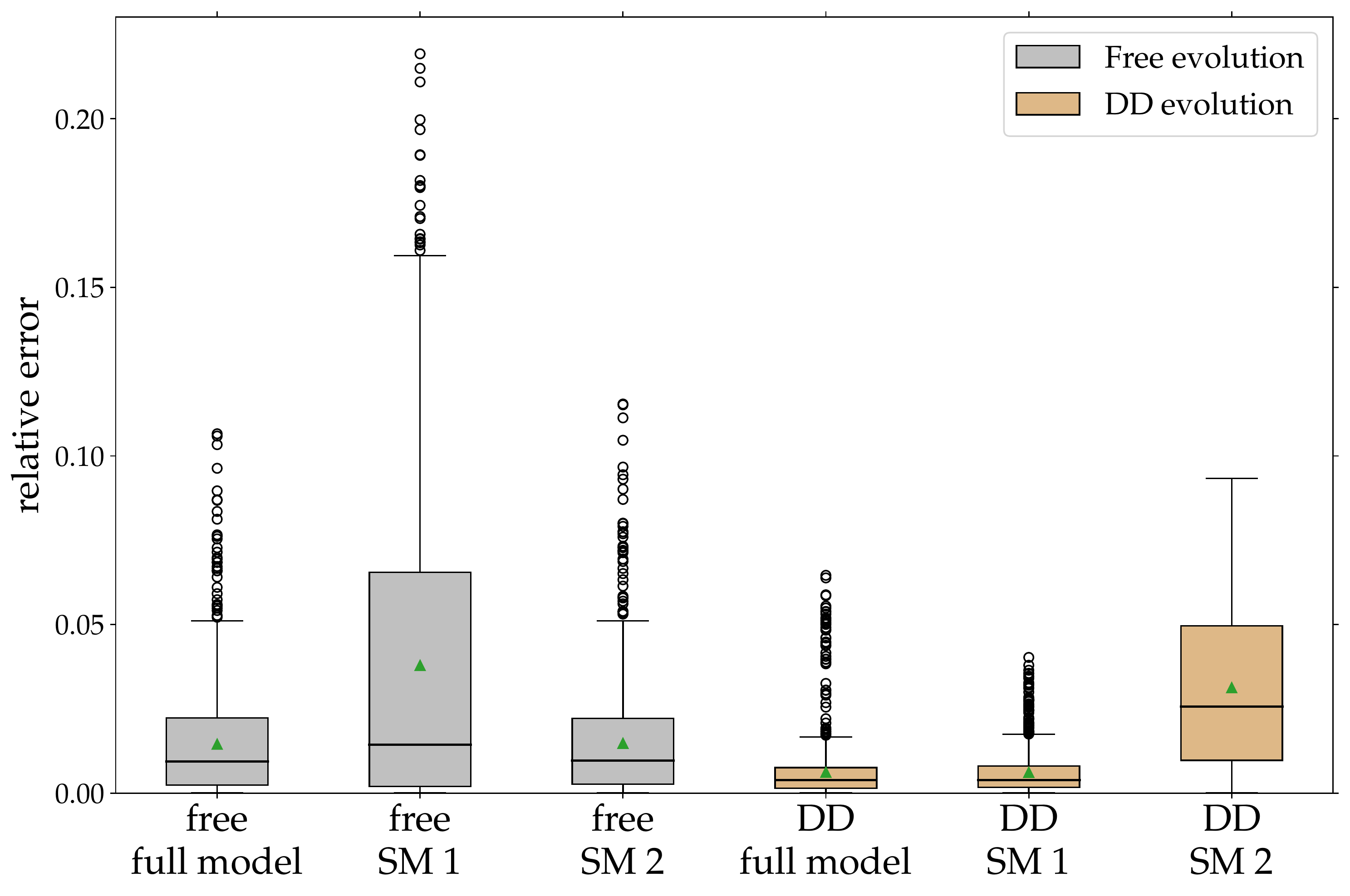}
	\caption{Relative error results for the Quito processor. We display a comparison of the relative errors between the full model (which uses a three-step learning procedure and consists of four energy levels per transmon and realistic pulses) with the simplified models SM1 and SM2 (which are based on a two-step learning procedure and use just two energy levels and instantaneous pulses) for free-evolution and DD-evolution experiments. SM1 (SM2) is trained on the DD (free) evolution experiments. Each box contains a total of $16$ initial states and all $70$ time instants varying from $0$ to $19.6\;\mu s$.}
	\label{fig-comparison}
\end{figure}

\section{Model Prediction Results}
\label{sec:results}

\subsection{Full Model}

We now test our model for different initial states of the main qubit of the Quito processor. Since we always apply the DD sequence to the spectator qubits during the free-evolution experiments, the initial states of the latter do not matter due to the suppressed $ZZ$ coupling. This section considers a total of $16$ initial states, consisting of the six Pauli states and ten Haar-random states. We model the experimental results using the bath parameters we extracted in the previous section (\cref{tab:params}). This serves as a stringent test of the model: we now use the previously fitted model to predict the outcome of experiments not included in Steps I-III of \cref{sec:fit}, i.e., the results with different initial states. The data for all the experiments (both fitting and testing) was obtained in one batch. We used only the data for the six Pauli initial states needed for Steps I-III to perform the fitting. We used the data for all $16$ initial states in the testing phase.

We consider the same two kinds of experiments: free-evolution and DD-evolution. \cref{fig-results} (top row) shows our model's prediction accuracy for the ten random and six Pauli states. The top left panel [\cref{fig-results}(a)] corresponds to the free-evolution case, where we present the relative error in the prediction of our model as a function of time compared with the experimental results. The relative error is defined as $({\rm mean_{exp}} - {\rm mean_{sim}})/{\rm mean_{exp}}$, where ${\rm mean_{exp}}$ is the bootstrapped average over $8192$ experimental repetition and  ${\rm mean_{sim}}$ is the average over $600$ trajectory simulations of the hybrid Redfield model for any given time instant.
The box plot contains the spread in the relative error over all $16$ states, showing that the relative error of the model is always below $8\%$ over the total time considered here. The median and the mean over the $16$ states are confined well below $3\%$ for every instant. 

The performance of our model for the DD-evolution experiments is shown in the top right panel [\cref{fig-results}(b)]. Here the relative error is always below $2\%$. The median and the mean are below $1\%$. The closer agreement of the model with the DD-evolution experiments is expected, given that in contrast to the free-evolution experiments, DD suppresses the low-frequency noise affecting the main qubit, and the limitations of our fluctuator model of this noise is a likely source of modeling error.

The first and fourth columns of \cref{fig-comparison} show the relative error of our model over the $16$ states and all $70$ instants of the free and DD-evolution experiments, respectively. The results of the latter are better, as expected from \cref{fig-results}. In both cases, however, we observe that the model has a relative prediction error of just a few percent.

\subsection{Simplified Models}
As discussed in \cref{Sec-numerics}, our numerical simulations use the circuit model Hamiltonian of a transmon qubit truncated to the four lowest transmon eigenstates. The gates are applied with time-dependent pulses of non-zero duration. To test the robustness of our detailed model and learning procedure, we compare it with two simpler models, SM1 and SM2, derived from our detailed model. The simpler models use a more straightforward qubit description where we truncate the transmon Hamiltonian to only two levels. The time-dependent gates are replaced with instantaneous (zero-duration), ideal gates. Moreover, we focus only on the Ohmic bath terms in \cref{eq:H_int}, thus simplifying the noise model by removing the classical fluctuators. 
To test these simpler models' predictive power, we follow the same procedure as in \cref{sec-methodology}, but using only Steps I and II. The difference between SM1 and SM2 lies in Step II, where SM1 uses the DD-evolution experiments for the six Pauli states, whereas SM2 uses the free-evolution experiments for the same states. In both cases, we extract the model parameters and then use the resulting learned models to predict the outcomes of both the free-evolution and the DD-evolution experiment.

\cref{fig-comparison} shows the comparison between our detailed model and the simpler models SM1 and SM2.
We observe that SM1 has the largest relative error for the free-evolution experiments, whereas SM2 has the largest relative error for the DD-evolution case. Our full model has the smallest relative error among the three models considered here for \emph{both} the free and {DD} evolution experiments. However, the performance of SM1 and SM2 is essentially indistinguishable from the full model results in the {DD} and free-evolution cases, respectively. This is not unexpected, given that SM1 (SM2) is trained on the DD (free) evolution experiments and predicts these well. In other words, SM1 (SM2) captures the high (low)-frequency noise well, as expected since for SM1, the use of DD suppresses most of the low-frequency noise, while for SM2, the use of free-evolution means that the low-frequency noise remains a dominant source of decoherence. The added value of the detailed model and the use of Step III is that this provides enough information to capture both the low and high-frequency components of the noise, which yields a more complete model with better predictive power. We do note that taking the qubit approximation and treating DD pulses as instantaneous does not seem to appreciably worsen the performance of the simple models in their regime of accuracy, as SM1 (SM2) are roughly as accurate as the full model in the DD (free) evolution case. This suggests that an intermediate model, taking the qubit and instantaneous-pulse approximations but retaining the fluctuators, may be accurate and computationally efficient.

\begin{table}[b]
\begin{tabular}{|c|c|c|}
\hline 
Params$/(2\pi)$  & Quito & Lima \\
\hline
$g_x$ [MHz] & $5.734$ &  $4.782$\\
$g_z$ [MHz] & $4.413$ & $9.393$\\ 
$\omega_x^c$ [GHz] & $1.948$ & $2.340$\\
$\omega_z^c$ [MHz] & $5.690$ & $5.979$\\
$b$ [MHz] & $0.598$ & $0.323$\\
$\gamma_{\max}$ [GHz] & $0.051$ & $0.083$\\
\hline
\end{tabular}
\caption{System-bath parameter values extracted using the fitting procedure of \cref{sec:fit}, and corresponding to the minima indicated by the green circles in \cref{fig-step_1} for Quito (top row) and Lima (bottom row).}
\label{tab:params}
\end{table}

\section{Calibration-independent learning}
\label{sec:lima-results}

For multi-qubit superconducting processors, calibrating single-qubit drive frequencies is crucial for gate operations. In the presence of $ZZ$ coupling, the state of the spectator qubits modifies the eigenfrequency of the main qubit. This results in different 
choices of calibration frequencies depending on the spectator qubits' state~\cite{tripathi2021suppression}.  So far, we have focused on one particular device (Quito), which is calibrated while keeping the spectator qubits in the $\ket{+}$ state (see \cref{app:B} and Ref.~\cite{tripathi2021suppression}). The all-$\ket{+}$ or all-$\ket{0}$ are usually the two preferred choices for the spectators' state while calibrating a given qubit in a multi-qubit processor. When we perform state protection experiments on the main qubit initialized in the $\ket{+}$ state while keeping all the spectator qubits in $\ket{0}$, there exists a frequency mismatch which results in $ZZ$ crosstalk oscillations (see \cref{app:B}); to remove these oscillations, we applied DD to the spectator qubits before starting our noise learning procedure.  When device calibration is performed while keeping the spectators in the $\ket{0}$ state, a similar state protection experiment does not result in any oscillations, as evidenced by our Lima results (see \cref{app:B}).

Therefore, we extend our noise learning method to Lima in this section. Following our procedure from \cref{sec-methodology}, we again perform free-evolution (no DD is applied to any qubit) and DD-evolution (XY4 is applied just to the main qubit) experiments. The only difference from the Quito case is that, for the reasons explained above, the free-evolution experiment does not require the application of DD to the spectator qubits to suppress crosstalk oscillations. \cref{fig-step_1} (bottom row) shows the contour plots for each of the three steps involved in our learning methodology as described in \cref{sec:fit}. We find the global minima at the parameter values given in \cref{tab:params}.
Comparing the Quito and Lima parameters in \cref{tab:params}, we observe that the coupling strength $g_z$ of the Ohmic bath along the $z$-axis is roughly double for Lima, whereas the strength of the fluctuators is roughly double for Quito. This indicates that Quito is more prone to low-frequency ($1/f$) noise.

\begin{figure}[t]
	\includegraphics[width=1\linewidth]{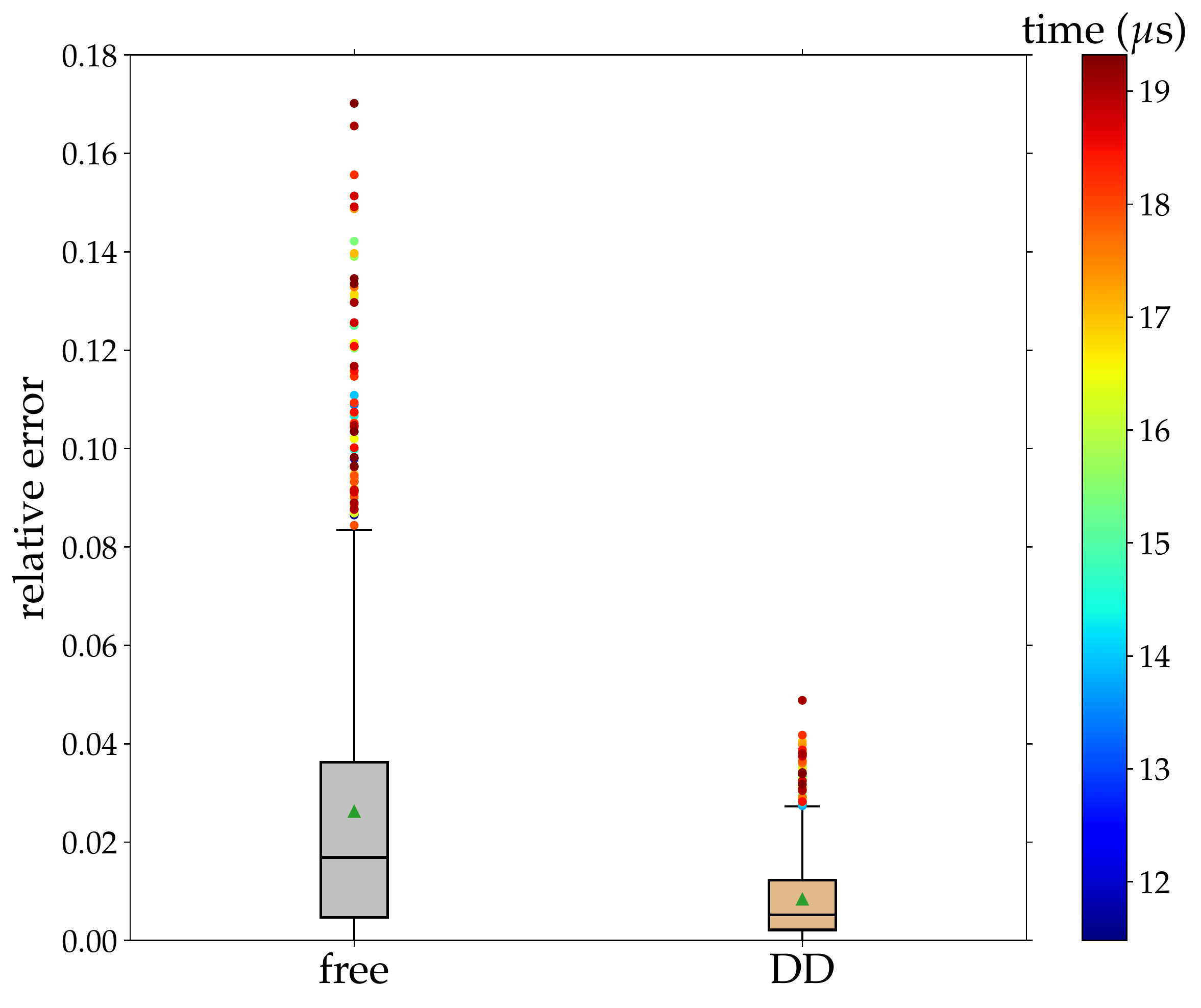}
	\caption{Integrated relative error results for the Lima processor. We compare the relative errors between the free-evolution and DD-evolution experiments. Each box contains $16$ initial states and all $70$ time instants varying from $0$ to $19.6\;\mu s$. The color bar indicates the time evolved for the outliers. All the outliers correspond to times longer than $11\mu$s. }
	\label{fig-lima-combined}
\end{figure}

\cref{fig-results} (bottom row) shows the Lima prediction results for the $16$ different initial states described above, using the learned noise parameters. The bottom left panel [\cref{fig-results}(c)] shows the relative error in the prediction of our model of the free-evolution experiments as a function of time compared to the experimental results for all $16$ states. The bottom right panel [\cref{fig-results}(d)] shows the same for the DD-evolution experiments. The relative error is always below $17\%$ and $4\%$ for free-evolution and DD-evolution, respectively. Similar to the Quito case, the relative error is significantly lower for the DD-evolution experiments. For longer evolution times ($\gtrsim 14 ~\mu$s), the agreement worsens for the free-evolution experiments. As for the Quito results, the closer agreement of the model with the DD-evolution experiments is likely because DD suppresses the low-frequency noise affecting the main qubit, which dominates the free-evolution case's simulation error.

\cref{fig-lima-combined} shows the time-integrated version of the Lima results of \cref{fig-results}(c,d), where we have combined all $16$ states and $70$ time instants into one box each for the free-evolution and DD-evolution experiments. Except for a few outliers, almost all the data points for the free-evolution case have relative errors below $9\%$. The outliers all correspond to evolution times longer than $11\mu$s, as indicated by the color bar. Similarly, for DD-evolution, all data points have relative errors below $3\%$, except for a few outliers. There are two main reasons for the larger errors at longer evolution times. First, the Redfield equation is based on the weak coupling approximation, and its accuracy degrades as we increase the simulation time (for rigorous error bounds, see Ref.~\cite{Mozgunov:2019aa}). Second, as time increases, the effect of distant fluctuators is increasingly felt, thus reducing the accuracy of our fluctuator model, which relies on a fixed number of fluctuators. Adding more fluctuators and exploring a distribution of fluctuator strengths, as opposed to our assumption of a fixed fluctuator strength $b$, is expected to improve the predictive power of our model.

Finally, as another test of our learning procedure, we computed $T_1$ from the learned models, using the values we report in \cref{table1}. We did this by simulating the fidelity decay of the $\ket{1}$ state for $19.6\mu$s and fitting $\exp(-t/T_1)$ to estimate $T_1$. 
We find $T_1=92.5\mu$s for Quito and $86.5\mu$s for Lima, compared to the reported $T_1=98.6\mu$s and $105\mu$s, respectively. Our result gives the correct order of magnitude and is particularly reasonable for Quito. The discrepancy may be in part due to the relatively short simulation time of $19.6\mu$s (larger times become prohibitively expensive). In addition, the discrepancy for Quito is smaller because we use three rounds of experiments within the same calibration to model the noise, which removes short-time fluctuations via bootstrap averaging, whereas for Lima, we used only one repetition. As $T_1$ often drifts significantly over hour-long time scales, we would not expect the Lima prediction to line up exactly with the reported $T_1$.

\section{Summary and Conclusions}
\label{Sec-conclusion}

This work presents a detailed noise model for transmon qubits consisting of both low ($1/f$-like) and high-frequency noise components based on a hybrid Redfield master equation. 
We designed an iterative three-step procedure to extract the unknown system-bath and bath parameters from a few simple ``free-evolution'' and ``DD-evolution'' experiments, as illustrated in \cref{fig-sequences}, using the six Pauli matrix eigenstates. In both cases, we used dynamical decoupling (DD) to suppress diagonal ($ZZ$) qubit crosstalk~\cite{tripathi2021suppression} so that the remaining dominant noise effect on the main qubit (the qubit of interest) is decoherence. Our model treats the transmon qubit as a four-level system based on the circuit model description of transmons (\cref{Sec-numerics}) and treats the DD pulses as realistic time-dependent gates subject to quantum control (DRAG). 

Once the unknown system-bath and bath parameters are extracted, we compare the model predictions with new experiments and a larger set of initial states and demonstrate that the model predicts the experimental results of free-evolution and DD-evolution with a relative error below $8\%$ and $2\%$ for Quito, and below $9\%$ and $3\%$ for Lima, respectively. This is based on a test with the six Pauli matrix eigenstates and ten random states for a total duration of up to $19.6\;\mu s$. The relative errors are higher for larger times (see \cref{fig-results}), as expected because the Redfield model is based on the weak-coupling approximation, and its accuracy degrades as the simulation time is increased~\cite{Mozgunov:2019aa}. 

To test the robustness of our model, we performed a comparison with two simpler, two-level models with instantaneous pulses; while these models capture either the low or the high-frequency noise, the full model captures both types of noise. Furthermore, our method is applicable independently of the particular device-calibration procedure followed, as witnessed by the agreement we find for both Quito and Lima -- devices with different drive-frequency calibrations. 

The low relative error we find in the case of DD-evolution experiments ($<2\%$ and $<3\%$ for Quito and Lima, respectively) suggests that our full noise model helps model gate dynamics under the influence of decoherence. The model can also be used to study several qubits in parallel as long as there is no direct crosstalk between these qubits. Extending our noise model beyond weak coupling and using it to analyze and improve entangling gates are promising future directions. 

We hope this work will benefit experimental groups working with superconducting qubits by helping them understand and learn experimental noise using a first principles approach, which only requires a set of straightforward experiments.

\acknowledgements
We thank Mostafa Khezri, Ka-Wa Yip, and Humberto Munoz Bauza for several helpful discussions. This material is based upon work supported by the National Science Foundation, the Quantum Leap Big Idea under Grant No.~OMA-1936388. We acknowledge the use of IBM Quantum services for this work. The views expressed are those of the authors and do not reflect the official policy or position of IBM or the IBM Quantum team.

\appendix

\section{Derivation of time-dependent drives}
\label{app:derivation}

We start with \cref{eq:transmon-drive} from the main text, and to obtain \cref{eq:hatp}, we focus on the charge coupling matrix elements for transmon qubits. The selection rules of the transmon qubit due to its cosine potential dictate that $g_{k,k\pm 2} = 0$ $\forall k$~\cite{transmon-invention}. The next order coupling terms, $g_{k,k\pm 3}$, are proportional to the ratio of the anharmonicity and the qubit frequency: $\eta_q/\omega_q$~\cite{Khezri_thesis}. Therefore $g_{k, k\pm 1}$ (the coupling between nearest levels) is dominant, and all other couplings can be ignored, giving \cref{eq:hatp}. 

We can write the approximated charge operator $\hat{p} \equiv \sum_{k\ge 0}\bra{k}\hat{n} \ket{k+1}\ketb{k}{k+1} + {\rm h.c.}$ of \cref{eq:hatp} in terms of effective creation and annihilation operators for the eigenstates of the transmon, i.e., $\hat{p} = i(a^\dag - a)$, where 
\beq
\label{eq:a-def}
a \equiv i\sum_{k\ge 0} g_{k, k+1}\ketb{k}{k+1} = i\sum_{k\ge 0} \tilde{g}_{k} \sqrt{k+1}\ketb{k}{k+1} ,
\eeq
with
\bes
\label{eq:g-coupling}
\begin{align}
\label{eq:g-coupling-a}
g_{k, k+1} &\equiv \bra{k}\hat{n}\ket{k+1} = g\frac{\bra{k}\hat{n}\ket{k+1}}{\bra{0}\hat{n}\ket{1}}  \\
\label{eq:g-coupling-b}
& \equiv \tilde{g}_k \sqrt{k+1} ,
\end{align} 
\ees
and $g\equiv g_{0,1} = \bra{0}\hat{n}\ket{1} = \tilde{g}_0$. We note that $g_{k, k+1}  = g_{k, k+1}^*$, which follows from the fact that $\hat{n}$ is the number operator for Cooper pairs.
We note further that to first order in $\eta_{q}/\omega_q$~\cite{Khezri_thesis} 
\begin{equation}
\tilde{g}_{k} \approx  g\left(1-\frac{k}{2} \frac{\eta_{q}}{\omega_{q}}\right) .
\label{eq:g-perturb}
\end{equation}
We defined $\tilde{g}_k(=\tilde{g}_k^*)$ in \cref{eq:g-coupling-b} to include all higher order perturbative corrections and do not use the approximation \cref{eq:g-perturb} in our numerical calculations. However, we note that the leading order correction to $\tilde{g}_k$ is of the same order as $g_{k,k\pm 3}$, i.e., proportional to $\eta_q/\omega_q$.

\cref{eq:transmon-drive} becomes
\bes
\label{eq:H-sys-eig2}
\begin{align}
H_{\mathrm{sys}} &= H_{\mathrm{trans}}^{\mathrm{eigen}}  + H_{\mathrm{drive}}
\label{eq:H-sys-eig2-a}\\
H_{\mathrm{drive}} &\equiv i\varepsilon(t) \cos \left(\omega_{\rm d} t+\phi_{\rm d}\right)\left(a^\dag - a\right) .
\label{eq:H-sys-eig2-b}
\end{align}
\ees

For simplicity, we consider the case when $\phi_{\rm d}=0$ in \cref{eq:H-sys-eig2-b},  
and transform the Hamiltonian $H_{\rm sys}^{\rm eigen}$ into a frame rotating with $U_{\rm d} = {e}^{i\omega_{\rm d}\hat{N}t}$. Here $\hat{N} = \sum_{k\ge 0} k\ketb{k}{k}
$ is the transmon number operator. In this frame, rotating  with the drive frequency, the effective Hamiltonian is given by 
\begin{subequations}
\label{eq:H-sys-drive}
\begin{align}
\tilde{H}_{\mathrm{sys}} & = U_{\rm d} H_{\mathrm{sys}} U_{\rm d}^{\dagger}+i  \dot{U}_{d}U_{\rm d}^{\dagger} \label{eq:drive-a} \\
 & = H_{\mathrm{trans}}^{\mathrm{eigen}}+U_{\rm d} H_{\mathrm{drive}} U_{\rm d}^{\dagger} +i  \dot{U}_{d}U_{\rm d}^{\dagger}\label{eq:drive-c}\\
 & =  \sum_{k\ge 0}\left(\omega_{k}-k \omega_{\rm d}\right)\ketb{k}{k} + U_{\rm d} H_{\mathrm{drive}} U_{\rm d}^{\dagger}  \label{eq:H-sys-drive-d}
\end{align}
\end{subequations}
where in going from \cref{eq:drive-a} to \cref{eq:drive-c}, we used the fact that $\hat{N}$ commutes with $H_{\mathrm{trans}}^{\mathrm{eigen}}$ [\cref{eq:eig-transmon}].

Note that $a$ and $a^{\dagger}$ should not be confused with the harmonic oscillator raising and lowering operators since 
\begin{align}
[a, a^{\dagger}] &=\sum_{k\ge 1} \left((k+1)\tilde{g}_{k}^{2} - k\tilde{g}_{k-1}^{2}\right)\ketb{k}{k} + \tilde{g}_0\ketb{0}{0} \notag \\
&\neq I.
\end{align}
For the harmonic oscillator case, $\tilde{g}_k=1$ for all $k$ and we obtain the usual commutation relation $\left[a, a^{\dagger}\right]= I$. Despite this, we have, as for the harmonic oscillator:
\begin{subequations}
\begin{align}
[a, \hat{N}]=&[i\sum_{k\ge 0} \tilde{g}_{k} \sqrt{k+1}\ketb{k}{k+1}, \sum_{k'\ge 0}k'\ketb{k'}{k'}]\\
=& i\sum_{k\ge 0} \tilde{g}_{k}(k+1) \sqrt{k+1}\ketb{k}{k+1}\nonumber\\
&  -i\sum_{k\ge 0} \tilde{g}_{k}k \sqrt{k+1}\ketb{k}{k+1}\\
=& i\sum_{k\ge 0} \tilde{g}_{k} \sqrt{k+1}\ketb{k}{k+1} ,
\label{eq:comm-1}
\end{align}
\end{subequations}
so that:
\begin{equation}
[a, \hat{N}]=a \ , \quad [a^{\dagger}, \hat{N}]=-a^{\dagger}
\label{eq:comm-2}
\end{equation}

Let us write the last term in \cref{eq:H-sys-drive-d} as
\begin{equation}
U_{\rm d} H_{\mathrm{drive}} U_{\rm d}^{\dagger}= i\frac{\varepsilon(t)}{2}\left(e^{i \omega_{\rm d} t}+e^{-i \omega_{\rm d} t}\right) U_{\rm d}\left(a^\dag - a\right) U_{\rm d}^{\dagger} .
\label{eq:H-eig-rot}
\end{equation}
Using \cref{eq:comm-2}, we then have:
\begin{equation}
U_{\rm d}\left(a^\dag - a\right) U_{\rm d}^{\dag}=e^{i \omega_{\rm d} t} a^{\dag} -e^{-i \omega_{\rm d} t} a .
\end{equation}
Making the rotating wave approximation, we ignore the fast-oscillating terms with frequencies $\pm 2\omega_{\rm d}$, and \cref{eq:H-eig-rot} reduces to
\begin{equation}
U_{\rm d} H_{\mathrm{drive}} U_{\rm d}^{\dagger} \approx i\frac{\varepsilon(t)}{2}\left(a^\dag - a\right) .
\end{equation}
Combining this with \cref{eq:a-def,eq:H-sys-drive}, the effective Hamiltonian for a drive that results in a rotation about the $x$-axis (in the qubit subspace) is given by 
\begin{align}
\tilde{H}_{\mathrm{sys}}&=\sum_{k\ge 0}\left(\omega_{k}-k \omega_{\rm d}\right)\ketb{k}{k} \\
& \quad +\frac{\varepsilon(t)}{2} \sum_{k\ge 0} \tilde{g}_{k} \sqrt{k+1}(\ketb{k}{k+1}+\ketb{k+1}{k}) , \nonumber
\end{align}
which is \cref{eq:Xgate} of the main text.
By tuning $\phi_{\rm d}$, we can implement a rotation about any axis in the $(x,y)$ plane in the qubit subspace.

\section{Data collection and analysis methodology}
\label{app:data-analysis}

\begin{figure}
\includegraphics[width=0.6\linewidth]{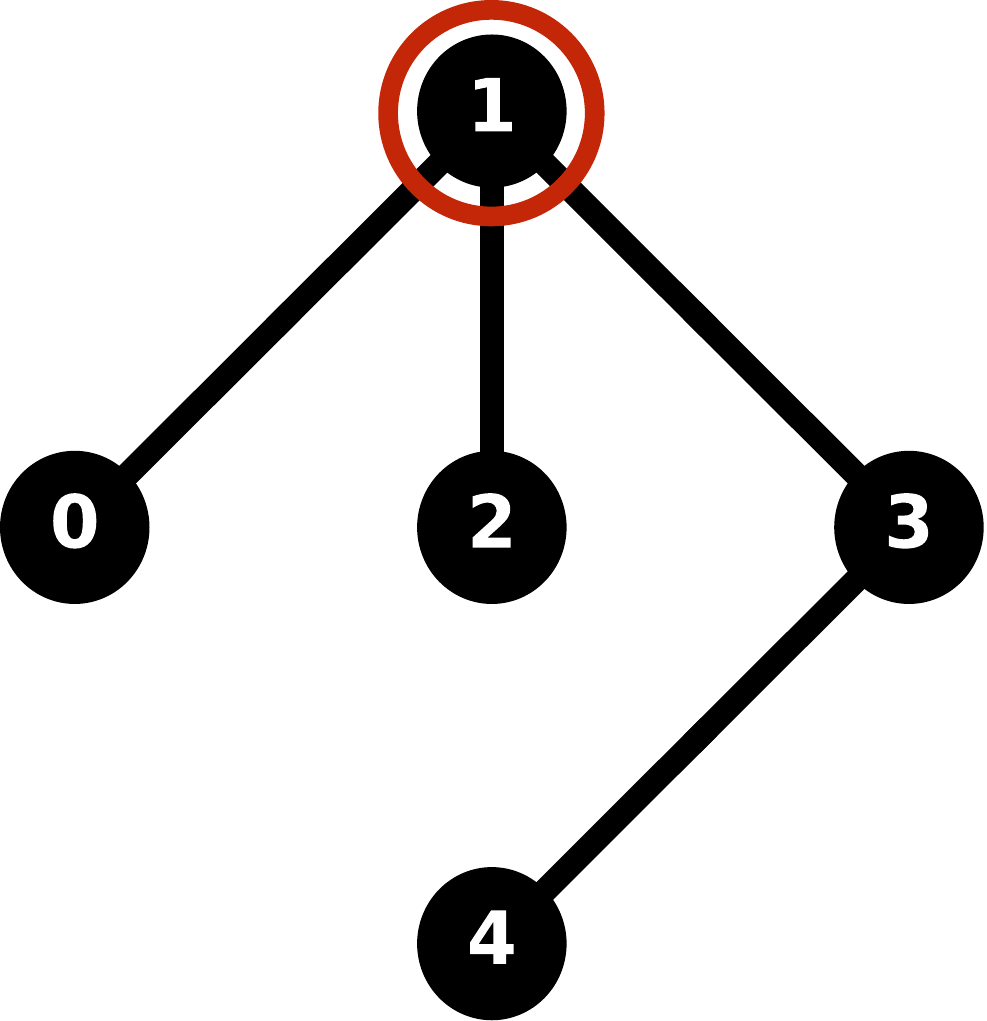}
\caption{Schematic layout of the Quito and Lima processors. The main qubit in our work is 1; we refer to the rest as spectators.}  
\label{fig:layout}
\end{figure}

We used the IBMQE processor ibmq$\_$quito (Quito) and ibmq$\_$lima (Lima), whose layout is shown schematically in \cref{fig:layout}. For both Quito and Lima, we use qubit 1 (Q1) as the main qubit. These are five-qubit processors consisting of superconducting transmon qubits. Various calibration details and hardware specifications relevant to the qubits and gates used in this work are provided in \cref{table1}. 

\begin{table}[t]
\begin{adjustbox}{width=1.01\columnwidth,center}
\centering
\begin{tabular}{ |p{3cm}||p{1.2 cm}|p{1.2 cm}|p{1.2 cm}|p{1.2 cm} | p{1.2 cm}|}
 \hline
 Quito  &  Q0 & Q1 & Q2 & Q3 & Q4\\
 \hline
Qubit freq. (GHz) & 5.3006 &5.0806 & 5.3220 &  5.1637 & 5.0524\\
$\eta_{q}$ (MHz)& 331.5 & 319.2 & 332.3 & 335.1 & 319.3\\
$T_1\;(\mu \rm s)$  & 86.7   &  98.6 & 61.5  & 111.5 & 85.7\\
$T_2\;(\mu \rm s)$   & 132.5  &  149.0 &  78.9 &  22.7 & 136.7\\
sx gate error [$10^{-2}$] &  0.0302  & 0.0243 & 0.1042 & 0.0629 & 0.0884\\
sx gate length\;(ns) & 35.556 & 35.556 & 35.556 & 35.556 & 35.556\\
readout error [$10^{-2}$] & 3.91  & 2.10   & 6.42 & 2.28 & 2.00\\
  \hline
  \hline
Lima & Q0 & Q1 & Q2 & Q3 & Q4\\
\hline
Qubit freq. (GHz) & 5.0297 &     5.1277 & 5.2474 &  5.3026 & 5.0920\\
$\eta_{q}$ (MHz)& 335.7 & 318.3 & 333.6 & 331.2 & 334.5\\
$T_1\;(\mu \rm s)$  & 125.2   &  105.7  & 88.1  & 59.9 & 23.2\\
$T_2\;(\mu \rm s)$   & 194.3  &  136.2 &  123.3 &  16.8 & 21.1\\
sx gate error [$10^{-2}$] &  0.0230    & 0.0189 & 0.0308 & 0.0251 & 0.0578\\
sx gate length\;(ns) & 35.556 & 35.556 & 35.556 & 35.556 & 35.556\\
readout error [$10^{-2}$] & 1.73  & 1.40  & 1.69 & 2.42 & 4.820\\
 \hline
 
\end{tabular}
\end{adjustbox}
\caption{Specifications of the Quito (top row) and Lima (bottom row) devices accessed on September 1, 2021, and January 1, 2023, respectively. The sx ($\sqrt{\s^x}$) gate forms the basis of all the single qubit gates, and any single qubit gate of the form $U3(\theta, \phi, \lambda)$ is composed of two sx and three rz$(\lambda) = {\rm exp}(-i\frac{\lambda}{2}\s^z)$ gates (which are error-free and take zero time, as they correspond to frame updates).}
\label{table1}
\end{table}

For each initial state, we selected $70$ equidistant time instants ($19.6~\mu$s$/70$), with each such instant corresponding to an integer number of cycles of the XY4 DD sequences. We generated a circuit according to the scheme given in \cref{fig-sequences} for each such initial state and each such instant. We sent all $70$ such circuits in one job (the maximum allowed number of circuits per job is $75$), and each job was repeated $8192$ times. We ensured that all the jobs were sent consecutively within the same calibration cycle to avoid charge-noise-dependent fluctuations and variations in critical features over different calibration cycles. 

We measured only the main qubit in the $Z$ basis, each measurement yielding either $0$ or $1$.
We computed the empirical fidelity $F^{(e)}$ as the number of favorable outcomes ($0$) to the total number of experiments ($8192$ per initial state and per measurement time instant). This is a proxy for the $F_{\ket{\psi}} = {\rm Tr}\left[ U^{-1} \mathcal{E}\left( U\ketb{0}{0}U^{-1}\right) U \right]$, where $U$ represents $U3(\theta, \phi, \lambda)$ and $\mathcal{E}$ represents the quantum map of the main qubit corresponding to any of the three types of experiments described in the main text and in \cref{app:B} below.
  
Error bars were then generated using the standard bootstrapping procedure, where we resample (with replacement) counts out of the experimental counts' dictionary (i.e., the list of $0/1$ measurement outcomes per state and instant) and create several new dictionaries. The final fidelity and error bars are obtained by calculating the mean and standard deviations over the fidelities of these newly resampled dictionaries. Using ten such resampled dictionaries of the counts sufficed to give small error bars. We report the final fidelity with $2\sigma$ error bars, corresponding to $95\%$ confidence intervals. 

\section{Experimental fidelity results}
\label{app:B}

\begin{figure*}[t]
\hspace*{-1.5cm}
\includegraphics[width=1.2\linewidth]{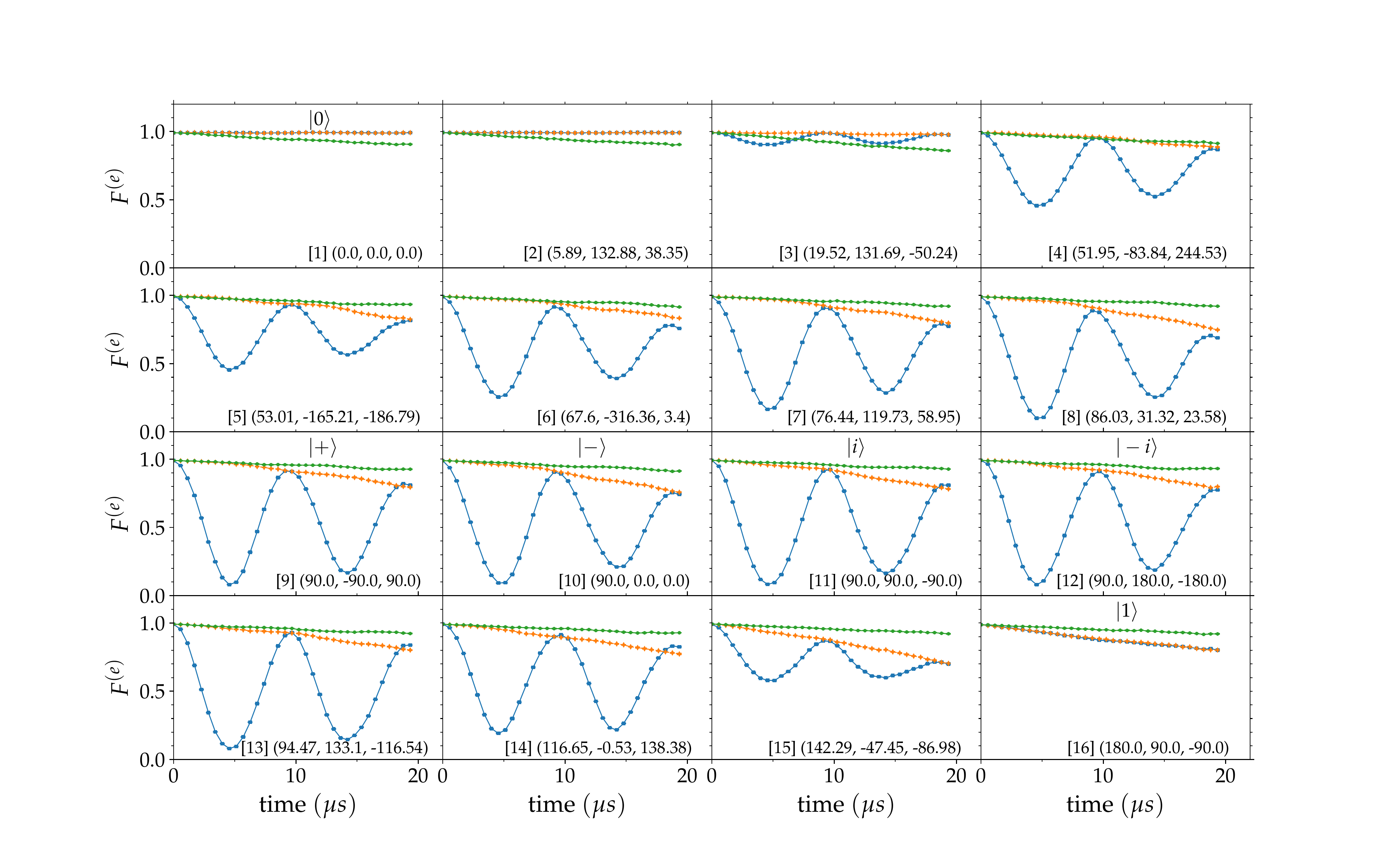}
\includegraphics[width=.6\linewidth]{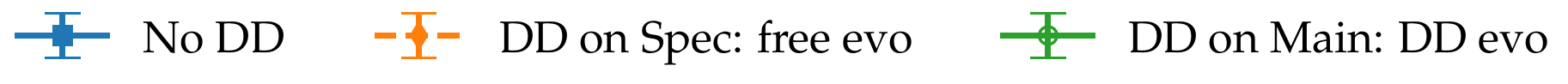}
\caption{Fidelity results for the Quito processor, for the $16$ different initial states of the main qubit. The caption of each of the panels gives $(\theta, \phi, \lambda)$ in degrees, parametrizing the initial state $\ket{\psi} = U3(\theta, \phi, \lambda)\ket{0}$ (panels are arranged in increasing order of $\theta$, the polar angle with the $z$-axis). These are the $6$ Pauli states (panels 1,9-12,16) and the $10$ Haar-random states. Blue curves (squares): no DD is applied, resulting in coherent oscillations due to crosstalk. Orange curves (diamonds): DD (XY4) is applied just to the spectator qubits; the resulting suppression of crosstalk between the main qubit and the spectator qubits removes the oscillations. These are what we call the free-evolution experiments in the main text. Green curves (circles): DD (XY4) is applied just to the main qubit, suppressing both crosstalk and errors due to environment-induced noise. Results are averaged over three different runs of experiments. All of the data was acquired on Sep. 1, 2021. Error bars are smaller than the markers.}
\label{fig-experiments}
\end{figure*}

For the Quito device, \cref{fig-experiments} shows the results of three different types of experiments for the $16$ states consisting of $6$ Pauli states and $10$ Haar-random states. In the first type of experiment, we apply a series of identity gates separated by barriers on all the qubits (main qubits and the spectator qubits). All the qubits always start in the $\ket{0}$ state. The second and third types are the experiments discussed in the main text: free-evolution, where we apply an XY4 DD sequence to the spectator qubits and identity gates on the main qubit, and DD-evolution, where we apply the XY4 DD sequences to the main qubit and identity gates to the spectator qubits (see \cref{fig-sequences}).

\begin{figure*}[t]
\hspace*{-1.5cm}
\includegraphics[width=1.2\linewidth]{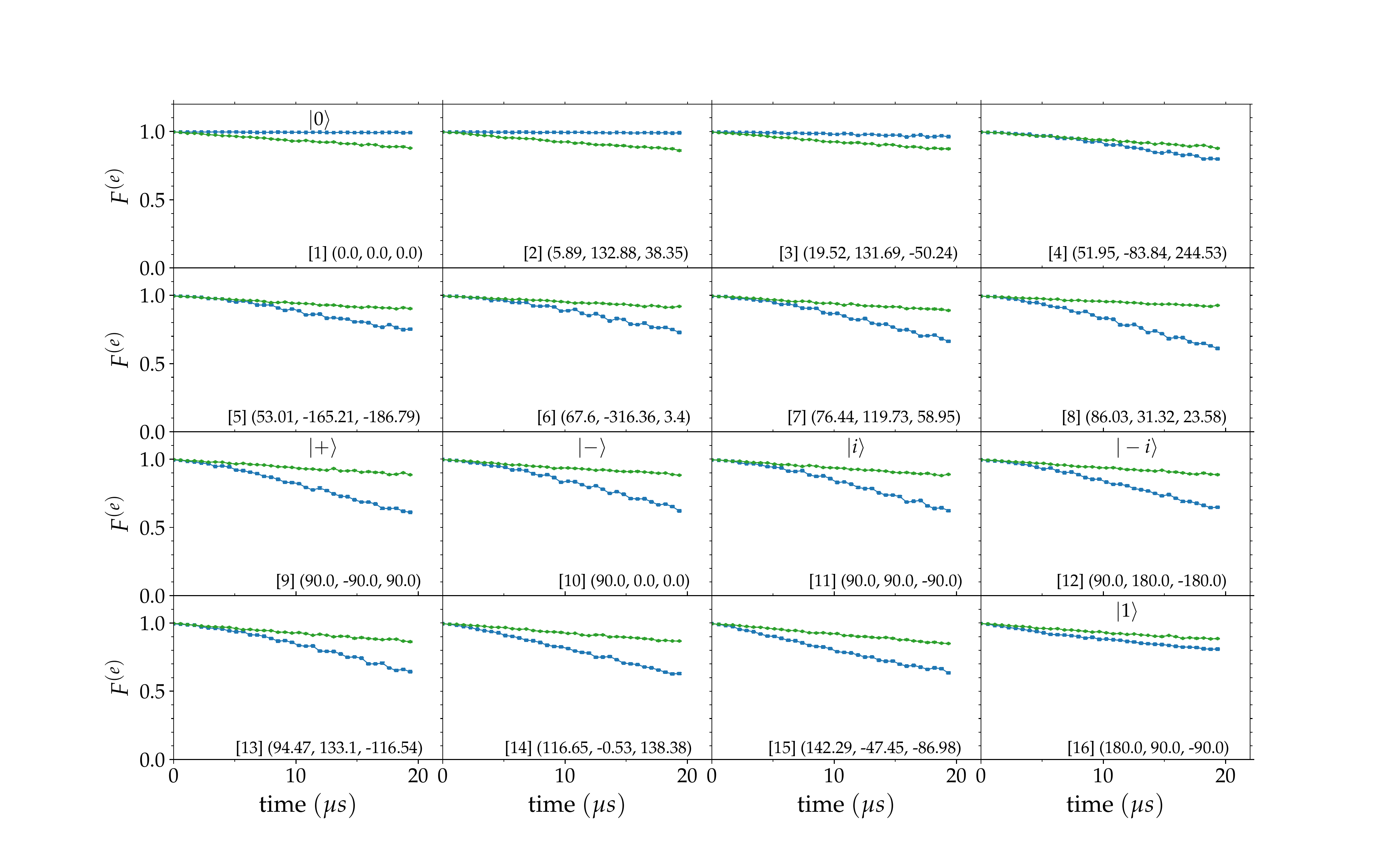}
\includegraphics[width=.5\linewidth]{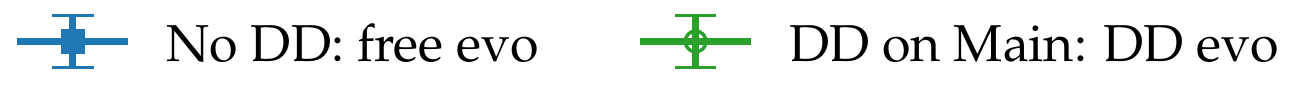}
\caption{Fidelity results for the Lima processor, for the $16$ different initial states of the main qubit. The caption of each of the panels gives $(\theta, \phi, \lambda)$ in degrees, parametrizing the initial state $\ket{\psi} = U3(\theta, \phi, \lambda)\ket{0}$ (panels are arranged in increasing order of $\theta$, the polar angle with the $z$-axis). These are the $6$ Pauli states (panels 1,9-12,16) and the $10$ Haar-random states. Blue curves (squares): no DD is applied. In contrast to \cref{fig-experiments}, there are no cross-talk oscillations. This is because all the spectator qubits are kept in the ground state $\ket{0}$, and calibration for Lima is done in the same spectators' state. These are what we call the free-evolution experiments for Lima in the main text. Green curves (circles): DD (XY4) is applied just to the main qubit, suppressing errors due to environment-induced noise. All of the data was acquired on Jan. 1, 2023. Error bars are smaller than the markers.}
\label{fig-experiments-lima}
\end{figure*}

For Lima, we only show two types of experiments in Fig.~\ref{fig-experiments-lima}. The first is the free-evolution experiment, where we prepare some initial state of the main qubit, apply a series of identity gates, and measure the computational basis. The second type is the usual DD experiment, where we apply a series of XY4 sequences to the main qubit and the identity operation to all the spectator qubits. In both types of experiments, the spectator qubits are always initialized in the ground state $\ket{0}$. In the Lima case, applying a DD sequence to the spectator qubits is unnecessary.

\begin{thebibliography}{58}%
\makeatletter
\providecommand \@ifxundefined [1]{%
 \@ifx{#1\undefined}
}%
\providecommand \@ifnum [1]{%
 \ifnum #1\expandafter \@firstoftwo
 \else \expandafter \@secondoftwo
 \fi
}%
\providecommand \@ifx [1]{%
 \ifx #1\expandafter \@firstoftwo
 \else \expandafter \@secondoftwo
 \fi
}%
\providecommand \natexlab [1]{#1}%
\providecommand \enquote  [1]{``#1''}%
\providecommand \bibnamefont  [1]{#1}%
\providecommand \bibfnamefont [1]{#1}%
\providecommand \citenamefont [1]{#1}%
\providecommand \href@noop [0]{\@secondoftwo}%
\providecommand \href [0]{\begingroup \@sanitize@url \@href}%
\providecommand \@href[1]{\@@startlink{#1}\@@href}%
\providecommand \@@href[1]{\endgroup#1\@@endlink}%
\providecommand \@sanitize@url [0]{\catcode `\\12\catcode `\$12\catcode
  `\&12\catcode `\#12\catcode `\^12\catcode `\_12\catcode `\%12\relax}%
\providecommand \@@startlink[1]{}%
\providecommand \@@endlink[0]{}%
\providecommand \url  [0]{\begingroup\@sanitize@url \@url }%
\providecommand \@url [1]{\endgroup\@href {#1}{\urlprefix }}%
\providecommand \urlprefix  [0]{URL }%
\providecommand \Eprint [0]{\href }%
\providecommand \doibase [0]{https://doi.org/}%
\providecommand \selectlanguage [0]{\@gobble}%
\providecommand \bibinfo  [0]{\@secondoftwo}%
\providecommand \bibfield  [0]{\@secondoftwo}%
\providecommand \translation [1]{[#1]}%
\providecommand \BibitemOpen [0]{}%
\providecommand \bibitemStop [0]{}%
\providecommand \bibitemNoStop [0]{.\EOS\space}%
\providecommand \EOS [0]{\spacefactor3000\relax}%
\providecommand \BibitemShut  [1]{\csname bibitem#1\endcsname}%
\let\auto@bib@innerbib\@empty
\bibitem [{\citenamefont {Nakamura}\ \emph {et~al.}(1999)\citenamefont
  {Nakamura}, \citenamefont {Pashkin},\ and\ \citenamefont
  {Tsai}}]{Nakamura:99}%
  \BibitemOpen
  \bibfield  {author} {\bibinfo {author} {\bibfnamefont {Y.}~\bibnamefont
  {Nakamura}}, \bibinfo {author} {\bibfnamefont {Y.~A.}\ \bibnamefont
  {Pashkin}},\ and\ \bibinfo {author} {\bibfnamefont {J.~S.}\ \bibnamefont
  {Tsai}},\ }\bibfield  {title} {\bibinfo {title} {Coherent control of
  macroscopic quantum states in a single-cooper-pair box},\ }\href
  {http://dx.doi.org/10.1038/19718} {\bibfield  {journal} {\bibinfo  {journal}
  {Nature}\ }\textbf {\bibinfo {volume} {398}},\ \bibinfo {pages} {786 EP }
  (\bibinfo {year} {1999})}\BibitemShut {NoStop}%
\bibitem [{\citenamefont {Mooij}\ \emph {et~al.}(1999)\citenamefont {Mooij},
  \citenamefont {Orlando}, \citenamefont {Levitov}, \citenamefont {Tian},
  \citenamefont {van~der Wal},\ and\ \citenamefont {Lloyd}}]{Mooij:99}%
  \BibitemOpen
  \bibfield  {author} {\bibinfo {author} {\bibfnamefont {J.~E.}\ \bibnamefont
  {Mooij}}, \bibinfo {author} {\bibfnamefont {T.~P.}\ \bibnamefont {Orlando}},
  \bibinfo {author} {\bibfnamefont {L.}~\bibnamefont {Levitov}}, \bibinfo
  {author} {\bibfnamefont {L.}~\bibnamefont {Tian}}, \bibinfo {author}
  {\bibfnamefont {C.~H.}\ \bibnamefont {van~der Wal}},\ and\ \bibinfo {author}
  {\bibfnamefont {S.}~\bibnamefont {Lloyd}},\ }\bibfield  {title} {\bibinfo
  {title} {Josephson persistent-current qubit},\ }\href
  {https://doi.org/10.1126/science.285.5430.1036} {\bibfield  {journal}
  {\bibinfo  {journal} {Science}\ }\textbf {\bibinfo {volume} {285}},\ \bibinfo
  {pages} {1036} (\bibinfo {year} {1999})}\BibitemShut {NoStop}%
\bibitem [{\citenamefont {Clarke}\ and\ \citenamefont
  {Wilhelm}(2008)}]{Clarke2008}%
  \BibitemOpen
  \bibfield  {author} {\bibinfo {author} {\bibfnamefont {J.}~\bibnamefont
  {Clarke}}\ and\ \bibinfo {author} {\bibfnamefont {F.~K.}\ \bibnamefont
  {Wilhelm}},\ }\bibfield  {title} {\bibinfo {title} {Superconducting quantum
  bits},\ }\href {https://doi.org/10.1038/nature07128} {\bibfield  {journal}
  {\bibinfo  {journal} {Nature}\ }\textbf {\bibinfo {volume} {453}},\ \bibinfo
  {pages} {1031} (\bibinfo {year} {2008})}\BibitemShut {NoStop}%
\bibitem [{\citenamefont {Hacohen-Gourgy}\ \emph {et~al.}(2016)\citenamefont
  {Hacohen-Gourgy}, \citenamefont {Martin}, \citenamefont {Flurin},
  \citenamefont {Ramasesh}, \citenamefont {Whaley},\ and\ \citenamefont
  {Siddiqi}}]{Hacohen2016}%
  \BibitemOpen
  \bibfield  {author} {\bibinfo {author} {\bibfnamefont {S.}~\bibnamefont
  {Hacohen-Gourgy}}, \bibinfo {author} {\bibfnamefont {L.~S.}\ \bibnamefont
  {Martin}}, \bibinfo {author} {\bibfnamefont {E.}~\bibnamefont {Flurin}},
  \bibinfo {author} {\bibfnamefont {V.~V.}\ \bibnamefont {Ramasesh}}, \bibinfo
  {author} {\bibfnamefont {K.~B.}\ \bibnamefont {Whaley}},\ and\ \bibinfo
  {author} {\bibfnamefont {I.}~\bibnamefont {Siddiqi}},\ }\bibfield  {title}
  {\bibinfo {title} {Quantum dynamics of simultaneously measured non-commuting
  observables},\ }\href {https://doi.org/10.1038/nature19762} {\bibfield
  {journal} {\bibinfo  {journal} {Nature}\ }\textbf {\bibinfo {volume} {538}},\
  \bibinfo {pages} {491} (\bibinfo {year} {2016})}\BibitemShut {NoStop}%
\bibitem [{\citenamefont {Kandala}\ \emph {et~al.}(2017)\citenamefont
  {Kandala}, \citenamefont {Mezzacapo}, \citenamefont {Temme}, \citenamefont
  {Takita}, \citenamefont {Brink}, \citenamefont {Chow},\ and\ \citenamefont
  {Gambetta}}]{Kandala:2017aa}%
  \BibitemOpen
  \bibfield  {author} {\bibinfo {author} {\bibfnamefont {A.}~\bibnamefont
  {Kandala}}, \bibinfo {author} {\bibfnamefont {A.}~\bibnamefont {Mezzacapo}},
  \bibinfo {author} {\bibfnamefont {K.}~\bibnamefont {Temme}}, \bibinfo
  {author} {\bibfnamefont {M.}~\bibnamefont {Takita}}, \bibinfo {author}
  {\bibfnamefont {M.}~\bibnamefont {Brink}}, \bibinfo {author} {\bibfnamefont
  {J.~M.}\ \bibnamefont {Chow}},\ and\ \bibinfo {author} {\bibfnamefont
  {J.~M.}\ \bibnamefont {Gambetta}},\ }\bibfield  {title} {\bibinfo {title}
  {Hardware-efficient variational quantum eigensolver for small molecules and
  quantum magnets},\ }\href {http://dx.doi.org/10.1038/nature23879} {\bibfield
  {journal} {\bibinfo  {journal} {Nature}\ }\textbf {\bibinfo {volume} {549}},\
  \bibinfo {pages} {242 EP } (\bibinfo {year} {2017})}\BibitemShut {NoStop}%
\bibitem [{\citenamefont {Minev}\ \emph {et~al.}(2019)\citenamefont {Minev},
  \citenamefont {Mundhada}, \citenamefont {Shankar}, \citenamefont {Reinhold},
  \citenamefont {Guti{\'e}rrez-J{\'a}uregui}, \citenamefont {Schoelkopf},
  \citenamefont {Mirrahimi}, \citenamefont {Carmichael},\ and\ \citenamefont
  {Devoret}}]{Minev2019}%
  \BibitemOpen
  \bibfield  {author} {\bibinfo {author} {\bibfnamefont {Z.~K.}\ \bibnamefont
  {Minev}}, \bibinfo {author} {\bibfnamefont {S.~O.}\ \bibnamefont {Mundhada}},
  \bibinfo {author} {\bibfnamefont {S.}~\bibnamefont {Shankar}}, \bibinfo
  {author} {\bibfnamefont {P.}~\bibnamefont {Reinhold}}, \bibinfo {author}
  {\bibfnamefont {R.}~\bibnamefont {Guti{\'e}rrez-J{\'a}uregui}}, \bibinfo
  {author} {\bibfnamefont {R.~J.}\ \bibnamefont {Schoelkopf}}, \bibinfo
  {author} {\bibfnamefont {M.}~\bibnamefont {Mirrahimi}}, \bibinfo {author}
  {\bibfnamefont {H.~J.}\ \bibnamefont {Carmichael}},\ and\ \bibinfo {author}
  {\bibfnamefont {M.~H.}\ \bibnamefont {Devoret}},\ }\bibfield  {title}
  {\bibinfo {title} {To catch and reverse a quantum jump mid-flight},\ }\href
  {https://doi.org/10.1038/s41586-019-1287-z} {\bibfield  {journal} {\bibinfo
  {journal} {Nature}\ }\textbf {\bibinfo {volume} {570}},\ \bibinfo {pages}
  {200} (\bibinfo {year} {2019})}\BibitemShut {NoStop}%
\bibitem [{\citenamefont {Arute}\ \emph {et~al.}(2019)\citenamefont {Arute},
  \citenamefont {Arya}, \citenamefont {Babbush}, \citenamefont {Bacon},
  \citenamefont {Bardin}, \citenamefont {Barends}, \citenamefont {Biswas},
  \citenamefont {Boixo}, \citenamefont {Brandao}, \citenamefont {Buell},
  \citenamefont {Burkett}, \citenamefont {Chen}, \citenamefont {Chen},
  \citenamefont {Chiaro}, \citenamefont {Collins}, \citenamefont {Courtney},
  \citenamefont {Dunsworth}, \citenamefont {Farhi}, \citenamefont {Foxen},
  \citenamefont {Fowler}, \citenamefont {Gidney}, \citenamefont {Giustina},
  \citenamefont {Graff}, \citenamefont {Guerin}, \citenamefont {Habegger},
  \citenamefont {Harrigan}, \citenamefont {Hartmann}, \citenamefont {Ho},
  \citenamefont {Hoffmann}, \citenamefont {Huang}, \citenamefont {Humble},
  \citenamefont {Isakov}, \citenamefont {Jeffrey}, \citenamefont {Jiang},
  \citenamefont {Kafri}, \citenamefont {Kechedzhi}, \citenamefont {Kelly},
  \citenamefont {Klimov}, \citenamefont {Knysh}, \citenamefont {Korotkov},
  \citenamefont {Kostritsa}, \citenamefont {Landhuis}, \citenamefont
  {Lindmark}, \citenamefont {Lucero}, \citenamefont {Lyakh}, \citenamefont
  {Mandr{\`a}}, \citenamefont {McClean}, \citenamefont {McEwen}, \citenamefont
  {Megrant}, \citenamefont {Mi}, \citenamefont {Michielsen}, \citenamefont
  {Mohseni}, \citenamefont {Mutus}, \citenamefont {Naaman}, \citenamefont
  {Neeley}, \citenamefont {Neill}, \citenamefont {Niu}, \citenamefont {Ostby},
  \citenamefont {Petukhov}, \citenamefont {Platt}, \citenamefont {Quintana},
  \citenamefont {Rieffel}, \citenamefont {Roushan}, \citenamefont {Rubin},
  \citenamefont {Sank}, \citenamefont {Satzinger}, \citenamefont {Smelyanskiy},
  \citenamefont {Sung}, \citenamefont {Trevithick}, \citenamefont
  {Vainsencher}, \citenamefont {Villalonga}, \citenamefont {White},
  \citenamefont {Yao}, \citenamefont {Yeh}, \citenamefont {Zalcman},
  \citenamefont {Neven},\ and\ \citenamefont {Martinis}}]{Arute:2019aa}%
  \BibitemOpen
  \bibfield  {author} {\bibinfo {author} {\bibfnamefont {F.}~\bibnamefont
  {Arute}}, \bibinfo {author} {\bibfnamefont {K.}~\bibnamefont {Arya}},
  \bibinfo {author} {\bibfnamefont {R.}~\bibnamefont {Babbush}}, \bibinfo
  {author} {\bibfnamefont {D.}~\bibnamefont {Bacon}}, \bibinfo {author}
  {\bibfnamefont {J.~C.}\ \bibnamefont {Bardin}}, \bibinfo {author}
  {\bibfnamefont {R.}~\bibnamefont {Barends}}, \bibinfo {author} {\bibfnamefont
  {R.}~\bibnamefont {Biswas}}, \bibinfo {author} {\bibfnamefont
  {S.}~\bibnamefont {Boixo}}, \bibinfo {author} {\bibfnamefont {F.~G. S.~L.}\
  \bibnamefont {Brandao}}, \bibinfo {author} {\bibfnamefont {D.~A.}\
  \bibnamefont {Buell}}, \bibinfo {author} {\bibfnamefont {B.}~\bibnamefont
  {Burkett}}, \bibinfo {author} {\bibfnamefont {Y.}~\bibnamefont {Chen}},
  \bibinfo {author} {\bibfnamefont {Z.}~\bibnamefont {Chen}}, \bibinfo {author}
  {\bibfnamefont {B.}~\bibnamefont {Chiaro}}, \bibinfo {author} {\bibfnamefont
  {R.}~\bibnamefont {Collins}}, \bibinfo {author} {\bibfnamefont
  {W.}~\bibnamefont {Courtney}}, \bibinfo {author} {\bibfnamefont
  {A.}~\bibnamefont {Dunsworth}}, \bibinfo {author} {\bibfnamefont
  {E.}~\bibnamefont {Farhi}}, \bibinfo {author} {\bibfnamefont
  {B.}~\bibnamefont {Foxen}}, \bibinfo {author} {\bibfnamefont
  {A.}~\bibnamefont {Fowler}}, \bibinfo {author} {\bibfnamefont
  {C.}~\bibnamefont {Gidney}}, \bibinfo {author} {\bibfnamefont
  {M.}~\bibnamefont {Giustina}}, \bibinfo {author} {\bibfnamefont
  {R.}~\bibnamefont {Graff}}, \bibinfo {author} {\bibfnamefont
  {K.}~\bibnamefont {Guerin}}, \bibinfo {author} {\bibfnamefont
  {S.}~\bibnamefont {Habegger}}, \bibinfo {author} {\bibfnamefont {M.~P.}\
  \bibnamefont {Harrigan}}, \bibinfo {author} {\bibfnamefont {M.~J.}\
  \bibnamefont {Hartmann}}, \bibinfo {author} {\bibfnamefont {A.}~\bibnamefont
  {Ho}}, \bibinfo {author} {\bibfnamefont {M.}~\bibnamefont {Hoffmann}},
  \bibinfo {author} {\bibfnamefont {T.}~\bibnamefont {Huang}}, \bibinfo
  {author} {\bibfnamefont {T.~S.}\ \bibnamefont {Humble}}, \bibinfo {author}
  {\bibfnamefont {S.~V.}\ \bibnamefont {Isakov}}, \bibinfo {author}
  {\bibfnamefont {E.}~\bibnamefont {Jeffrey}}, \bibinfo {author} {\bibfnamefont
  {Z.}~\bibnamefont {Jiang}}, \bibinfo {author} {\bibfnamefont
  {D.}~\bibnamefont {Kafri}}, \bibinfo {author} {\bibfnamefont
  {K.}~\bibnamefont {Kechedzhi}}, \bibinfo {author} {\bibfnamefont
  {J.}~\bibnamefont {Kelly}}, \bibinfo {author} {\bibfnamefont {P.~V.}\
  \bibnamefont {Klimov}}, \bibinfo {author} {\bibfnamefont {S.}~\bibnamefont
  {Knysh}}, \bibinfo {author} {\bibfnamefont {A.}~\bibnamefont {Korotkov}},
  \bibinfo {author} {\bibfnamefont {F.}~\bibnamefont {Kostritsa}}, \bibinfo
  {author} {\bibfnamefont {D.}~\bibnamefont {Landhuis}}, \bibinfo {author}
  {\bibfnamefont {M.}~\bibnamefont {Lindmark}}, \bibinfo {author}
  {\bibfnamefont {E.}~\bibnamefont {Lucero}}, \bibinfo {author} {\bibfnamefont
  {D.}~\bibnamefont {Lyakh}}, \bibinfo {author} {\bibfnamefont
  {S.}~\bibnamefont {Mandr{\`a}}}, \bibinfo {author} {\bibfnamefont {J.~R.}\
  \bibnamefont {McClean}}, \bibinfo {author} {\bibfnamefont {M.}~\bibnamefont
  {McEwen}}, \bibinfo {author} {\bibfnamefont {A.}~\bibnamefont {Megrant}},
  \bibinfo {author} {\bibfnamefont {X.}~\bibnamefont {Mi}}, \bibinfo {author}
  {\bibfnamefont {K.}~\bibnamefont {Michielsen}}, \bibinfo {author}
  {\bibfnamefont {M.}~\bibnamefont {Mohseni}}, \bibinfo {author} {\bibfnamefont
  {J.}~\bibnamefont {Mutus}}, \bibinfo {author} {\bibfnamefont
  {O.}~\bibnamefont {Naaman}}, \bibinfo {author} {\bibfnamefont
  {M.}~\bibnamefont {Neeley}}, \bibinfo {author} {\bibfnamefont
  {C.}~\bibnamefont {Neill}}, \bibinfo {author} {\bibfnamefont {M.~Y.}\
  \bibnamefont {Niu}}, \bibinfo {author} {\bibfnamefont {E.}~\bibnamefont
  {Ostby}}, \bibinfo {author} {\bibfnamefont {A.}~\bibnamefont {Petukhov}},
  \bibinfo {author} {\bibfnamefont {J.~C.}\ \bibnamefont {Platt}}, \bibinfo
  {author} {\bibfnamefont {C.}~\bibnamefont {Quintana}}, \bibinfo {author}
  {\bibfnamefont {E.~G.}\ \bibnamefont {Rieffel}}, \bibinfo {author}
  {\bibfnamefont {P.}~\bibnamefont {Roushan}}, \bibinfo {author} {\bibfnamefont
  {N.~C.}\ \bibnamefont {Rubin}}, \bibinfo {author} {\bibfnamefont
  {D.}~\bibnamefont {Sank}}, \bibinfo {author} {\bibfnamefont {K.~J.}\
  \bibnamefont {Satzinger}}, \bibinfo {author} {\bibfnamefont {V.}~\bibnamefont
  {Smelyanskiy}}, \bibinfo {author} {\bibfnamefont {K.~J.}\ \bibnamefont
  {Sung}}, \bibinfo {author} {\bibfnamefont {M.~D.}\ \bibnamefont
  {Trevithick}}, \bibinfo {author} {\bibfnamefont {A.}~\bibnamefont
  {Vainsencher}}, \bibinfo {author} {\bibfnamefont {B.}~\bibnamefont
  {Villalonga}}, \bibinfo {author} {\bibfnamefont {T.}~\bibnamefont {White}},
  \bibinfo {author} {\bibfnamefont {Z.~J.}\ \bibnamefont {Yao}}, \bibinfo
  {author} {\bibfnamefont {P.}~\bibnamefont {Yeh}}, \bibinfo {author}
  {\bibfnamefont {A.}~\bibnamefont {Zalcman}}, \bibinfo {author} {\bibfnamefont
  {H.}~\bibnamefont {Neven}},\ and\ \bibinfo {author} {\bibfnamefont {J.~M.}\
  \bibnamefont {Martinis}},\ }\bibfield  {title} {\bibinfo {title} {Quantum
  supremacy using a programmable superconducting processor},\ }\href
  {https://doi.org/10.1038/s41586-019-1666-5} {\bibfield  {journal} {\bibinfo
  {journal} {Nature}\ }\textbf {\bibinfo {volume} {574}},\ \bibinfo {pages}
  {505} (\bibinfo {year} {2019})}\BibitemShut {NoStop}%
\bibitem [{\citenamefont {Havl{\'\i}{\v c}ek}\ \emph
  {et~al.}(2019)\citenamefont {Havl{\'\i}{\v c}ek}, \citenamefont
  {C{\'o}rcoles}, \citenamefont {Temme}, \citenamefont {Harrow}, \citenamefont
  {Kandala}, \citenamefont {Chow},\ and\ \citenamefont
  {Gambetta}}]{Havlicek2019}%
  \BibitemOpen
  \bibfield  {author} {\bibinfo {author} {\bibfnamefont {V.}~\bibnamefont
  {Havl{\'\i}{\v c}ek}}, \bibinfo {author} {\bibfnamefont {A.~D.}\ \bibnamefont
  {C{\'o}rcoles}}, \bibinfo {author} {\bibfnamefont {K.}~\bibnamefont {Temme}},
  \bibinfo {author} {\bibfnamefont {A.~W.}\ \bibnamefont {Harrow}}, \bibinfo
  {author} {\bibfnamefont {A.}~\bibnamefont {Kandala}}, \bibinfo {author}
  {\bibfnamefont {J.~M.}\ \bibnamefont {Chow}},\ and\ \bibinfo {author}
  {\bibfnamefont {J.~M.}\ \bibnamefont {Gambetta}},\ }\bibfield  {title}
  {\bibinfo {title} {Supervised learning with quantum-enhanced feature
  spaces},\ }\href {https://doi.org/10.1038/s41586-019-0980-2} {\bibfield
  {journal} {\bibinfo  {journal} {Nature}\ }\textbf {\bibinfo {volume} {567}},\
  \bibinfo {pages} {209} (\bibinfo {year} {2019})}\BibitemShut {NoStop}%
\bibitem [{\citenamefont {Kandala}\ \emph {et~al.}(2019)\citenamefont
  {Kandala}, \citenamefont {Temme}, \citenamefont {C{\'o}rcoles}, \citenamefont
  {Mezzacapo}, \citenamefont {Chow},\ and\ \citenamefont
  {Gambetta}}]{Kandala2019}%
  \BibitemOpen
  \bibfield  {author} {\bibinfo {author} {\bibfnamefont {A.}~\bibnamefont
  {Kandala}}, \bibinfo {author} {\bibfnamefont {K.}~\bibnamefont {Temme}},
  \bibinfo {author} {\bibfnamefont {A.~D.}\ \bibnamefont {C{\'o}rcoles}},
  \bibinfo {author} {\bibfnamefont {A.}~\bibnamefont {Mezzacapo}}, \bibinfo
  {author} {\bibfnamefont {J.~M.}\ \bibnamefont {Chow}},\ and\ \bibinfo
  {author} {\bibfnamefont {J.~M.}\ \bibnamefont {Gambetta}},\ }\bibfield
  {title} {\bibinfo {title} {Error mitigation extends the computational reach
  of a noisy quantum processor},\ }\href
  {https://doi.org/10.1038/s41586-019-1040-7} {\bibfield  {journal} {\bibinfo
  {journal} {Nature}\ }\textbf {\bibinfo {volume} {567}},\ \bibinfo {pages}
  {491} (\bibinfo {year} {2019})}\BibitemShut {NoStop}%
\bibitem [{\citenamefont {Campagne-Ibarcq}\ \emph {et~al.}(2020)\citenamefont
  {Campagne-Ibarcq}, \citenamefont {Eickbusch}, \citenamefont {Touzard},
  \citenamefont {Zalys-Geller}, \citenamefont {Frattini}, \citenamefont
  {Sivak}, \citenamefont {Reinhold}, \citenamefont {Puri}, \citenamefont
  {Shankar}, \citenamefont {Schoelkopf}, \citenamefont {Frunzio}, \citenamefont
  {Mirrahimi},\ and\ \citenamefont {Devoret}}]{Campagne2020}%
  \BibitemOpen
  \bibfield  {author} {\bibinfo {author} {\bibfnamefont {P.}~\bibnamefont
  {Campagne-Ibarcq}}, \bibinfo {author} {\bibfnamefont {A.}~\bibnamefont
  {Eickbusch}}, \bibinfo {author} {\bibfnamefont {S.}~\bibnamefont {Touzard}},
  \bibinfo {author} {\bibfnamefont {E.}~\bibnamefont {Zalys-Geller}}, \bibinfo
  {author} {\bibfnamefont {N.~E.}\ \bibnamefont {Frattini}}, \bibinfo {author}
  {\bibfnamefont {V.~V.}\ \bibnamefont {Sivak}}, \bibinfo {author}
  {\bibfnamefont {P.}~\bibnamefont {Reinhold}}, \bibinfo {author}
  {\bibfnamefont {S.}~\bibnamefont {Puri}}, \bibinfo {author} {\bibfnamefont
  {S.}~\bibnamefont {Shankar}}, \bibinfo {author} {\bibfnamefont {R.~J.}\
  \bibnamefont {Schoelkopf}}, \bibinfo {author} {\bibfnamefont
  {L.}~\bibnamefont {Frunzio}}, \bibinfo {author} {\bibfnamefont
  {M.}~\bibnamefont {Mirrahimi}},\ and\ \bibinfo {author} {\bibfnamefont
  {M.~H.}\ \bibnamefont {Devoret}},\ }\bibfield  {title} {\bibinfo {title}
  {Quantum error correction of a qubit encoded in grid states of an
  oscillator},\ }\href {https://doi.org/10.1038/s41586-020-2603-3} {\bibfield
  {journal} {\bibinfo  {journal} {Nature}\ }\textbf {\bibinfo {volume} {584}},\
  \bibinfo {pages} {368} (\bibinfo {year} {2020})}\BibitemShut {NoStop}%
\bibitem [{\citenamefont {Andersen}\ \emph {et~al.}(2020)\citenamefont
  {Andersen}, \citenamefont {Remm}, \citenamefont {Lazar}, \citenamefont
  {Krinner}, \citenamefont {Lacroix}, \citenamefont {Norris}, \citenamefont
  {Gabureac}, \citenamefont {Eichler},\ and\ \citenamefont
  {Wallraff}}]{Andersen2020}%
  \BibitemOpen
  \bibfield  {author} {\bibinfo {author} {\bibfnamefont {C.~K.}\ \bibnamefont
  {Andersen}}, \bibinfo {author} {\bibfnamefont {A.}~\bibnamefont {Remm}},
  \bibinfo {author} {\bibfnamefont {S.}~\bibnamefont {Lazar}}, \bibinfo
  {author} {\bibfnamefont {S.}~\bibnamefont {Krinner}}, \bibinfo {author}
  {\bibfnamefont {N.}~\bibnamefont {Lacroix}}, \bibinfo {author} {\bibfnamefont
  {G.~J.}\ \bibnamefont {Norris}}, \bibinfo {author} {\bibfnamefont
  {M.}~\bibnamefont {Gabureac}}, \bibinfo {author} {\bibfnamefont
  {C.}~\bibnamefont {Eichler}},\ and\ \bibinfo {author} {\bibfnamefont
  {A.}~\bibnamefont {Wallraff}},\ }\bibfield  {title} {\bibinfo {title}
  {Repeated quantum error detection in a surface code},\ }\href
  {https://doi.org/10.1038/s41567-020-0920-y} {\bibfield  {journal} {\bibinfo
  {journal} {Nature Physics}\ }\textbf {\bibinfo {volume} {16}},\ \bibinfo
  {pages} {875} (\bibinfo {year} {2020})}\BibitemShut {NoStop}%
\bibitem [{\citenamefont {Arute}\ \emph {et~al.}(2020)\citenamefont {Arute},
  \citenamefont {Arya}, \citenamefont {Babbush}, \citenamefont {Bacon},
  \citenamefont {Bardin}, \citenamefont {Barends}, \citenamefont {Boixo},
  \citenamefont {Broughton}, \citenamefont {Buckley}, \citenamefont {Buell},
  \citenamefont {Burkett}, \citenamefont {Bushnell}, \citenamefont {Chen},
  \citenamefont {Chen}, \citenamefont {Chiaro}, \citenamefont {Collins},
  \citenamefont {Courtney}, \citenamefont {Demura}, \citenamefont {Dunsworth},
  \citenamefont {Farhi}, \citenamefont {Fowler}, \citenamefont {Foxen},
  \citenamefont {Gidney}, \citenamefont {Giustina}, \citenamefont {Graff},
  \citenamefont {Habegger}, \citenamefont {Harrigan}, \citenamefont {Ho},
  \citenamefont {Hong}, \citenamefont {Huang}, \citenamefont {Huggins},
  \citenamefont {Ioffe}, \citenamefont {Isakov}, \citenamefont {Jeffrey},
  \citenamefont {Jiang}, \citenamefont {Jones}, \citenamefont {Kafri},
  \citenamefont {Kechedzhi}, \citenamefont {Kelly}, \citenamefont {Kim},
  \citenamefont {Klimov}, \citenamefont {Korotkov}, \citenamefont {Kostritsa},
  \citenamefont {Landhuis}, \citenamefont {Laptev}, \citenamefont {Lindmark},
  \citenamefont {Lucero}, \citenamefont {Martin}, \citenamefont {Martinis},
  \citenamefont {McClean}, \citenamefont {McEwen}, \citenamefont {Megrant},
  \citenamefont {Mi}, \citenamefont {Mohseni}, \citenamefont {Mruczkiewicz},
  \citenamefont {Mutus}, \citenamefont {Naaman}, \citenamefont {Neeley},
  \citenamefont {Neill}, \citenamefont {Neven}, \citenamefont {Niu},
  \citenamefont {O{\textquoteright}Brien}, \citenamefont {Ostby}, \citenamefont
  {Petukhov}, \citenamefont {Putterman}, \citenamefont {Quintana},
  \citenamefont {Roushan}, \citenamefont {Rubin}, \citenamefont {Sank},
  \citenamefont {Satzinger}, \citenamefont {Smelyanskiy}, \citenamefont
  {Strain}, \citenamefont {Sung}, \citenamefont {Szalay}, \citenamefont
  {Takeshita}, \citenamefont {Vainsencher}, \citenamefont {White},
  \citenamefont {Wiebe}, \citenamefont {Yao}, \citenamefont {Yeh},\ and\
  \citenamefont {Zalcman}}]{Arute2020}%
  \BibitemOpen
  \bibfield  {author} {\bibinfo {author} {\bibfnamefont {F.}~\bibnamefont
  {Arute}}, \bibinfo {author} {\bibfnamefont {K.}~\bibnamefont {Arya}},
  \bibinfo {author} {\bibfnamefont {R.}~\bibnamefont {Babbush}}, \bibinfo
  {author} {\bibfnamefont {D.}~\bibnamefont {Bacon}}, \bibinfo {author}
  {\bibfnamefont {J.~C.}\ \bibnamefont {Bardin}}, \bibinfo {author}
  {\bibfnamefont {R.}~\bibnamefont {Barends}}, \bibinfo {author} {\bibfnamefont
  {S.}~\bibnamefont {Boixo}}, \bibinfo {author} {\bibfnamefont
  {M.}~\bibnamefont {Broughton}}, \bibinfo {author} {\bibfnamefont {B.~B.}\
  \bibnamefont {Buckley}}, \bibinfo {author} {\bibfnamefont {D.~A.}\
  \bibnamefont {Buell}}, \bibinfo {author} {\bibfnamefont {B.}~\bibnamefont
  {Burkett}}, \bibinfo {author} {\bibfnamefont {N.}~\bibnamefont {Bushnell}},
  \bibinfo {author} {\bibfnamefont {Y.}~\bibnamefont {Chen}}, \bibinfo {author}
  {\bibfnamefont {Z.}~\bibnamefont {Chen}}, \bibinfo {author} {\bibfnamefont
  {B.}~\bibnamefont {Chiaro}}, \bibinfo {author} {\bibfnamefont
  {R.}~\bibnamefont {Collins}}, \bibinfo {author} {\bibfnamefont
  {W.}~\bibnamefont {Courtney}}, \bibinfo {author} {\bibfnamefont
  {S.}~\bibnamefont {Demura}}, \bibinfo {author} {\bibfnamefont
  {A.}~\bibnamefont {Dunsworth}}, \bibinfo {author} {\bibfnamefont
  {E.}~\bibnamefont {Farhi}}, \bibinfo {author} {\bibfnamefont
  {A.}~\bibnamefont {Fowler}}, \bibinfo {author} {\bibfnamefont
  {B.}~\bibnamefont {Foxen}}, \bibinfo {author} {\bibfnamefont
  {C.}~\bibnamefont {Gidney}}, \bibinfo {author} {\bibfnamefont
  {M.}~\bibnamefont {Giustina}}, \bibinfo {author} {\bibfnamefont
  {R.}~\bibnamefont {Graff}}, \bibinfo {author} {\bibfnamefont
  {S.}~\bibnamefont {Habegger}}, \bibinfo {author} {\bibfnamefont {M.~P.}\
  \bibnamefont {Harrigan}}, \bibinfo {author} {\bibfnamefont {A.}~\bibnamefont
  {Ho}}, \bibinfo {author} {\bibfnamefont {S.}~\bibnamefont {Hong}}, \bibinfo
  {author} {\bibfnamefont {T.}~\bibnamefont {Huang}}, \bibinfo {author}
  {\bibfnamefont {W.~J.}\ \bibnamefont {Huggins}}, \bibinfo {author}
  {\bibfnamefont {L.}~\bibnamefont {Ioffe}}, \bibinfo {author} {\bibfnamefont
  {S.~V.}\ \bibnamefont {Isakov}}, \bibinfo {author} {\bibfnamefont
  {E.}~\bibnamefont {Jeffrey}}, \bibinfo {author} {\bibfnamefont
  {Z.}~\bibnamefont {Jiang}}, \bibinfo {author} {\bibfnamefont
  {C.}~\bibnamefont {Jones}}, \bibinfo {author} {\bibfnamefont
  {D.}~\bibnamefont {Kafri}}, \bibinfo {author} {\bibfnamefont
  {K.}~\bibnamefont {Kechedzhi}}, \bibinfo {author} {\bibfnamefont
  {J.}~\bibnamefont {Kelly}}, \bibinfo {author} {\bibfnamefont
  {S.}~\bibnamefont {Kim}}, \bibinfo {author} {\bibfnamefont {P.~V.}\
  \bibnamefont {Klimov}}, \bibinfo {author} {\bibfnamefont {A.}~\bibnamefont
  {Korotkov}}, \bibinfo {author} {\bibfnamefont {F.}~\bibnamefont {Kostritsa}},
  \bibinfo {author} {\bibfnamefont {D.}~\bibnamefont {Landhuis}}, \bibinfo
  {author} {\bibfnamefont {P.}~\bibnamefont {Laptev}}, \bibinfo {author}
  {\bibfnamefont {M.}~\bibnamefont {Lindmark}}, \bibinfo {author}
  {\bibfnamefont {E.}~\bibnamefont {Lucero}}, \bibinfo {author} {\bibfnamefont
  {O.}~\bibnamefont {Martin}}, \bibinfo {author} {\bibfnamefont {J.~M.}\
  \bibnamefont {Martinis}}, \bibinfo {author} {\bibfnamefont {J.~R.}\
  \bibnamefont {McClean}}, \bibinfo {author} {\bibfnamefont {M.}~\bibnamefont
  {McEwen}}, \bibinfo {author} {\bibfnamefont {A.}~\bibnamefont {Megrant}},
  \bibinfo {author} {\bibfnamefont {X.}~\bibnamefont {Mi}}, \bibinfo {author}
  {\bibfnamefont {M.}~\bibnamefont {Mohseni}}, \bibinfo {author} {\bibfnamefont
  {W.}~\bibnamefont {Mruczkiewicz}}, \bibinfo {author} {\bibfnamefont
  {J.}~\bibnamefont {Mutus}}, \bibinfo {author} {\bibfnamefont
  {O.}~\bibnamefont {Naaman}}, \bibinfo {author} {\bibfnamefont
  {M.}~\bibnamefont {Neeley}}, \bibinfo {author} {\bibfnamefont
  {C.}~\bibnamefont {Neill}}, \bibinfo {author} {\bibfnamefont
  {H.}~\bibnamefont {Neven}}, \bibinfo {author} {\bibfnamefont {M.~Y.}\
  \bibnamefont {Niu}}, \bibinfo {author} {\bibfnamefont {T.~E.}\ \bibnamefont
  {O{\textquoteright}Brien}}, \bibinfo {author} {\bibfnamefont
  {E.}~\bibnamefont {Ostby}}, \bibinfo {author} {\bibfnamefont
  {A.}~\bibnamefont {Petukhov}}, \bibinfo {author} {\bibfnamefont
  {H.}~\bibnamefont {Putterman}}, \bibinfo {author} {\bibfnamefont
  {C.}~\bibnamefont {Quintana}}, \bibinfo {author} {\bibfnamefont
  {P.}~\bibnamefont {Roushan}}, \bibinfo {author} {\bibfnamefont {N.~C.}\
  \bibnamefont {Rubin}}, \bibinfo {author} {\bibfnamefont {D.}~\bibnamefont
  {Sank}}, \bibinfo {author} {\bibfnamefont {K.~J.}\ \bibnamefont {Satzinger}},
  \bibinfo {author} {\bibfnamefont {V.}~\bibnamefont {Smelyanskiy}}, \bibinfo
  {author} {\bibfnamefont {D.}~\bibnamefont {Strain}}, \bibinfo {author}
  {\bibfnamefont {K.~J.}\ \bibnamefont {Sung}}, \bibinfo {author}
  {\bibfnamefont {M.}~\bibnamefont {Szalay}}, \bibinfo {author} {\bibfnamefont
  {T.~Y.}\ \bibnamefont {Takeshita}}, \bibinfo {author} {\bibfnamefont
  {A.}~\bibnamefont {Vainsencher}}, \bibinfo {author} {\bibfnamefont
  {T.}~\bibnamefont {White}}, \bibinfo {author} {\bibfnamefont
  {N.}~\bibnamefont {Wiebe}}, \bibinfo {author} {\bibfnamefont {Z.~J.}\
  \bibnamefont {Yao}}, \bibinfo {author} {\bibfnamefont {P.}~\bibnamefont
  {Yeh}},\ and\ \bibinfo {author} {\bibfnamefont {A.}~\bibnamefont {Zalcman}},\
  }\bibfield  {title} {\bibinfo {title} {Hartree-fock on a superconducting
  qubit quantum computer},\ }\href
  {https://science.sciencemag.org/content/369/6507/1084} {\bibfield  {journal}
  {\bibinfo  {journal} {Science}\ }\textbf {\bibinfo {volume} {369}},\ \bibinfo
  {pages} {1084} (\bibinfo {year} {2020})}\BibitemShut {NoStop}%
\bibitem [{\citenamefont {Wu}\ \emph {et~al.}(2021)\citenamefont {Wu},
  \citenamefont {Bao}, \citenamefont {Cao}, \citenamefont {Chen}, \citenamefont
  {Chen}, \citenamefont {Chen}, \citenamefont {Chung}, \citenamefont {Deng},
  \citenamefont {Du}, \citenamefont {Fan}, \citenamefont {Gong}, \citenamefont
  {Guo}, \citenamefont {Guo}, \citenamefont {Guo}, \citenamefont {Han},
  \citenamefont {Hong}, \citenamefont {Huang}, \citenamefont {Huo},
  \citenamefont {Li}, \citenamefont {Li}, \citenamefont {Li}, \citenamefont
  {Li}, \citenamefont {Liang}, \citenamefont {Lin}, \citenamefont {Lin},
  \citenamefont {Qian}, \citenamefont {Qiao}, \citenamefont {Rong},
  \citenamefont {Su}, \citenamefont {Sun}, \citenamefont {Wang}, \citenamefont
  {Wang}, \citenamefont {Wu}, \citenamefont {Xu}, \citenamefont {Yan},
  \citenamefont {Yang}, \citenamefont {Yang}, \citenamefont {Ye}, \citenamefont
  {Yin}, \citenamefont {Ying}, \citenamefont {Yu}, \citenamefont {Zha},
  \citenamefont {Zhang}, \citenamefont {Zhang}, \citenamefont {Zhang},
  \citenamefont {Zhang}, \citenamefont {Zhao}, \citenamefont {Zhao},
  \citenamefont {Zhou}, \citenamefont {Zhu}, \citenamefont {Lu}, \citenamefont
  {Peng}, \citenamefont {Zhu},\ and\ \citenamefont {Pan}}]{wu2021strong}%
  \BibitemOpen
  \bibfield  {author} {\bibinfo {author} {\bibfnamefont {Y.}~\bibnamefont
  {Wu}}, \bibinfo {author} {\bibfnamefont {W.-S.}\ \bibnamefont {Bao}},
  \bibinfo {author} {\bibfnamefont {S.}~\bibnamefont {Cao}}, \bibinfo {author}
  {\bibfnamefont {F.}~\bibnamefont {Chen}}, \bibinfo {author} {\bibfnamefont
  {M.-C.}\ \bibnamefont {Chen}}, \bibinfo {author} {\bibfnamefont
  {X.}~\bibnamefont {Chen}}, \bibinfo {author} {\bibfnamefont {T.-H.}\
  \bibnamefont {Chung}}, \bibinfo {author} {\bibfnamefont {H.}~\bibnamefont
  {Deng}}, \bibinfo {author} {\bibfnamefont {Y.}~\bibnamefont {Du}}, \bibinfo
  {author} {\bibfnamefont {D.}~\bibnamefont {Fan}}, \bibinfo {author}
  {\bibfnamefont {M.}~\bibnamefont {Gong}}, \bibinfo {author} {\bibfnamefont
  {C.}~\bibnamefont {Guo}}, \bibinfo {author} {\bibfnamefont {C.}~\bibnamefont
  {Guo}}, \bibinfo {author} {\bibfnamefont {S.}~\bibnamefont {Guo}}, \bibinfo
  {author} {\bibfnamefont {L.}~\bibnamefont {Han}}, \bibinfo {author}
  {\bibfnamefont {L.}~\bibnamefont {Hong}}, \bibinfo {author} {\bibfnamefont
  {H.-L.}\ \bibnamefont {Huang}}, \bibinfo {author} {\bibfnamefont {Y.-H.}\
  \bibnamefont {Huo}}, \bibinfo {author} {\bibfnamefont {L.}~\bibnamefont
  {Li}}, \bibinfo {author} {\bibfnamefont {N.}~\bibnamefont {Li}}, \bibinfo
  {author} {\bibfnamefont {S.}~\bibnamefont {Li}}, \bibinfo {author}
  {\bibfnamefont {Y.}~\bibnamefont {Li}}, \bibinfo {author} {\bibfnamefont
  {F.}~\bibnamefont {Liang}}, \bibinfo {author} {\bibfnamefont
  {C.}~\bibnamefont {Lin}}, \bibinfo {author} {\bibfnamefont {J.}~\bibnamefont
  {Lin}}, \bibinfo {author} {\bibfnamefont {H.}~\bibnamefont {Qian}}, \bibinfo
  {author} {\bibfnamefont {D.}~\bibnamefont {Qiao}}, \bibinfo {author}
  {\bibfnamefont {H.}~\bibnamefont {Rong}}, \bibinfo {author} {\bibfnamefont
  {H.}~\bibnamefont {Su}}, \bibinfo {author} {\bibfnamefont {L.}~\bibnamefont
  {Sun}}, \bibinfo {author} {\bibfnamefont {L.}~\bibnamefont {Wang}}, \bibinfo
  {author} {\bibfnamefont {S.}~\bibnamefont {Wang}}, \bibinfo {author}
  {\bibfnamefont {D.}~\bibnamefont {Wu}}, \bibinfo {author} {\bibfnamefont
  {Y.}~\bibnamefont {Xu}}, \bibinfo {author} {\bibfnamefont {K.}~\bibnamefont
  {Yan}}, \bibinfo {author} {\bibfnamefont {W.}~\bibnamefont {Yang}}, \bibinfo
  {author} {\bibfnamefont {Y.}~\bibnamefont {Yang}}, \bibinfo {author}
  {\bibfnamefont {Y.}~\bibnamefont {Ye}}, \bibinfo {author} {\bibfnamefont
  {J.}~\bibnamefont {Yin}}, \bibinfo {author} {\bibfnamefont {C.}~\bibnamefont
  {Ying}}, \bibinfo {author} {\bibfnamefont {J.}~\bibnamefont {Yu}}, \bibinfo
  {author} {\bibfnamefont {C.}~\bibnamefont {Zha}}, \bibinfo {author}
  {\bibfnamefont {C.}~\bibnamefont {Zhang}}, \bibinfo {author} {\bibfnamefont
  {H.}~\bibnamefont {Zhang}}, \bibinfo {author} {\bibfnamefont
  {K.}~\bibnamefont {Zhang}}, \bibinfo {author} {\bibfnamefont
  {Y.}~\bibnamefont {Zhang}}, \bibinfo {author} {\bibfnamefont
  {H.}~\bibnamefont {Zhao}}, \bibinfo {author} {\bibfnamefont {Y.}~\bibnamefont
  {Zhao}}, \bibinfo {author} {\bibfnamefont {L.}~\bibnamefont {Zhou}}, \bibinfo
  {author} {\bibfnamefont {Q.}~\bibnamefont {Zhu}}, \bibinfo {author}
  {\bibfnamefont {C.-Y.}\ \bibnamefont {Lu}}, \bibinfo {author} {\bibfnamefont
  {C.-Z.}\ \bibnamefont {Peng}}, \bibinfo {author} {\bibfnamefont
  {X.}~\bibnamefont {Zhu}},\ and\ \bibinfo {author} {\bibfnamefont {J.-W.}\
  \bibnamefont {Pan}},\ }\bibfield  {title} {\bibinfo {title} {Strong quantum
  computational advantage using a superconducting quantum processor},\ }\href
  {https://link.aps.org/doi/10.1103/PhysRevLett.127.180501} {\bibfield
  {journal} {\bibinfo  {journal} {Phys. Rev. Lett.}\ }\textbf {\bibinfo
  {volume} {127}},\ \bibinfo {pages} {180501} (\bibinfo {year}
  {2021})}\BibitemShut {NoStop}%
\bibitem [{\citenamefont {Koch}\ \emph {et~al.}(2007)\citenamefont {Koch},
  \citenamefont {Yu}, \citenamefont {Gambetta}, \citenamefont {Houck},
  \citenamefont {Schuster}, \citenamefont {Majer}, \citenamefont {Blais},
  \citenamefont {Devoret}, \citenamefont {Girvin},\ and\ \citenamefont
  {Schoelkopf}}]{transmon-invention}%
  \BibitemOpen
  \bibfield  {author} {\bibinfo {author} {\bibfnamefont {J.}~\bibnamefont
  {Koch}}, \bibinfo {author} {\bibfnamefont {T.~M.}\ \bibnamefont {Yu}},
  \bibinfo {author} {\bibfnamefont {J.}~\bibnamefont {Gambetta}}, \bibinfo
  {author} {\bibfnamefont {A.~A.}\ \bibnamefont {Houck}}, \bibinfo {author}
  {\bibfnamefont {D.~I.}\ \bibnamefont {Schuster}}, \bibinfo {author}
  {\bibfnamefont {J.}~\bibnamefont {Majer}}, \bibinfo {author} {\bibfnamefont
  {A.}~\bibnamefont {Blais}}, \bibinfo {author} {\bibfnamefont {M.~H.}\
  \bibnamefont {Devoret}}, \bibinfo {author} {\bibfnamefont {S.~M.}\
  \bibnamefont {Girvin}},\ and\ \bibinfo {author} {\bibfnamefont {R.~J.}\
  \bibnamefont {Schoelkopf}},\ }\bibfield  {title} {\bibinfo {title}
  {Charge-insensitive qubit design derived from the cooper pair box},\ }\href
  {https://doi.org/10.1103/PhysRevA.76.042319} {\bibfield  {journal} {\bibinfo
  {journal} {Physical Review A}\ }\textbf {\bibinfo {volume} {76}},\ \bibinfo
  {pages} {042319} (\bibinfo {year} {2007})}\BibitemShut {NoStop}%
\bibitem [{\citenamefont {Devitt}(2016)}]{Devitt2016}%
  \BibitemOpen
  \bibfield  {author} {\bibinfo {author} {\bibfnamefont {S.~J.}\ \bibnamefont
  {Devitt}},\ }\bibfield  {title} {\bibinfo {title} {Performing quantum
  computing experiments in the cloud},\ }\href
  {https://doi.org/10.1103/PhysRevA.94.032329} {\bibfield  {journal} {\bibinfo
  {journal} {Phys. Rev. A}\ }\textbf {\bibinfo {volume} {94}},\ \bibinfo
  {pages} {032329} (\bibinfo {year} {2016})}\BibitemShut {NoStop}%
\bibitem [{\citenamefont {Wootton}\ and\ \citenamefont
  {Loss}(2018)}]{Wootton:2018aa}%
  \BibitemOpen
  \bibfield  {author} {\bibinfo {author} {\bibfnamefont {J.~R.}\ \bibnamefont
  {Wootton}}\ and\ \bibinfo {author} {\bibfnamefont {D.}~\bibnamefont {Loss}},\
  }\bibfield  {title} {\bibinfo {title} {Repetition code of 15 qubits},\ }\href
  {https://doi.org/10.1103/PhysRevA.97.052313} {\bibfield  {journal} {\bibinfo
  {journal} {Physical Review A}\ }\textbf {\bibinfo {volume} {97}},\ \bibinfo
  {pages} {052313} (\bibinfo {year} {2018})}\BibitemShut {NoStop}%
\bibitem [{\citenamefont {Vuillot}(2018)}]{Vuillot2018}%
  \BibitemOpen
  \bibfield  {author} {\bibinfo {author} {\bibfnamefont {C.}~\bibnamefont
  {Vuillot}},\ }\bibfield  {title} {\bibinfo {title} {Is error detection
  helpful on ibm 5q chips?},\ }\href {https://doi.org/10.26421/QIC18.11-12}
  {\bibfield  {journal} {\bibinfo  {journal} {Quantum Inf. Comput.}\ }\textbf
  {\bibinfo {volume} {18}},\ \bibinfo {pages} {0949} (\bibinfo {year}
  {2018})}\BibitemShut {NoStop}%
\bibitem [{\citenamefont {Roffe}\ \emph {et~al.}(2018)\citenamefont {Roffe},
  \citenamefont {Headley}, \citenamefont {Chancellor}, \citenamefont
  {Horsman},\ and\ \citenamefont {Kendon}}]{Roffe:2018aa}%
  \BibitemOpen
  \bibfield  {author} {\bibinfo {author} {\bibfnamefont {J.}~\bibnamefont
  {Roffe}}, \bibinfo {author} {\bibfnamefont {D.}~\bibnamefont {Headley}},
  \bibinfo {author} {\bibfnamefont {N.}~\bibnamefont {Chancellor}}, \bibinfo
  {author} {\bibfnamefont {D.}~\bibnamefont {Horsman}},\ and\ \bibinfo {author}
  {\bibfnamefont {V.}~\bibnamefont {Kendon}},\ }\bibfield  {title} {\bibinfo
  {title} {Protecting quantum memories using coherent parity check codes},\
  }\href {https://doi.org/10.1088/2058-9565/aac64e} {\bibfield  {journal}
  {\bibinfo  {journal} {Quantum Science and Technology}\ }\textbf {\bibinfo
  {volume} {3}},\ \bibinfo {pages} {035010} (\bibinfo {year}
  {2018})}\BibitemShut {NoStop}%
\bibitem [{\citenamefont {Willsch}\ \emph {et~al.}(2018)\citenamefont
  {Willsch}, \citenamefont {Willsch}, \citenamefont {Jin}, \citenamefont
  {De~Raedt},\ and\ \citenamefont {Michielsen}}]{Willsch2018}%
  \BibitemOpen
  \bibfield  {author} {\bibinfo {author} {\bibfnamefont {D.}~\bibnamefont
  {Willsch}}, \bibinfo {author} {\bibfnamefont {M.}~\bibnamefont {Willsch}},
  \bibinfo {author} {\bibfnamefont {F.}~\bibnamefont {Jin}}, \bibinfo {author}
  {\bibfnamefont {H.}~\bibnamefont {De~Raedt}},\ and\ \bibinfo {author}
  {\bibfnamefont {K.}~\bibnamefont {Michielsen}},\ }\bibfield  {title}
  {\bibinfo {title} {Testing quantum fault tolerance on small systems},\ }\href
  {https://doi.org/10.1103/PhysRevA.98.052348} {\bibfield  {journal} {\bibinfo
  {journal} {Phys. Rev. A}\ }\textbf {\bibinfo {volume} {98}},\ \bibinfo
  {pages} {052348} (\bibinfo {year} {2018})}\BibitemShut {NoStop}%
\bibitem [{\citenamefont {Pokharel}\ \emph {et~al.}(2018)\citenamefont
  {Pokharel}, \citenamefont {Anand}, \citenamefont {Fortman},\ and\
  \citenamefont {Lidar}}]{Pokharel2018}%
  \BibitemOpen
  \bibfield  {author} {\bibinfo {author} {\bibfnamefont {B.}~\bibnamefont
  {Pokharel}}, \bibinfo {author} {\bibfnamefont {N.}~\bibnamefont {Anand}},
  \bibinfo {author} {\bibfnamefont {B.}~\bibnamefont {Fortman}},\ and\ \bibinfo
  {author} {\bibfnamefont {D.~A.}\ \bibnamefont {Lidar}},\ }\bibfield  {title}
  {\bibinfo {title} {Demonstration of fidelity improvement using dynamical
  decoupling with superconducting qubits},\ }\href
  {https://link.aps.org/doi/10.1103/PhysRevLett.121.220502} {\bibfield
  {journal} {\bibinfo  {journal} {Phys. Rev. Lett.}\ }\textbf {\bibinfo
  {volume} {121}},\ \bibinfo {pages} {220502} (\bibinfo {year}
  {2018})}\BibitemShut {NoStop}%
\bibitem [{\citenamefont {Harper}\ and\ \citenamefont
  {Flammia}(2019)}]{Harper:2019aa}%
  \BibitemOpen
  \bibfield  {author} {\bibinfo {author} {\bibfnamefont {R.}~\bibnamefont
  {Harper}}\ and\ \bibinfo {author} {\bibfnamefont {S.~T.}\ \bibnamefont
  {Flammia}},\ }\bibfield  {title} {\bibinfo {title} {Fault-tolerant logical
  gates in the ibm quantum experience},\ }\href
  {https://doi.org/10.1103/PhysRevLett.122.080504} {\bibfield  {journal}
  {\bibinfo  {journal} {Physical Review Letters}\ }\textbf {\bibinfo {volume}
  {122}},\ \bibinfo {pages} {080504} (\bibinfo {year} {2019})}\BibitemShut
  {NoStop}%
\bibitem [{\citenamefont {Maslov}\ \emph {et~al.}(2021)\citenamefont {Maslov},
  \citenamefont {Kim}, \citenamefont {Bravyi}, \citenamefont {Yoder},\ and\
  \citenamefont {Sheldon}}]{Maslov:2021aa}%
  \BibitemOpen
  \bibfield  {author} {\bibinfo {author} {\bibfnamefont {D.}~\bibnamefont
  {Maslov}}, \bibinfo {author} {\bibfnamefont {J.-S.}\ \bibnamefont {Kim}},
  \bibinfo {author} {\bibfnamefont {S.}~\bibnamefont {Bravyi}}, \bibinfo
  {author} {\bibfnamefont {T.~J.}\ \bibnamefont {Yoder}},\ and\ \bibinfo
  {author} {\bibfnamefont {S.}~\bibnamefont {Sheldon}},\ }\bibfield  {title}
  {\bibinfo {title} {Quantum advantage for computations with limited space},\
  }\href {https://doi.org/10.1038/s41567-021-01271-7} {\bibfield  {journal}
  {\bibinfo  {journal} {Nature Physics}\ }\textbf {\bibinfo {volume} {17}},\
  \bibinfo {pages} {894} (\bibinfo {year} {2021})}\BibitemShut {NoStop}%
\bibitem [{\citenamefont {Zhang}\ \emph {et~al.}(2021)\citenamefont {Zhang},
  \citenamefont {Rao}, \citenamefont {Yu}, \citenamefont {Lim},\ and\
  \citenamefont {Korepin}}]{Zhang_2021}%
  \BibitemOpen
  \bibfield  {author} {\bibinfo {author} {\bibfnamefont {K.}~\bibnamefont
  {Zhang}}, \bibinfo {author} {\bibfnamefont {P.}~\bibnamefont {Rao}}, \bibinfo
  {author} {\bibfnamefont {K.}~\bibnamefont {Yu}}, \bibinfo {author}
  {\bibfnamefont {H.}~\bibnamefont {Lim}},\ and\ \bibinfo {author}
  {\bibfnamefont {V.}~\bibnamefont {Korepin}},\ }\bibfield  {title} {\bibinfo
  {title} {{Implementation of efficient quantum search algorithms on NISQ
  computers}},\ }\href {https://doi.org/10.1007%2Fs11128-021-03165-2}
  {\bibfield  {journal} {\bibinfo  {journal} {{Quant. Inf. Proc.}}\ }\textbf
  {\bibinfo {volume} {20}},\ \bibinfo {pages} {233} (\bibinfo {year}
  {2021})}\BibitemShut {NoStop}%
\bibitem [{\citenamefont {Pokharel}\ and\ \citenamefont
  {Lidar}(2022{\natexlab{a}})}]{pokharel2022demonstration}%
  \BibitemOpen
  \bibfield  {author} {\bibinfo {author} {\bibfnamefont {B.}~\bibnamefont
  {Pokharel}}\ and\ \bibinfo {author} {\bibfnamefont {D.~A.}\ \bibnamefont
  {Lidar}},\ }\href {https://arxiv.org/abs/2207.07647} {\bibinfo {title}
  {Demonstration of algorithmic quantum speedup}} (\bibinfo {year}
  {2022}{\natexlab{a}}),\ \Eprint {https://arxiv.org/abs/2207.07647}
  {arXiv:2207.07647 [quant-ph]} \BibitemShut {NoStop}%
\bibitem [{\citenamefont {Pokharel}\ and\ \citenamefont
  {Lidar}(2022{\natexlab{b}})}]{Pokharel:better-than-classical-Grover}%
  \BibitemOpen
  \bibfield  {author} {\bibinfo {author} {\bibfnamefont {B.}~\bibnamefont
  {Pokharel}}\ and\ \bibinfo {author} {\bibfnamefont {D.}~\bibnamefont
  {Lidar}},\ }\href {https://arxiv.org/abs/2211.04543} {\bibinfo {title}
  {{Better-than-classical Grover search via quantum error detection and
  suppression}}} (\bibinfo {year} {2022}{\natexlab{b}}),\ \Eprint
  {https://arxiv.org/abs/2211.04543} {arXiv:2211.04543 [quant-ph]} \BibitemShut
  {NoStop}%
\bibitem [{\citenamefont {Aliferis}\ \emph {et~al.}(2006)\citenamefont
  {Aliferis}, \citenamefont {Gottesman},\ and\ \citenamefont
  {Preskill}}]{Aliferis:05}%
  \BibitemOpen
  \bibfield  {author} {\bibinfo {author} {\bibfnamefont {P.}~\bibnamefont
  {Aliferis}}, \bibinfo {author} {\bibfnamefont {D.}~\bibnamefont
  {Gottesman}},\ and\ \bibinfo {author} {\bibfnamefont {J.}~\bibnamefont
  {Preskill}},\ }\bibfield  {title} {\bibinfo {title} {Quantum accuracy
  threshold for concatenated distance-3 codes},\ }\href
  {http://www.rintonpress.com/xqic6/qic-6-2/097-165.pdf} {\bibfield  {journal}
  {\bibinfo  {journal} {Quantum Inf. Comput.}\ }\textbf {\bibinfo {volume}
  {6}},\ \bibinfo {pages} {97} (\bibinfo {year} {2006})}\BibitemShut {NoStop}%
\bibitem [{\citenamefont {Chao}\ and\ \citenamefont
  {Reichardt}(2018)}]{Chao:2017ab}%
  \BibitemOpen
  \bibfield  {author} {\bibinfo {author} {\bibfnamefont {R.}~\bibnamefont
  {Chao}}\ and\ \bibinfo {author} {\bibfnamefont {B.~W.}\ \bibnamefont
  {Reichardt}},\ }\bibfield  {title} {\bibinfo {title} {Quantum error
  correction with only two extra qubits},\ }\href
  {https://link.aps.org/doi/10.1103/PhysRevLett.121.050502} {\bibfield
  {journal} {\bibinfo  {journal} {Physical Review Letters}\ }\textbf {\bibinfo
  {volume} {121}},\ \bibinfo {pages} {050502} (\bibinfo {year}
  {2018})}\BibitemShut {NoStop}%
\bibitem [{\citenamefont {Campbell}\ \emph {et~al.}(2017)\citenamefont
  {Campbell}, \citenamefont {Terhal},\ and\ \citenamefont
  {Vuillot}}]{Campbell:2017aa}%
  \BibitemOpen
  \bibfield  {author} {\bibinfo {author} {\bibfnamefont {E.~T.}\ \bibnamefont
  {Campbell}}, \bibinfo {author} {\bibfnamefont {B.~M.}\ \bibnamefont
  {Terhal}},\ and\ \bibinfo {author} {\bibfnamefont {C.}~\bibnamefont
  {Vuillot}},\ }\bibfield  {title} {\bibinfo {title} {Roads towards
  fault-tolerant universal quantum computation},\ }\href
  {http://dx.doi.org/10.1038/nature23460} {\bibfield  {journal} {\bibinfo
  {journal} {Nature}\ }\textbf {\bibinfo {volume} {549}},\ \bibinfo {pages}
  {172 EP } (\bibinfo {year} {2017})}\BibitemShut {NoStop}%
\bibitem [{\citenamefont {Souza}(2021)}]{souza2020process}%
  \BibitemOpen
  \bibfield  {author} {\bibinfo {author} {\bibfnamefont {A.~M.}\ \bibnamefont
  {Souza}},\ }\bibfield  {title} {\bibinfo {title} {Process tomography of
  robust dynamical decoupling with superconducting qubits},\ }\href
  {https://doi.org/10.1007/s11128-021-03176-z} {\bibfield  {journal} {\bibinfo
  {journal} {Quantum Information Processing}\ }\textbf {\bibinfo {volume}
  {20}},\ \bibinfo {pages} {237} (\bibinfo {year} {2021})}\BibitemShut
  {NoStop}%
\bibitem [{\citenamefont {Harper}\ \emph {et~al.}(2020)\citenamefont {Harper},
  \citenamefont {Flammia},\ and\ \citenamefont {Wallman}}]{Harper2020}%
  \BibitemOpen
  \bibfield  {author} {\bibinfo {author} {\bibfnamefont {R.}~\bibnamefont
  {Harper}}, \bibinfo {author} {\bibfnamefont {S.~T.}\ \bibnamefont
  {Flammia}},\ and\ \bibinfo {author} {\bibfnamefont {J.~J.}\ \bibnamefont
  {Wallman}},\ }\bibfield  {title} {\bibinfo {title} {Efficient learning of
  quantum noise},\ }\href {https://doi.org/10.1038/s41567-020-0992-8}
  {\bibfield  {journal} {\bibinfo  {journal} {Nature Physics}\ }\textbf
  {\bibinfo {volume} {16}},\ \bibinfo {pages} {1184} (\bibinfo {year}
  {2020})}\BibitemShut {NoStop}%
\bibitem [{\citenamefont {Georgopoulos}\ \emph {et~al.}(2021)\citenamefont
  {Georgopoulos}, \citenamefont {Emary},\ and\ \citenamefont
  {Zuliani}}]{Georgopoulos2021}%
  \BibitemOpen
  \bibfield  {author} {\bibinfo {author} {\bibfnamefont {K.}~\bibnamefont
  {Georgopoulos}}, \bibinfo {author} {\bibfnamefont {C.}~\bibnamefont
  {Emary}},\ and\ \bibinfo {author} {\bibfnamefont {P.}~\bibnamefont
  {Zuliani}},\ }\bibfield  {title} {\bibinfo {title} {Modeling and simulating
  the noisy behavior of near-term quantum computers},\ }\href
  {https://doi.org/10.1103/PhysRevA.104.062432} {\bibfield  {journal} {\bibinfo
   {journal} {Physical Review A}\ }\textbf {\bibinfo {volume} {104}},\ \bibinfo
  {pages} {062432} (\bibinfo {year} {2021})}\BibitemShut {NoStop}%
\bibitem [{\citenamefont {McCourt}\ \emph {et~al.}(2022)\citenamefont
  {McCourt}, \citenamefont {Neill}, \citenamefont {Lee}, \citenamefont
  {Quintana}, \citenamefont {Chen}, \citenamefont {Kelly}, \citenamefont
  {Smelyanskiy}, \citenamefont {Dykman}, \citenamefont {Korotkov},
  \citenamefont {Chuang},\ and\ \citenamefont {Petukhov}}]{McCourt2022}%
  \BibitemOpen
  \bibfield  {author} {\bibinfo {author} {\bibfnamefont {T.}~\bibnamefont
  {McCourt}}, \bibinfo {author} {\bibfnamefont {C.}~\bibnamefont {Neill}},
  \bibinfo {author} {\bibfnamefont {K.}~\bibnamefont {Lee}}, \bibinfo {author}
  {\bibfnamefont {C.}~\bibnamefont {Quintana}}, \bibinfo {author}
  {\bibfnamefont {Y.}~\bibnamefont {Chen}}, \bibinfo {author} {\bibfnamefont
  {J.}~\bibnamefont {Kelly}}, \bibinfo {author} {\bibfnamefont {V.~N.}\
  \bibnamefont {Smelyanskiy}}, \bibinfo {author} {\bibfnamefont {M.~I.}\
  \bibnamefont {Dykman}}, \bibinfo {author} {\bibfnamefont {A.}~\bibnamefont
  {Korotkov}}, \bibinfo {author} {\bibfnamefont {I.~L.}\ \bibnamefont
  {Chuang}},\ and\ \bibinfo {author} {\bibfnamefont {A.~G.}\ \bibnamefont
  {Petukhov}},\ }\href {https://arxiv.org/abs/2201.11173} {\bibinfo {title}
  {Learning noise via dynamical decoupling of entangled qubits}} (\bibinfo
  {year} {2022}),\ \Eprint {https://arxiv.org/abs/2201.11173} {arXiv:2201.11173
  [quant-ph]} \BibitemShut {NoStop}%
\bibitem [{\citenamefont {Viola}\ and\ \citenamefont {Lloyd}(1998)}]{Viola:98}%
  \BibitemOpen
  \bibfield  {author} {\bibinfo {author} {\bibfnamefont {L.}~\bibnamefont
  {Viola}}\ and\ \bibinfo {author} {\bibfnamefont {S.}~\bibnamefont {Lloyd}},\
  }\bibfield  {title} {\bibinfo {title} {Dynamical suppression of decoherence
  in two-state quantum systems},\ }\href
  {https://link.aps.org/doi/10.1103/PhysRevA.58.2733} {\bibfield  {journal}
  {\bibinfo  {journal} {Phys. Rev. A}\ }\textbf {\bibinfo {volume} {58}},\
  \bibinfo {pages} {2733} (\bibinfo {year} {1998})}\BibitemShut {NoStop}%
\bibitem [{\citenamefont {Duan}\ and\ \citenamefont {Guo}(1999)}]{Duan:98e}%
  \BibitemOpen
  \bibfield  {author} {\bibinfo {author} {\bibfnamefont {L.-M.}\ \bibnamefont
  {Duan}}\ and\ \bibinfo {author} {\bibfnamefont {G.-C.}\ \bibnamefont {Guo}},\
  }\bibfield  {title} {\bibinfo {title} {Suppressing environmental noise in
  quantum computation through pulse control},\ }\href
  {https://www.sciencedirect.com/science/article/pii/S0375960199005927}
  {\bibfield  {journal} {\bibinfo  {journal} {Physics Letters A}\ }\textbf
  {\bibinfo {volume} {261}},\ \bibinfo {pages} {139} (\bibinfo {year}
  {1999})}\BibitemShut {NoStop}%
\bibitem [{\citenamefont {Vitali}\ and\ \citenamefont
  {Tombesi}(1999)}]{Vitali:99}%
  \BibitemOpen
  \bibfield  {author} {\bibinfo {author} {\bibfnamefont {D.}~\bibnamefont
  {Vitali}}\ and\ \bibinfo {author} {\bibfnamefont {P.}~\bibnamefont
  {Tombesi}},\ }\bibfield  {title} {\bibinfo {title} {Using parity kicks for
  decoherence control},\ }\href
  {https://link.aps.org/doi/10.1103/PhysRevA.59.4178} {\bibfield  {journal}
  {\bibinfo  {journal} {Physical Review A}\ }\textbf {\bibinfo {volume} {59}},\
  \bibinfo {pages} {4178} (\bibinfo {year} {1999})}\BibitemShut {NoStop}%
\bibitem [{\citenamefont {Zanardi}(1999)}]{Zanardi:1999fk}%
  \BibitemOpen
  \bibfield  {author} {\bibinfo {author} {\bibfnamefont {P.}~\bibnamefont
  {Zanardi}},\ }\bibfield  {title} {\bibinfo {title} {Symmetrizing
  evolutions},\ }\href
  {http://www.sciencedirect.com/science/article/pii/S0375960199003655}
  {\bibfield  {journal} {\bibinfo  {journal} {Physics Letters A}\ }\textbf
  {\bibinfo {volume} {258}},\ \bibinfo {pages} {77} (\bibinfo {year}
  {1999})}\BibitemShut {NoStop}%
\bibitem [{\citenamefont {Viola}\ \emph {et~al.}(1999)\citenamefont {Viola},
  \citenamefont {Knill},\ and\ \citenamefont {Lloyd}}]{Viola:99}%
  \BibitemOpen
  \bibfield  {author} {\bibinfo {author} {\bibfnamefont {L.}~\bibnamefont
  {Viola}}, \bibinfo {author} {\bibfnamefont {E.}~\bibnamefont {Knill}},\ and\
  \bibinfo {author} {\bibfnamefont {S.}~\bibnamefont {Lloyd}},\ }\bibfield
  {title} {\bibinfo {title} {Dynamical decoupling of open quantum systems},\
  }\href {http://link.aps.org/doi/10.1103/PhysRevLett.82.2417} {\bibfield
  {journal} {\bibinfo  {journal} {Physical Review Letters}\ }\textbf {\bibinfo
  {volume} {82}},\ \bibinfo {pages} {2417} (\bibinfo {year}
  {1999})}\BibitemShut {NoStop}%
\bibitem [{\citenamefont {Bylander}\ \emph {et~al.}(2011)\citenamefont
  {Bylander}, \citenamefont {Gustavsson}, \citenamefont {Yan}, \citenamefont
  {Yoshihara}, \citenamefont {Harrabi}, \citenamefont {Fitch}, \citenamefont
  {Cory}, \citenamefont {Nakamura}, \citenamefont {Tsai},\ and\ \citenamefont
  {Oliver}}]{Byl11a}%
  \BibitemOpen
  \bibfield  {author} {\bibinfo {author} {\bibfnamefont {J.}~\bibnamefont
  {Bylander}}, \bibinfo {author} {\bibfnamefont {S.}~\bibnamefont
  {Gustavsson}}, \bibinfo {author} {\bibfnamefont {F.}~\bibnamefont {Yan}},
  \bibinfo {author} {\bibfnamefont {F.}~\bibnamefont {Yoshihara}}, \bibinfo
  {author} {\bibfnamefont {K.}~\bibnamefont {Harrabi}}, \bibinfo {author}
  {\bibfnamefont {G.}~\bibnamefont {Fitch}}, \bibinfo {author} {\bibfnamefont
  {D.}~\bibnamefont {Cory}}, \bibinfo {author} {\bibfnamefont {Y.}~\bibnamefont
  {Nakamura}}, \bibinfo {author} {\bibfnamefont {J.}~\bibnamefont {Tsai}},\
  and\ \bibinfo {author} {\bibfnamefont {W.}~\bibnamefont {Oliver}},\
  }\bibfield  {title} {\bibinfo {title} {Noise spectroscopy through dynamical
  decoupling with a superconducting flux qubit},\ }\href
  {https://www.nature.com/articles/nphys1994} {\bibfield  {journal} {\bibinfo
  {journal} {Nature Phys.}\ }\textbf {\bibinfo {volume} {7}},\ \bibinfo {pages}
  {565} (\bibinfo {year} {2011})}\BibitemShut {NoStop}%
\bibitem [{\citenamefont {Motzoi}\ \emph {et~al.}(2009)\citenamefont {Motzoi},
  \citenamefont {Gambetta}, \citenamefont {Rebentrost},\ and\ \citenamefont
  {Wilhelm}}]{Motzoi2009}%
  \BibitemOpen
  \bibfield  {author} {\bibinfo {author} {\bibfnamefont {F.}~\bibnamefont
  {Motzoi}}, \bibinfo {author} {\bibfnamefont {J.~M.}\ \bibnamefont
  {Gambetta}}, \bibinfo {author} {\bibfnamefont {P.}~\bibnamefont
  {Rebentrost}},\ and\ \bibinfo {author} {\bibfnamefont {F.~K.}\ \bibnamefont
  {Wilhelm}},\ }\bibfield  {title} {\bibinfo {title} {Simple pulses for
  elimination of leakage in weakly nonlinear qubits},\ }\href
  {https://doi.org/10.1103/PhysRevLett.103.110501} {\bibfield  {journal}
  {\bibinfo  {journal} {Phys. Rev. Lett.}\ }\textbf {\bibinfo {volume} {103}},\
  \bibinfo {pages} {110501} (\bibinfo {year} {2009})}\BibitemShut {NoStop}%
\bibitem [{\citenamefont {Freeman}(1998)}]{freeman1998spin}%
  \BibitemOpen
  \bibfield  {author} {\bibinfo {author} {\bibfnamefont {R.}~\bibnamefont
  {Freeman}},\ }\href@noop {} {\emph {\bibinfo {title} {Spin choreography}}}\
  (\bibinfo  {publisher} {Oxford University Press Oxford},\ \bibinfo {year}
  {1998})\BibitemShut {NoStop}%
\bibitem [{\citenamefont {Motzoi}\ and\ \citenamefont
  {Wilhelm}(2013)}]{Motzoi2013-sx}%
  \BibitemOpen
  \bibfield  {author} {\bibinfo {author} {\bibfnamefont {F.}~\bibnamefont
  {Motzoi}}\ and\ \bibinfo {author} {\bibfnamefont {F.~K.}\ \bibnamefont
  {Wilhelm}},\ }\bibfield  {title} {\bibinfo {title} {Improving frequency
  selection of driven pulses using derivative-based transition suppression},\
  }\href {https://doi.org/10.1103/PhysRevA.88.062318} {\bibfield  {journal}
  {\bibinfo  {journal} {Phys. Rev. A}\ }\textbf {\bibinfo {volume} {88}},\
  \bibinfo {pages} {062318} (\bibinfo {year} {2013})}\BibitemShut {NoStop}%
\bibitem [{\citenamefont {Bowdrey}\ \emph {et~al.}(2002)\citenamefont
  {Bowdrey}, \citenamefont {Oi}, \citenamefont {Short}, \citenamefont
  {Banaszek},\ and\ \citenamefont {Jones}}]{Bowdrey:2002aa}%
  \BibitemOpen
  \bibfield  {author} {\bibinfo {author} {\bibfnamefont {M.~D.}\ \bibnamefont
  {Bowdrey}}, \bibinfo {author} {\bibfnamefont {D.~K.~L.}\ \bibnamefont {Oi}},
  \bibinfo {author} {\bibfnamefont {A.~J.}\ \bibnamefont {Short}}, \bibinfo
  {author} {\bibfnamefont {K.}~\bibnamefont {Banaszek}},\ and\ \bibinfo
  {author} {\bibfnamefont {J.~A.}\ \bibnamefont {Jones}},\ }\bibfield  {title}
  {\bibinfo {title} {Fidelity of single qubit maps},\ }\href
  {https://doi.org/https://doi.org/10.1016/S0375-9601(02)00069-5} {\bibfield
  {journal} {\bibinfo  {journal} {Physics Letters A}\ }\textbf {\bibinfo
  {volume} {294}},\ \bibinfo {pages} {258} (\bibinfo {year}
  {2002})}\BibitemShut {NoStop}%
\bibitem [{\citenamefont {McKay}\ \emph {et~al.}(2017)\citenamefont {McKay},
  \citenamefont {Wood}, \citenamefont {Sheldon}, \citenamefont {Chow},\ and\
  \citenamefont {Gambetta}}]{Mckay2017}%
  \BibitemOpen
  \bibfield  {author} {\bibinfo {author} {\bibfnamefont {D.~C.}\ \bibnamefont
  {McKay}}, \bibinfo {author} {\bibfnamefont {C.~J.}\ \bibnamefont {Wood}},
  \bibinfo {author} {\bibfnamefont {S.}~\bibnamefont {Sheldon}}, \bibinfo
  {author} {\bibfnamefont {J.~M.}\ \bibnamefont {Chow}},\ and\ \bibinfo
  {author} {\bibfnamefont {J.~M.}\ \bibnamefont {Gambetta}},\ }\bibfield
  {title} {\bibinfo {title} {Efficient $z$ gates for quantum computing},\
  }\href {https://doi.org/10.1103/PhysRevA.96.022330} {\bibfield  {journal}
  {\bibinfo  {journal} {Phys. Rev. A}\ }\textbf {\bibinfo {volume} {96}},\
  \bibinfo {pages} {022330} (\bibinfo {year} {2017})}\BibitemShut {NoStop}%
\bibitem [{\citenamefont {McKay}\ \emph {et~al.}(2018)\citenamefont {McKay},
  \citenamefont {Alexander}, \citenamefont {Bello}, \citenamefont {Biercuk},
  \citenamefont {Bishop}, \citenamefont {Chen}, \citenamefont {Chow},
  \citenamefont {C{\'o}rcoles}, \citenamefont {Egger}, \citenamefont {Filipp},
  \citenamefont {Gomez}, \citenamefont {Hush}, \citenamefont {Javadi-Abhari},
  \citenamefont {Moreda}, \citenamefont {Nation}, \citenamefont {Paulovicks},
  \citenamefont {Winston}, \citenamefont {Wood}, \citenamefont {Wootton},\ and\
  \citenamefont {Gambetta}}]{Mckay2018}%
  \BibitemOpen
  \bibfield  {author} {\bibinfo {author} {\bibfnamefont {D.~C.}\ \bibnamefont
  {McKay}}, \bibinfo {author} {\bibfnamefont {T.}~\bibnamefont {Alexander}},
  \bibinfo {author} {\bibfnamefont {L.}~\bibnamefont {Bello}}, \bibinfo
  {author} {\bibfnamefont {M.~J.}\ \bibnamefont {Biercuk}}, \bibinfo {author}
  {\bibfnamefont {L.}~\bibnamefont {Bishop}}, \bibinfo {author} {\bibfnamefont
  {J.}~\bibnamefont {Chen}}, \bibinfo {author} {\bibfnamefont {J.~M.}\
  \bibnamefont {Chow}}, \bibinfo {author} {\bibfnamefont {A.~D.}\ \bibnamefont
  {C{\'o}rcoles}}, \bibinfo {author} {\bibfnamefont {D.}~\bibnamefont {Egger}},
  \bibinfo {author} {\bibfnamefont {S.}~\bibnamefont {Filipp}}, \bibinfo
  {author} {\bibfnamefont {J.}~\bibnamefont {Gomez}}, \bibinfo {author}
  {\bibfnamefont {M.}~\bibnamefont {Hush}}, \bibinfo {author} {\bibfnamefont
  {A.}~\bibnamefont {Javadi-Abhari}}, \bibinfo {author} {\bibfnamefont
  {D.}~\bibnamefont {Moreda}}, \bibinfo {author} {\bibfnamefont
  {P.}~\bibnamefont {Nation}}, \bibinfo {author} {\bibfnamefont
  {B.}~\bibnamefont {Paulovicks}}, \bibinfo {author} {\bibfnamefont
  {E.}~\bibnamefont {Winston}}, \bibinfo {author} {\bibfnamefont {C.~J.}\
  \bibnamefont {Wood}}, \bibinfo {author} {\bibfnamefont {J.}~\bibnamefont
  {Wootton}},\ and\ \bibinfo {author} {\bibfnamefont {J.~M.}\ \bibnamefont
  {Gambetta}},\ }\href@noop {} {\bibinfo {title} {Qiskit backend specifications
  for openqasm and openpulse experiments}} (\bibinfo {year} {2018}),\ \Eprint
  {https://arxiv.org/abs/1809.03452} {arXiv:1809.03452 [quant-ph]} \BibitemShut
  {NoStop}%
\bibitem [{\citenamefont {Babu}\ \emph {et~al.}(2021)\citenamefont {Babu},
  \citenamefont {Tuorila},\ and\ \citenamefont {Ala-Nissila}}]{Babu2020}%
  \BibitemOpen
  \bibfield  {author} {\bibinfo {author} {\bibfnamefont {A.~P.}\ \bibnamefont
  {Babu}}, \bibinfo {author} {\bibfnamefont {J.}~\bibnamefont {Tuorila}},\ and\
  \bibinfo {author} {\bibfnamefont {T.}~\bibnamefont {Ala-Nissila}},\
  }\bibfield  {title} {\bibinfo {title} {State leakage during fast decay and
  control of a superconducting transmon qubit},\ }\href
  {https://doi.org/10.1038/s41534-020-00357-z} {\bibfield  {journal} {\bibinfo
  {journal} {npj Quantum Information}\ }\textbf {\bibinfo {volume} {7}},\
  \bibinfo {pages} {30} (\bibinfo {year} {2021})}\BibitemShut {NoStop}%
\bibitem [{\citenamefont {Gambetta}\ \emph {et~al.}(2011)\citenamefont
  {Gambetta}, \citenamefont {Motzoi}, \citenamefont {Merkel},\ and\
  \citenamefont {Wilhelm}}]{Gambetta2011}%
  \BibitemOpen
  \bibfield  {author} {\bibinfo {author} {\bibfnamefont {J.~M.}\ \bibnamefont
  {Gambetta}}, \bibinfo {author} {\bibfnamefont {F.}~\bibnamefont {Motzoi}},
  \bibinfo {author} {\bibfnamefont {S.~T.}\ \bibnamefont {Merkel}},\ and\
  \bibinfo {author} {\bibfnamefont {F.~K.}\ \bibnamefont {Wilhelm}},\
  }\bibfield  {title} {\bibinfo {title} {Analytic control methods for
  high-fidelity unitary operations in a weakly nonlinear oscillator},\ }\href
  {https://doi.org/10.1103/PhysRevA.83.012308} {\bibfield  {journal} {\bibinfo
  {journal} {Phys. Rev. A}\ }\textbf {\bibinfo {volume} {83}},\ \bibinfo
  {pages} {012308} (\bibinfo {year} {2011})}\BibitemShut {NoStop}%
\bibitem [{\citenamefont {Chen}\ \emph {et~al.}(2016)\citenamefont {Chen},
  \citenamefont {Kelly}, \citenamefont {Quintana}, \citenamefont {Barends},
  \citenamefont {Campbell}, \citenamefont {Chen}, \citenamefont {Chiaro},
  \citenamefont {Dunsworth}, \citenamefont {Fowler}, \citenamefont {Lucero},
  \citenamefont {Jeffrey}, \citenamefont {Megrant}, \citenamefont {Mutus},
  \citenamefont {Neeley}, \citenamefont {Neill}, \citenamefont {O'Malley},
  \citenamefont {Roushan}, \citenamefont {Sank}, \citenamefont {Vainsencher},
  \citenamefont {Wenner}, \citenamefont {White}, \citenamefont {Korotkov},\
  and\ \citenamefont {Martinis}}]{Zijun2016}%
  \BibitemOpen
  \bibfield  {author} {\bibinfo {author} {\bibfnamefont {Z.}~\bibnamefont
  {Chen}}, \bibinfo {author} {\bibfnamefont {J.}~\bibnamefont {Kelly}},
  \bibinfo {author} {\bibfnamefont {C.}~\bibnamefont {Quintana}}, \bibinfo
  {author} {\bibfnamefont {R.}~\bibnamefont {Barends}}, \bibinfo {author}
  {\bibfnamefont {B.}~\bibnamefont {Campbell}}, \bibinfo {author}
  {\bibfnamefont {Y.}~\bibnamefont {Chen}}, \bibinfo {author} {\bibfnamefont
  {B.}~\bibnamefont {Chiaro}}, \bibinfo {author} {\bibfnamefont
  {A.}~\bibnamefont {Dunsworth}}, \bibinfo {author} {\bibfnamefont {A.~G.}\
  \bibnamefont {Fowler}}, \bibinfo {author} {\bibfnamefont {E.}~\bibnamefont
  {Lucero}}, \bibinfo {author} {\bibfnamefont {E.}~\bibnamefont {Jeffrey}},
  \bibinfo {author} {\bibfnamefont {A.}~\bibnamefont {Megrant}}, \bibinfo
  {author} {\bibfnamefont {J.}~\bibnamefont {Mutus}}, \bibinfo {author}
  {\bibfnamefont {M.}~\bibnamefont {Neeley}}, \bibinfo {author} {\bibfnamefont
  {C.}~\bibnamefont {Neill}}, \bibinfo {author} {\bibfnamefont {P.~J.~J.}\
  \bibnamefont {O'Malley}}, \bibinfo {author} {\bibfnamefont {P.}~\bibnamefont
  {Roushan}}, \bibinfo {author} {\bibfnamefont {D.}~\bibnamefont {Sank}},
  \bibinfo {author} {\bibfnamefont {A.}~\bibnamefont {Vainsencher}}, \bibinfo
  {author} {\bibfnamefont {J.}~\bibnamefont {Wenner}}, \bibinfo {author}
  {\bibfnamefont {T.~C.}\ \bibnamefont {White}}, \bibinfo {author}
  {\bibfnamefont {A.~N.}\ \bibnamefont {Korotkov}},\ and\ \bibinfo {author}
  {\bibfnamefont {J.~M.}\ \bibnamefont {Martinis}},\ }\bibfield  {title}
  {\bibinfo {title} {Measuring and suppressing quantum state leakage in a
  superconducting qubit},\ }\href
  {https://doi.org/10.1103/PhysRevLett.116.020501} {\bibfield  {journal}
  {\bibinfo  {journal} {Phys. Rev. Lett.}\ }\textbf {\bibinfo {volume} {116}},\
  \bibinfo {pages} {020501} (\bibinfo {year} {2016})}\BibitemShut {NoStop}%
\bibitem [{\citenamefont {Emerson}\ \emph {et~al.}(2005)\citenamefont
  {Emerson}, \citenamefont {Alicki},\ and\ \citenamefont
  {{\.Z}yczkowski}}]{Emerson:2005sf}%
  \BibitemOpen
  \bibfield  {author} {\bibinfo {author} {\bibfnamefont {J.}~\bibnamefont
  {Emerson}}, \bibinfo {author} {\bibfnamefont {R.}~\bibnamefont {Alicki}},\
  and\ \bibinfo {author} {\bibfnamefont {K.}~\bibnamefont {{\.Z}yczkowski}},\
  }\bibfield  {title} {\bibinfo {title} {Scalable noise estimation with random
  unitary operators},\ }\href {http://stacks.iop.org/1464-4266/7/i=10/a=021}
  {\bibfield  {journal} {\bibinfo  {journal} {Journal of Optics B: Quantum and
  Semiclassical Optics}\ }\textbf {\bibinfo {volume} {7}},\ \bibinfo {pages}
  {S347} (\bibinfo {year} {2005})}\BibitemShut {NoStop}%
\bibitem [{\citenamefont {Magesan}\ \emph {et~al.}(2011)\citenamefont
  {Magesan}, \citenamefont {Gambetta},\ and\ \citenamefont
  {Emerson}}]{Magesan:2011kx}%
  \BibitemOpen
  \bibfield  {author} {\bibinfo {author} {\bibfnamefont {E.}~\bibnamefont
  {Magesan}}, \bibinfo {author} {\bibfnamefont {J.~M.}\ \bibnamefont
  {Gambetta}},\ and\ \bibinfo {author} {\bibfnamefont {J.}~\bibnamefont
  {Emerson}},\ }\bibfield  {title} {\bibinfo {title} {Scalable and robust
  randomized benchmarking of quantum processes},\ }\href
  {http://link.aps.org/doi/10.1103/PhysRevLett.106.180504} {\bibfield
  {journal} {\bibinfo  {journal} {Phys. Rev. Lett.}\ }\textbf {\bibinfo
  {volume} {106}},\ \bibinfo {pages} {180504} (\bibinfo {year}
  {2011})}\BibitemShut {NoStop}%
\bibitem [{\citenamefont {Proctor}\ \emph {et~al.}(2017)\citenamefont
  {Proctor}, \citenamefont {Rudinger}, \citenamefont {Young}, \citenamefont
  {Sarovar},\ and\ \citenamefont {Blume-Kohout}}]{Proctor:2017uq}%
  \BibitemOpen
  \bibfield  {author} {\bibinfo {author} {\bibfnamefont {T.}~\bibnamefont
  {Proctor}}, \bibinfo {author} {\bibfnamefont {K.}~\bibnamefont {Rudinger}},
  \bibinfo {author} {\bibfnamefont {K.}~\bibnamefont {Young}}, \bibinfo
  {author} {\bibfnamefont {M.}~\bibnamefont {Sarovar}},\ and\ \bibinfo {author}
  {\bibfnamefont {R.}~\bibnamefont {Blume-Kohout}},\ }\bibfield  {title}
  {\bibinfo {title} {What randomized benchmarking actually measures},\ }\href
  {https://doi.org/10.1103/PhysRevLett.119.130502} {\bibfield  {journal}
  {\bibinfo  {journal} {Physical Review Letters}\ }\textbf {\bibinfo {volume}
  {119}},\ \bibinfo {pages} {130502} (\bibinfo {year} {2017})}\BibitemShut
  {NoStop}%
\bibitem [{\citenamefont {Quintana}\ \emph {et~al.}(2017)\citenamefont
  {Quintana}, \citenamefont {Chen}, \citenamefont {Sank}, \citenamefont
  {Petukhov}, \citenamefont {White}, \citenamefont {Kafri}, \citenamefont
  {Chiaro}, \citenamefont {Megrant}, \citenamefont {Barends}, \citenamefont
  {Campbell}, \citenamefont {Chen}, \citenamefont {Dunsworth}, \citenamefont
  {Fowler}, \citenamefont {Graff}, \citenamefont {Jeffrey}, \citenamefont
  {Kelly}, \citenamefont {Lucero}, \citenamefont {Mutus}, \citenamefont
  {Neeley}, \citenamefont {Neill}, \citenamefont {O'Malley}, \citenamefont
  {Roushan}, \citenamefont {Shabani}, \citenamefont {Smelyanskiy},
  \citenamefont {Vainsencher}, \citenamefont {Wenner}, \citenamefont {Neven},\
  and\ \citenamefont {Martinis}}]{Quintana:2017aa}%
  \BibitemOpen
  \bibfield  {author} {\bibinfo {author} {\bibfnamefont {C.~M.}\ \bibnamefont
  {Quintana}}, \bibinfo {author} {\bibfnamefont {Y.}~\bibnamefont {Chen}},
  \bibinfo {author} {\bibfnamefont {D.}~\bibnamefont {Sank}}, \bibinfo {author}
  {\bibfnamefont {A.~G.}\ \bibnamefont {Petukhov}}, \bibinfo {author}
  {\bibfnamefont {T.~C.}\ \bibnamefont {White}}, \bibinfo {author}
  {\bibfnamefont {D.}~\bibnamefont {Kafri}}, \bibinfo {author} {\bibfnamefont
  {B.}~\bibnamefont {Chiaro}}, \bibinfo {author} {\bibfnamefont
  {A.}~\bibnamefont {Megrant}}, \bibinfo {author} {\bibfnamefont
  {R.}~\bibnamefont {Barends}}, \bibinfo {author} {\bibfnamefont
  {B.}~\bibnamefont {Campbell}}, \bibinfo {author} {\bibfnamefont
  {Z.}~\bibnamefont {Chen}}, \bibinfo {author} {\bibfnamefont {A.}~\bibnamefont
  {Dunsworth}}, \bibinfo {author} {\bibfnamefont {A.~G.}\ \bibnamefont
  {Fowler}}, \bibinfo {author} {\bibfnamefont {R.}~\bibnamefont {Graff}},
  \bibinfo {author} {\bibfnamefont {E.}~\bibnamefont {Jeffrey}}, \bibinfo
  {author} {\bibfnamefont {J.}~\bibnamefont {Kelly}}, \bibinfo {author}
  {\bibfnamefont {E.}~\bibnamefont {Lucero}}, \bibinfo {author} {\bibfnamefont
  {J.~Y.}\ \bibnamefont {Mutus}}, \bibinfo {author} {\bibfnamefont
  {M.}~\bibnamefont {Neeley}}, \bibinfo {author} {\bibfnamefont
  {C.}~\bibnamefont {Neill}}, \bibinfo {author} {\bibfnamefont {P.~J.~J.}\
  \bibnamefont {O'Malley}}, \bibinfo {author} {\bibfnamefont {P.}~\bibnamefont
  {Roushan}}, \bibinfo {author} {\bibfnamefont {A.}~\bibnamefont {Shabani}},
  \bibinfo {author} {\bibfnamefont {V.~N.}\ \bibnamefont {Smelyanskiy}},
  \bibinfo {author} {\bibfnamefont {A.}~\bibnamefont {Vainsencher}}, \bibinfo
  {author} {\bibfnamefont {J.}~\bibnamefont {Wenner}}, \bibinfo {author}
  {\bibfnamefont {H.}~\bibnamefont {Neven}},\ and\ \bibinfo {author}
  {\bibfnamefont {J.~M.}\ \bibnamefont {Martinis}},\ }\bibfield  {title}
  {\bibinfo {title} {Observation of classical-quantum crossover of $1/f$ flux
  noise and its paramagnetic temperature dependence},\ }\href
  {https://doi.org/10.1103/PhysRevLett.118.057702} {\bibfield  {journal}
  {\bibinfo  {journal} {Physical Review Letters}\ }\textbf {\bibinfo {volume}
  {118}},\ \bibinfo {pages} {057702} (\bibinfo {year} {2017})}\BibitemShut
  {NoStop}%
\bibitem [{\citenamefont {Paladino}\ \emph {et~al.}(2014)\citenamefont
  {Paladino}, \citenamefont {Galperin}, \citenamefont {Falci},\ and\
  \citenamefont {Altshuler}}]{RevModPhys.86.361}%
  \BibitemOpen
  \bibfield  {author} {\bibinfo {author} {\bibfnamefont {E.}~\bibnamefont
  {Paladino}}, \bibinfo {author} {\bibfnamefont {Y.~M.}\ \bibnamefont
  {Galperin}}, \bibinfo {author} {\bibfnamefont {G.}~\bibnamefont {Falci}},\
  and\ \bibinfo {author} {\bibfnamefont {B.~L.}\ \bibnamefont {Altshuler}},\
  }\bibfield  {title} {\bibinfo {title} {$1/f$ noise: Implications for
  solid-state quantum information},\ }\href
  {http://link.aps.org/doi/10.1103/RevModPhys.86.361} {\bibfield  {journal}
  {\bibinfo  {journal} {Rev. Mod. Phys.}\ }\textbf {\bibinfo {volume} {86}},\
  \bibinfo {pages} {361} (\bibinfo {year} {2014})}\BibitemShut {NoStop}%
\bibitem [{\citenamefont {Yip}(2021)}]{KaWa-Yip-PhDthesis}%
  \BibitemOpen
  \bibfield  {author} {\bibinfo {author} {\bibfnamefont {K.~W.}\ \bibnamefont
  {Yip}},\ }\emph {\bibinfo {title} {Open-system modeling of quantum annealing:
  theory and applications}},\ \href {http://arxiv.org/abs/2107.07231} {Ph.D.
  thesis},\ \bibinfo  {school} {{University of Southern California}} (\bibinfo
  {year} {2021})\BibitemShut {NoStop}%
\bibitem [{\citenamefont {Tripathi}\ \emph {et~al.}(2022)\citenamefont
  {Tripathi}, \citenamefont {Chen}, \citenamefont {Khezri}, \citenamefont
  {Yip}, \citenamefont {Levenson-Falk},\ and\ \citenamefont
  {Lidar}}]{tripathi2021suppression}%
  \BibitemOpen
  \bibfield  {author} {\bibinfo {author} {\bibfnamefont {V.}~\bibnamefont
  {Tripathi}}, \bibinfo {author} {\bibfnamefont {H.}~\bibnamefont {Chen}},
  \bibinfo {author} {\bibfnamefont {M.}~\bibnamefont {Khezri}}, \bibinfo
  {author} {\bibfnamefont {K.-W.}\ \bibnamefont {Yip}}, \bibinfo {author}
  {\bibfnamefont {E.~M.}\ \bibnamefont {Levenson-Falk}},\ and\ \bibinfo
  {author} {\bibfnamefont {D.~A.}\ \bibnamefont {Lidar}},\ }\bibfield  {title}
  {\bibinfo {title} {Suppression of crosstalk in superconducting qubits using
  dynamical decoupling},\ }\href
  {https://doi.org/10.1103/PhysRevApplied.18.024068} {\bibfield  {journal}
  {\bibinfo  {journal} {Physical Review Applied}\ }\textbf {\bibinfo {volume}
  {18}},\ \bibinfo {pages} {024068} (\bibinfo {year} {2022})}\BibitemShut
  {NoStop}%
\bibitem [{\citenamefont {Redfield}(1965)}]{Redfield:66}%
  \BibitemOpen
  \bibfield  {author} {\bibinfo {author} {\bibfnamefont {A.~G.}\ \bibnamefont
  {Redfield}},\ }\bibfield  {title} {\bibinfo {title} {The theory of relaxation
  processes},\ }in\ \href
  {http://www.sciencedirect.com/science/article/pii/B9781483231143500076}
  {\emph {\bibinfo {booktitle} {Advances in Magnetic and Optical Resonance}}},\
  Vol.~\bibinfo {volume} {1},\ \bibinfo {editor} {edited by\ \bibinfo {editor}
  {\bibfnamefont {J.~S.}\ \bibnamefont {Waugh}}}\ (\bibinfo  {publisher}
  {Academic Press},\ \bibinfo {year} {1965})\ pp.\ \bibinfo {pages}
  {1--32}\BibitemShut {NoStop}%
\bibitem [{\citenamefont {Chen}\ and\ \citenamefont
  {Lidar}(2022)}]{chen2020hoqst}%
  \BibitemOpen
  \bibfield  {author} {\bibinfo {author} {\bibfnamefont {H.}~\bibnamefont
  {Chen}}\ and\ \bibinfo {author} {\bibfnamefont {D.~A.}\ \bibnamefont
  {Lidar}},\ }\bibfield  {title} {\bibinfo {title} {Hamiltonian open quantum
  system toolkit},\ }\href {https://doi.org/10.1038/s42005-022-00887-2}
  {\bibfield  {journal} {\bibinfo  {journal} {Communications Physics}\ }\textbf
  {\bibinfo {volume} {5}},\ \bibinfo {pages} {112} (\bibinfo {year}
  {2022})}\BibitemShut {NoStop}%
\bibitem{IBMQ-U3}
{qiskit.circuit.library.U3Gate} [online].
\newblock 2021.
\newblock URL:
  \url{https://qiskit.org/documentation/stubs/qiskit.circuit.library.U3Gate.html}.

  
\bibitem [{\citenamefont {Mozgunov}\ and\ \citenamefont
  {Lidar}(2020)}]{Mozgunov:2019aa}%
  \BibitemOpen
  \bibfield  {author} {\bibinfo {author} {\bibfnamefont {E.}~\bibnamefont
  {Mozgunov}}\ and\ \bibinfo {author} {\bibfnamefont {D.}~\bibnamefont
  {Lidar}},\ }\bibfield  {title} {\bibinfo {title} {Completely positive master
  equation for arbitrary driving and small level spacing},\ }\href
  {https://doi.org/10.22331/q-2020-02-06-227} {\bibfield  {journal} {\bibinfo
  {journal} {{Quantum}}\ }\textbf {\bibinfo {volume} {4}},\ \bibinfo {pages}
  {227} (\bibinfo {year} {2020})}\BibitemShut {NoStop}%
\bibitem [{\citenamefont {Khezri}(2018)}]{Khezri_thesis}%
  \BibitemOpen
  \bibfield  {author} {\bibinfo {author} {\bibfnamefont {M.}~\bibnamefont
  {Khezri}},\ }\emph {\bibinfo {title} {{Dispersive Measurement of
  Superconducting Qubits}}},\ \href {https://escholarship.org/uc/item/96x1x5rk}
  {Ph.D. thesis},\ \bibinfo  {school} {University of California, Riverside}
  (\bibinfo {year} {2018})\BibitemShut {NoStop}%
\end{thebibliography}

%

\end{document}